\documentclass[letterpaper,12pt]{article}\usepackage{test,graphicx,subcaption,caption,framed}
\usepackage[colorlinks,citecolor=blue,urlcolor=magenta]{hyperref}

\usepackage[style=authoryear-comp, 
sorting=ynt, 
dashed=false, 
maxcitenames=2, 
maxbibnames=10,   
uniquelist=false,
uniquename=init,
natbib, 
date=year 
]{biblatex}
\DeclareFieldFormat{pages}{#1} 
\renewbibmacro{in:}{\ifentrytype{article}{}{\printtext{\bibstring{in}\intitlepunct}}} 

\usepackage[inline,shortlabels]{enumitem}
\setlist[enumerate,1]{label=(\roman*)}
\usepackage{booktabs}

\usepackage[left=1in, right=1in, top=1.5in, bottom=1.5in]{geometry}
\usepackage{setspace}
\usepackage{dcolumn}
\newcolumntype{d}[1]{D{.}{.}{#1}} 
\usepackage[official]{eurosym} 

\usepackage{placeins} 

\newcommand{\lr}[1]{\left[#1 \right]} 
\newcommand{\lro}[1]{\left(#1 \right)} 

\newcommand{\cA}{\mathcal{A}}
\newcommand{\cZ}{\mathcal{Z}}

\newcommand{\cF}{\mathcal{F}}
\newcommand{\cM}{\mathcal{M}}

\newcommand{\pmax}[1]{\lro{#1}^+}

\newcommand{\amin}{a_{\min}}
\newcommand{\amax}{a_{\max}}

\newcommand{\Qdollar}{Q^{\$}}
\newcommand{\ttt}{t \to T}
\newcommand{\PL}{\mathrm{PL}}

\title{Recovering Risk-Neutral Moments from Options}
\author{Tjeerd De Vries\thanks{Department of Finance, HEC Paris. Email: \href{mailto:de-vries@hec.fr}{de-vries@hec.fr}. I thank Nikolay Kudrin, Evgenii Vladimirov,  Olivier Scaillet, Casper de Vries, Irina Zviadadze, Gustavo Freire, Ian Martin, Bruno Biais, Caio Almeida and participants of the HEC Brownbag seminar, NV FEB conference, SGF conference and SoFiE conference for helpful feedback.}}

\begin{document}
\maketitle

\begin{abstract}
Extracting risk-neutral dependence from option prices has remained an open problem since \citet{ross1976options}. We propose a projection estimator that uses portfolios of observed options to approximate payoffs depending on multiple assets. The method delivers estimates of risk-neutral dependence in incomplete markets, improves univariate estimates, and yields a finite-sample error bound. Applying the method to two unexpected Swiss National Bank announcements about the EUR/CHF floor, we find that dependence accounts for two-thirds of the change in the probability that EUR/CHF and USD/CHF both fall sharply. A policy targeting EUR/CHF thus reshaped dependence in the CHF market.

\medskip

\noindent
\textbf{Keywords:} risk-neutral moments, market completeness, FX options, joint tail risk.

\medskip

\noindent
\textbf{JEL codes:} G12, G13, C14, F31.

\end{abstract}

\section{Introduction}
Option prices reveal the risk-neutral distribution of individual asset returns. This fundamental insight, due to \citet{ross1976options} and \citet{breeden1978prices}, underpins a wide range of applications, from the construction of the VIX to estimates of the equity premium, risk aversion, and pricing kernels.\footnote{See, for example, \citet{bates1991crash,martin2017expected,kremens2019theory,schneider2019almost,andersen2017short} for estimating the equity premium; \citet{britten2000option,Carr_Madan_2001,carr2009variance,bollerslev2009expected,jiang2005model} for volatility estimation; and \citet{bakshi2003stock,chabi2020conditional,kozhan2013skew} for higher-order moment estimation.} Yet for many questions in finance, the object of interest is not the marginal distribution of a single return but risk-neutral dependence across returns. Under log utility, for example, the risk-neutral covariance between a stock and the market enters directly into the stock's expected return \citep{martin2025information}. The joint crash probability of two exchange rates determines the cost of hedging cross-currency tail risk. Risk-neutral dependence is also central to portfolio construction, hedging, and the pricing of multi-asset derivatives. Despite five decades of work since \citet{ross1976options}, extracting risk-neutral dependence from traded options has remained an open problem.

The difficulty is fundamental. Options on individual assets identify marginal distributions but are silent about dependence. To identify joint dependence, one needs payoffs that depend on all assets jointly, such as options on an index. \citet{ross1976options} showed that options on a single suitably chosen linear combination of two stocks can complete the market in a finite-state setting, provided that the portfolio payoff separates states. \citet{martin2018options} extends this logic, showing that full recovery of the joint density requires options on a \emph{continuum} of portfolio directions, which are not available in practice. \citet{bondarenko2024option} take a different approach, combining estimated marginal densities with a maximum-entropy criterion, but their method requires intermediate density estimation and does not yield a tradable portfolio.

This paper proposes a projection-based estimator that provides a practical solution for estimating risk-neutral dependence from traded options. The core idea is to project the target payoff (for example, the product of two returns) onto the linear span of observed option payoffs across multiple assets and portfolios. Because the risk-neutral expectation of each traded option is directly observable from market prices, the projection coefficient vector immediately delivers an estimate of the desired risk-neutral expectation. The resulting estimate corresponds to an investable portfolio of traded options that best replicates the target payoff over the entire state space, not only at the observed strikes. It admits an explicit finite-sample error bound, which quantifies market incompleteness by showing how closely available options span the target payoff. The bound is small precisely when replication is tight. It also avoids intermediate density estimation or parametric assumptions about the copula.

Because projection finds the replicating portfolio of options that best approximates the target payoff across all states, it identifies any risk-neutral moment in the spanning case, that is, whenever options span all continuous multivariate payoffs. A natural question
is therefore when this spanning property holds. Using results on ridge functions from approximation theory, we derive necessary and sufficient conditions for the spanning property and show that, with two or more assets, the conditions require 
options on infinitely many distinct portfolios, a condition that cannot be met in practice. Since each market index is itself a portfolio of the underlying assets, options on multiple indexes enrich the basis, but a finite collection is not enough to identify the full joint distribution. In two dimensions, however, the covariance can be expressed exactly as a linear combination of variances on the market and on each individual asset, and projection recovers it from the corresponding options. In higher dimensions, this exact identification generally fails, but projection can still deliver an accurate approximation.

Because dependence can only be estimated from payoffs that depend on more than one underlying, FX markets provide a particularly clean application. Triangular no-arbitrage makes the EUR/USD rate, for example, the ratio of EUR/CHF and USD/CHF. Options on the cross rate therefore contain direct information about dependence between the two bilateral rates, whereas options on EUR/CHF and USD/CHF alone identify only the marginals. In this setting, we characterize the class of joint payoffs spanned by options on the three bilateral rates and show that, although this market is richer than what marginals alone deliver, the FX market remains incomplete: the full joint distribution is not identified. Projection nonetheless exploits the cross-rate information efficiently and delivers highly accurate approximations of joint risk-neutral probabilities in simulations.

We apply the projection approach in an event-study setting around two unexpected Swiss National Bank (SNB) announcements concerning the EUR/CHF floor: its introduction in September 2011 and its removal in January 2015. These episodes allow us to trace how a policy targeting a single exchange rate affects the joint tail risk of EUR/CHF and USD/CHF returns. Although the floor was set for one currency pair, it may have altered broader currency-market tail risk by changing the price of states in which CHF exchange rates move together. Changes in joint crash prices can reflect both movements in the marginal risk-neutral distribution of the targeted rate and shifts in its dependence with other rates; standard univariate methods cannot disentangle these channels.

We therefore decompose the change in joint crash risk between EUR/CHF and USD/CHF returns into a marginal channel and a dependence channel. For both the introduction and the removal, roughly two-thirds of the change comes from the dependence channel, showing that the SNB policy fundamentally altered the pricing of joint tail risk rather than merely shifting the  marginal distributions. The risk-neutral correlation changes in the same direction: it falls from 0.82 to 0.38 after the floor introduction and rises from 0.24 to 0.91 when the floor is removed. We further document a \emph{joint honeymoon effect}: the floor suppresses not only EUR/CHF volatility, as predicted by the univariate target-zone model of \citet{krugman1991target}, but also cross-currency tail dependence; the removal reverses this pattern almost immediately. These results connect to \citet{haddad2025whatever}, who show that conditional policy promises redistribute Arrow prices across states; we provide complementary multivariate evidence that a single-rate commitment changes both option-implied joint tail prices and cross-currency dependence.

Although the multivariate problem has no direct empirical benchmark in the existing literature, the univariate case provides a natural setting in which to compare projection with \citet{Carr_Madan_2001} (henceforth CM). Their approach is close in spirit to ours: it also represents the target payoff by a tradable portfolio of options and the underlying. The CM approximation starts from a Taylor expansion that expresses the target payoff as an integral over option payoffs (the CM formula); in practice, this integral is discretized using the observed strike grid (the CM approximation). Projection instead treats the observed options as a finite set of basis functions and asks which tradable portfolio best approximates the target payoff over the positive real line.

A key benefit of our approach is that the replicating portfolio is tuned to perform well over the entire positive real line despite a limited range of strike prices. We achieve this through a weighted projection, with the weighting density chosen as a tractable proxy for the risk-neutral distribution. We also prove that the projection and CM portfolios are asymptotically equivalent when the strike grid is dense and regularly spaced (Proposition~\ref{prop:cm_proj_weights}); the practical advantage of univariate projection is therefore a finite-sample one. Consistent with this, simulations show that projection delivers estimation errors an order of magnitude smaller than the CM approximation, especially in sparse-strike 
settings. This advantage is particularly relevant in our FX application, where only five strikes are available per currency pair. Out-of-sample, the projection-based variance portfolio used in the estimation of the risk-neutral correlation is more accurate than CM on 85\% of trading days.

Beyond accuracy, our framework yields three theoretical benefits. First, the projection estimator admits a finite-sample error bound and thus provides a quantitative measure of market incompleteness (Proposition~\ref{prop:proj_errorBound}). No analogous bound is available for the CM approximation. Second, combined with a spanning result, this bound delivers a new version of the second Fundamental Theorem of Asset Pricing (Theorem~\ref{thm:ftap_II}): spanning of continuous payoffs by simple options and uniqueness of the risk-neutral measure are equivalent. Third, because projection can be applied to any target payoff, it provides a unified way to estimate the risk-neutral distribution itself. Applying the method to indicator payoffs yields a projection-implied CDF that is pricing-consistent in finite samples, i.e., integrating a target payoff against the CDF returns the same value as projecting the payoff directly (Proposition~\ref{prop:rn_expectation}).

The rest of the paper is structured as follows. Section~\ref{sec:idea_projection} reviews the CM approach and introduces the projection estimator. Section~\ref{sec:pricing_spanning} derives the finite-sample pricing error bound, establishes a new version of the second Fundamental Theorem of Asset Pricing, and characterizes the projection-implied risk-neutral distribution. Section~\ref{sec:multivariate} extends the method to multiple underlyings and studies identification of joint risk-neutral moments. Section~\ref{sec:simulate} presents simulation evidence on finite-sample performance, and Section~\ref{sec:target_zone} presents the FX application. Section~\ref{sec:conclusion} concludes.

\section{Estimating nonlinear payoffs using projection}\label{sec:idea_projection}
In this section, we introduce the projection method to estimate risk-neutral moments. We first review \citet{Carr_Madan_2001} as a benchmark, then illustrate the projection idea on a simple finite-state example, and finally develop the general projection estimator with a continuum of states.

\subsection{Carr-Madan approach}
Let $g(S_T)$ denote a payoff at maturity $T$ as a function of the realized stock price $S_T$. Our object of interest is the conditional risk-neutral expectation $\Expect_t^Q[g(S_T)]$. The CM formula constructs a portfolio of puts and calls that replicates $g(S_T)$ state by state. By the law of one price, $\Expect_t^Q[g(S_T)]$ equals the time-$t$ value of this replicating portfolio, which can be computed from observed option prices.

To implement this idea, CM start from an exact second-order Taylor expansion to obtain a replicating portfolio,
\begin{align}\label{eq:cm_taylor}
g(S_T) &= g(F_{\ttt}) + g'(F_{\ttt}) (S_T - F_{\ttt}) \nonumber \\
&+\int_0^{F_{\ttt}} g''(K) \pmax{K - S_T} \diff K + \int_{F_{\ttt}}^\infty g''(K) \pmax{S_T - K} \diff K,
\end{align}
where $F_{\ttt}$ is the time-$t$ forward price for maturity $T$. Using risk-neutral valuation, it follows that 
\begin{equation}\label{eq:carr-mad}
 \Expect_t^Q g(S_T) =  g(F_{\ttt }) + R_{f,\ttt} \int_0^{F_{\ttt}} g''(K) P_{\ttt}(K) \diff K + R_{f,\ttt} \int_{F_{\ttt}}^\infty g''(K) C_{\ttt}(K) \diff K,
\end{equation}
where $R_{f,\ttt}$ is the gross risk-free rate from $t$ to $T$, and $P_{\ttt}(K)$ and $C_{\ttt}(K)$ denote European put and call option prices with strike $K$ and maturity $T$.

In practice, option prices are observed only at a discrete set of strikes, so the integrals in \eqref{eq:carr-mad} are approximated by a trapezoidal rule. For example, for observed put strikes $K_0<\cdots<K_J\le F_{\ttt}$,
\begin{align}\label{eq:cm_approx}
&\int_0^{F_{\ttt}} g''(K) P_{\ttt}(K) \diff K 
\approx \sum_{j=0}^{J} g''(K_j) P_{\ttt}(K_j)\,\Delta K_j, \\
\Delta K_0 &\coloneqq K_1-K_0,\quad 
\Delta K_J \coloneqq K_J-K_{J-1},\quad 
\Delta K_j \coloneqq \frac{K_{j+1}-K_{j-1}}{2}\ (1\le j\le J-1). \nonumber
\end{align}
This is the trapezoidal discretization used in the CBOE’s VIX methodology and in related model-free moment estimators. We refer to \eqref{eq:cm_approx} as the CM \emph{approximation or discretization}, to distinguish it from the exact CM formula in \eqref{eq:carr-mad}. Before introducing our projection-based alternative, we illustrate how  \eqref{eq:cm_approx} is used in two canonical applications.

\begin{exmp}[Risk-neutral variance (SVIX)]\label{exmp:var}
\citet{martin2017expected} derives a bound on the conditional expected market return using the risk-neutral variance:
\begin{equation*}
\Expect_t R_{\ttt} - R_{f,\ttt} \ge \frac{1}{R_{f,\ttt}} \Var_t^Q R_{\ttt},
\end{equation*}
where $R_{\ttt} = S_T/S_t$ is the return on the stock. To compute this bound from the data, it is necessary to calculate $\Expect_t^Q S_T^2$. The CM approximation can then be used with $g(S_T) = S_T^2$ and $g''(S_T) = 2$. 
\end{exmp}

\begin{exmp}[Risk-neutral entropy (VIX)]\label{exmp:vix}
The VIX is a popular measure of market uncertainty and is defined by the risk-neutral entropy of returns \citep{martin2017expected}:
\begin{equation}\label{eq:rn_ent}
\mathrm{VIX}_{\ttt}^2 = \frac{2}{T-t} \lro{\log R_{f,\ttt} - \Expect_t^Q \log R_{\ttt}}.
\end{equation}
Entropy, just like variance, is a measure of variability of a random variable. In this case it is necessary to calculate the expectation of a $\log$-return, which can be accomplished with the CM approximation using $g(S_T) = \log(S_T)$ and $g''(S_T) = -1/S_T^2$. \citet{britten2000option} further show that the VIX measures the  risk-neutral expected volatility from time $t$ to $t+T$. 


\end{exmp}

In addition to these examples, there are important settings in which the CM formula does not directly apply. The next two examples illustrate cases that are central for empirical work.

\begin{exmp}[Risk-neutral distribution]\label{exmp:density}
The estimation of the risk-neutral density is not covered by the CM formula because the payoff function necessary to calculate the PDF corresponds to a ``discontinuous payoff''. However, \citet{breeden1978prices} show that the risk-neutral CDF and PDF can be derived from
\begin{align*}
F_{\ttt}^Q(K) &= \Expect_t^Q \ind{\{S_T \le K\}} = 1 + R_{f,\ttt} \frac{\partial}{\partial K} C_{\ttt}(K)\\
f_{\ttt}^Q(K) &= \frac{\partial}{\partial K} F_{\ttt}^Q(K) = R_{f,\ttt} \frac{\partial^2}{\partial K^2} C_{\ttt}(K).
\end{align*}
These formulas are widely used to estimate risk-neutral densities and, when combined with additional information on physical probabilities, to infer pricing kernels and risk aversion. We will show that projection can also be used to estimate the risk-neutral distribution, thereby treating Examples~\ref{exmp:var}--\ref{exmp:density} in a unified manner.
\end{exmp}

\begin{exmp}[Risk-neutral covariance and correlation]\label{exmp:correlation}
For hedging purposes, it is often useful to estimate the risk-neutral covariance between two stock returns (see, e.g., \citet{lustig2014countercyclical}). In a different direction, the risk-neutral covariance between the market return and an individual stock also allows us to infer that stock’s equity premium when the representative investor has log utility (\citet{martin2019expected,martin2025information}):
\begin{equation*}
\Expect_t R_{i,\ttt} - R_{f,\ttt} = \frac{1}{R_{f,\ttt}} \Cov_t^Q\lro{R_{i,\ttt},R_{\ttt}}.
\end{equation*}
In this case, the CM formula does not apply because it is inherently univariate. Generally, estimating a covariance from options remains an open problem.\footnote{In certain settings the covariance is identifiable from option prices, e.g., for quanto options \citep{kremens2019theory}, or one can estimate it by imposing additional constraints, such as maximizing entropy (see \citet{bondarenko2024option}).}
Section~\ref{sec:multivariate} shows how the projection approach extends to the multivariate setting, and establishes when a correlation can be identified from options.

It can also be of interest to estimate the joint risk-neutral distribution. However, there is no higher-dimensional analogue of \citet{breeden1978prices}. We derive necessary and sufficient conditions on the option market that guarantee a unique multivariate risk-neutral measure. Although these conditions are typically not met in practice, the projection approach can nonetheless yield accurate approximations.
\end{exmp}

\subsection{A simple illustration of the projection method}
To illustrate the projection approach to estimating risk-neutral expectations of nonlinear payoffs, consider the following simple example.

\begin{exmp}[Projection approach]
Suppose the stock price at time $T$ can take four possible values: $S_T = [10,11,12,13]'$. We aim to replicate the payoff of the squared stock value, $S_T^2$. Assume we can trade a risk-free asset with return $R_{f,\ttt}$, the stock itself, and a call option on the stock with strike $K=12$. The squared stock value and the payoffs of the tradable assets, denoted by the matrix $X$, are given by
\begin{equation*}
S_T^2 = \begin{pmatrix}
100\\
121\\
144\\
169
\end{pmatrix}, \qquad 
X = 
\begin{pmatrix}
1 & 10 & 0\\
1 & 11 & 0\\
1 & 12 & 0\\
1 & 13 & 1
\end{pmatrix}. 
\end{equation*}
Clearly the market in this example is not complete because the value of $S_T^2$ cannot be replicated perfectly by a portfolio of tradable assets. To find a portfolio that comes closest to replicating $S_T^2$, a natural idea is to project $S_T^2$ onto the space spanned by $X$:
\begin{equation*}
    S_T^2 \approx X \hat{\beta}, \quad \text{where } \hat{\beta} = \lro{X'X}^{-1} X' S_T^2. 
\end{equation*}
Because the prices of the tradable assets are observable, we can estimate the risk-neutral expectation of $S_T^2$ via
\begin{equation*}
\Expect_t^Q S_T^2 \approx [1,F_{\ttt},R_{f,\ttt} C_{\ttt}(12)] \hat{\beta}.
\end{equation*}
This approximation follows from risk-neutral pricing because $F_{\ttt} = \Expect_t^Q[S_T]$ and $C_{\ttt}(12) = (1/R_{f,\ttt})\Expect_t^Q[\max(S_T - 12,0)]$. In general, the projection estimate will differ from the CM estimate, because CM assigns each option a weight proportional to $g''(K) = 2$ regardless of the strike, whereas projection optimizes the option weight directly to minimize replication error. The projection approach also generalizes the familiar put–call parity. For example, if we replace $S_T^2$ with the payoff of a put option, $\max(12-S_T,0)$, the projection on $X$ yields zero error, thereby recovering the classical parity relation. By contrast, put–call parity is not covered by the CM formula because the payoff functions are not twice differentiable.
\end{exmp}

\subsection{General projection approach}\label{sec:gen_proj}
This section generalizes the example above and introduces notation. Let the observed (ordered) out-of-the-money put and call strikes be
\begin{equation*}
\mathbf K^P \coloneqq [K_1^P,\dots,K_{n_k^P}^P]',\qquad
\mathbf K^C \coloneqq [K_1^C,\dots,K_{n_k^C}^C]',
\end{equation*}
with $K_{n_k^P}^P \le F_{\ttt}$ and $K_1^C > F_{\ttt}$, and define the total number of strikes by $n_k \coloneqq n_k^P+n_k^C$. Let
\begin{equation*}
\mathbf s \coloneqq [s_1,\dots,s_{n_s}]'
\end{equation*}
denote a researcher-chosen grid of stock prices at maturity $T$. Define the payoff design matrices for puts and calls on the grid $\mathbf s$ by
\begin{equation*}
X^P_{ij} \coloneqq \pmax{K_j^P-s_i},\qquad
X^C_{ij} \coloneqq \pmax{s_i-K_j^C},
\qquad i=1,\dots,n_s.
\end{equation*}
When it creates no confusion, we drop the superscripts $P$ and $C$ on strikes. Let $\mathbf 1_{n_s}$ denote an $n_s$-vector of ones and define the state-by-state payoff matrix
\begin{equation*}
X \coloneqq \bigl[\, \mathbf 1_{n_s}\ \ \mathbf s\ \ X^P\ \ X^C \,\bigr]\in\mathbb R^{n_s\times(2+n_k)}.
\end{equation*}
If a put and a call share the same strike, including both is redundant given put--call parity and the presence of the bond and stock columns. Let $Y\in\mathbb R^{n_s}$ be the payoff evaluated on the grid, $Y_i\coloneqq g(s_i)$. We compute the projection of $Y$ onto the column span of $X$:
\begin{equation*}
Y = X\widehat\beta + \widehat\varepsilon,\qquad
\widehat\beta \coloneqq (X' X)^{-1}X' Y.
\end{equation*}
Equivalently, this yields the approximation
\begin{equation}\label{eq:g_hat}
g(S_T)\approx 
\hat\beta_1  + \hat\beta_2 S_T
+ \sum_{j=1}^{n_k^P}\hat\beta_j^P \pmax{K_j-S_T}
+ \sum_{j=1}^{n_k^C}\hat\beta_j^C \pmax{S_T-K_j}
\eqqcolon \hat g(S_T).
\end{equation}
Taking risk-neutral expectations on both sides, we obtain a projection estimate of the risk-neutral expectation.

The replicating portfolio in \eqref{eq:g_hat} penalizes deviations equally across states (stock prices). In applications it can be preferable to penalize errors more heavily near the forward price---where the risk-neutral measure places more mass--—and less heavily in the tails. This can be implemented via weighted least squares:
\begin{equation}\label{eq:wls}
\hat{\beta}_{\mathrm{wls}}
= (X' \Omega X)^{-1} X' \Omega Y,
\end{equation}
where $\Omega=\mathrm{diag}(\omega_1,\dots,\omega_{n_s})$ collects state weights. The (infeasible) theoretically optimal choice sets weights equal to the risk-neutral density, $\omega_i = f_{\ttt}^{Q}(s_i)$. A practical alternative is to use the Variance Gamma process of \citet{madan1998variance}, calibrated so that its distribution provides the closest match to the observed option prices. Weighting can improve the accuracy of the replicating portfolio with finitely many options, and yields a replicating portfolio that remains accurate over the unbounded state space $\R_{++}$. This matters because in most applications the risk-neutral measure is supported on $\R_{++}$, while the available strikes cover only a limited range. We therefore use the weighted projection in the definition of the estimator. The unweighted projection can also be understood as setting the weights equal to the uniform density on a bounded domain. As the strike grid becomes dense, however, the choice of weighting affects the replicating portfolio only at second order, so projections with different weighting schemes coincide asymptotically. To simplify the notation, in the rest of the paper we drop the subscript in \eqref{eq:wls}.

\begin{defn}[Projection estimator]
Let $X$ collect terminal payoffs at $T$ (cash, the underlying, and options) evaluated on a state grid, and let $\hat\beta$ be the weighted least-squares solution in \eqref{eq:wls} from projecting the target payoff $Y$ on $X$.  Then the projection estimator is defined by 
\begin{equation}\label{eq:E_estimate}
\Expect_t^Q \hat{g}(S_T)  \coloneqq \hat{\beta}_1 + \hat{\beta}_2 F_{\ttt} +  R_{f,\ttt} \lro{\sum_{j=1}^{n_k^P} \hat{\beta}_j^P P_{\ttt}(K_j) + \sum_{j=1}^{n_k^C} \hat{\beta}_j^C C_{\ttt}(K_j)}.
\end{equation}  
\end{defn}

\begin{rem}[Constrained least squares]
In some applications, such as estimating risk-neutral variance, it is natural to impose that the estimate be nonnegative. With very few options, the least-squares replicating portfolio implied by $\hat\beta$ can produce a payoff that is negative over parts of the state space, which in turn can yield a negative variance estimate. In such cases, it is natural to require the replicating payoff to be nonnegative pointwise. This is achieved by solving the constrained least-squares problem 
\begin{equation*}   
\min_{\beta}\ \|\sqrt{\Omega} Y - \sqrt{\Omega} X\beta\|_2^2 \quad \text{subject to}\quad X\beta \ge 0,
\end{equation*}
where the inequality is interpreted componentwise on the chosen state grid. This convex quadratic program enforces a nonnegative replication in every state and, hence, a nonnegative variance estimate. Similarly, one may impose direct restrictions on the portfolio weights, for example, the componentwise bound $\beta \ge -c$ for some $c > 0$ to reflect borrowing constraints.
\end{rem}

\begin{rem}[Redundancy of option-implied regressors]
Because the projection estimator is a linear projection of the target payoff onto the span of the option basis functions, the Frisch–Waugh–Lovell theorem implies that adding any payoff that already lies in this span does not change the fitted values. For example, the CBOE VIX (Example \ref{exmp:vix}) corresponds to a log contract that is replicated from options. Hence adding the VIX payoff does not improve the estimation of a general payoff. By contrast, if there were a genuinely tradable claim delivering the log payoff (or a variance claim) whose price were not implied by the options in the basis, then adding $\log(S_T)$ would enlarge the span and improve estimation. Notice that the CM formula does not provide a generic way to exploit information from non-option payoffs.
\end{rem}

\begin{figure}[htb!]
\centering
\includegraphics[width=0.6\linewidth]{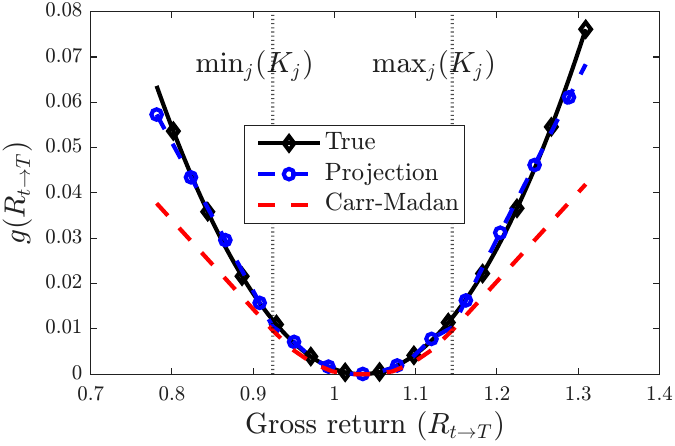}
\caption{\textbf{Replication of SVIX}. The figure shows the SVIX payoff $g(R_{\ttt}) = (R_{\ttt} - R_{f,\ttt})^2$ (black), together with the projection-based portfolio (blue) and CM portfolio (red). The projection uses the Variance Gamma weighting density. The portfolios are based on 10 equally spaced strike prices. Dashed vertical lines indicate the minimum and 
maximum strike values used.}\label{fig:approx_svix}
\end{figure}

To illustrate the benefits of the replicating portfolio obtained by projection in \eqref{eq:wls} relative to the CM approximation in \eqref{eq:cm_approx}, Figure~\ref{fig:approx_svix} plots both replicating portfolios corresponding to the SVIX payoff in Example \ref{exmp:var}. The projection portfolio tracks the true payoff almost perfectly,    
even outside the observed strike range. The CM portfolio is exact at the forward by construction (see \eqref{eq:cm_taylor}) but loses accuracy as $S_T$ moves further away, and it has no flexibility to fit the payoff outside the observed strike range. The upshot is that the CM approximation can produce substantially biased estimates of the risk-neutral expectation. The projection, in contrast, picks coefficients to minimize the weighted approximation error across the whole state space, so it stays close to the payoff even where strikes are absent.


\subsection{Continuous-state limit}
To implement the projection method, the researcher needs to choose a grid of possible future stock values, $\mathbf s$. This is analogous to specifying the up and down states in the binomial option pricing model. Since the grid can be made arbitrarily fine, a natural question is what the discrete projection converges to as the mesh size tends to zero.

Throughout, we denote the set of basis functions used for portfolio replication by
\begin{equation*}
\cF_{2+n_k} = \set{1,S_T,\pmax{K_1 - S_T},\dots,\pmax{K_{n_k^P} - S_T},\pmax{S_T - K_1},\dots,\pmax{S_T - K_{n_k^C}}}.    
\end{equation*}
Without loss of generality, we assume that all strike prices are positive and unique. When convenient, we index the basis as $\phi_i \in \cF_{2+n_k}$ for $i = 1, \dots, 2 + n_k $.  To derive the limiting value as the state space becomes continuous, we make the following assumption.  
\begin{asmp}\label{asmp:compact}
Let $\omega(s): \R_{++} \to \R_{++}$ be a continuous density function with $\int_0^\infty s^2 \omega(s) \diff s < \infty$. Let $g \in L^2(\R_{++}, \omega)$: $\int_0^\infty g(s)^2 \omega(s) \diff s < \infty$.  
\end{asmp}

Assumption \ref{asmp:compact} guarantees that the projection estimator is well defined when $n_s$ is sufficiently large. In particular, because the strike prices are assumed to be unique, all basis functions are linearly independent over $L^2(\R_{++}, \omega)$. Implicitly, the assumption also puts some parameter restrictions on $\omega$. For example, when $g(s) = s^2$, the assumption implies that the fourth moment under $\omega$ must exist. The next result establishes the continuous-grid limit. By slight abuse of notation, let $\hat\beta_{n_s}$ denote the projection coefficients obtained from a grid of size $n_s$.

\begin{prop}\label{prop:ols_cont}
Let Assumption \ref{asmp:compact} hold and define an inner product on $L^2(\R_{++}, \omega)$ by
\begin{equation*}
\seq{\phi_i,\phi_j} = \int_0^\infty \phi_i(s) \phi_j(s) \omega(s) \diff s.
\end{equation*}
Let $s_1^{(n)} < s_2^{(n)} < \cdots < s_{n_s}^{(n)}$ be a sequence of state grids such that $\max_i |s_{i+1}^{(n)} - s_i^{(n)}| \to 0, s_1^{(n)} \to 0$, and $s_{n_s}^{(n)} \to \infty$ as $n_s \to \infty$. Then $\hat\beta_{n_s}\to\hat\beta$, where
\begin{equation}\label{eq:ols_cont}
\hat{\beta}_{n_s} \to  
\begin{bmatrix}
\seq{\phi_1,\phi_1} & \dots & \seq{\phi_{1},\phi_{2 + n_k}} \\
\vdots & \ddots &\vdots \\
\seq{\phi_{2 + n_k},\phi_1} & \dots & \seq{\phi_{2 + n_k},\phi_{2 + n_k}}
\end{bmatrix}^{-1} 
\begin{bmatrix}
\seq{\phi_1,g}\\
\vdots\\
\seq{\phi_{2 + n_k},g}
\end{bmatrix}
\eqqcolon \hat{\beta}.
\end{equation}
Moreover, $\hat{\beta}$ solves the minimization problem 
\begin{equation}\label{eq:beta_continuous}
\hat{\beta} = \argmin_{\beta \in \R^{2+n_k}} \int_{0}^\infty \lro{g(s) - \sum_{j=1}^{2+n_k}\beta_j \phi_j(s) }^2 \omega(s) \diff s.
\end{equation}
\end{prop}
All proofs are in Internet Appendix \ref{app:proofs}. The minimization property in \eqref{eq:beta_continuous} states that $\hat{\beta}$ minimizes the weighted $L^2$-distance between $g(\cdot)$ and the option basis functions, with weight $\omega$. In this sense, the basis functions optimally replicate $g(\cdot)$ over the entire state space. This property is attractive because the replication continues to be accurate outside the range of observed strike prices, in regions where $\omega$ places mass. The continuous-state limit is also a convenient tool in some of the proofs. However, for practical computations we will mostly rely on the discrete approximation, as it is faster and numerically more stable.

\section{Pricing errors, spanning, and the projection-implied distribution}\label{sec:pricing_spanning}

This section asks how much information finitely many options contain about risk-neutral expectations. We first derive a finite-sample bound on the pricing error when only finitely many strikes are observed. Combined with a spanning result, this bound delivers a finite-strike version of the second Fundamental Theorem of Asset Pricing (FTAP), in which spanning of continuous payoffs by simple options and uniqueness of the risk-neutral measure are equivalent. We then compare the resulting projection portfolio with the CM portfolio. Finally, we characterize the risk-neutral distribution implied by projection.

\subsection{Finite-sample pricing error bound}\label{sec:proj_errorBound}
Market completeness, or the spanning property, is the requirement that every contingent claim can be hedged. Under no-arbitrage, the second FTAP states that this is equivalent to having a unique risk-neutral measure. As is well known, options complete the market for a single underlying security \citep{ross1976options,breeden1978prices,nachman1988spanning,bakshi2000spanning}. The classical results either assume that options are observable on a continuum of strikes \citep{breeden1978prices,nachman1988spanning}, or build the   
spanning argument using the characteristic function of the state-price density \citep{bakshi2000spanning}, which is tractable under specific dynamics on the underlying. Our projection framework is finite-sample and model-free: we bound the pricing error for any \emph{finite} collection of strikes without imposing structure on the underlying dynamics. This bound is a quantitative refinement of the classical completeness result: as we show in Proposition~\ref{prop:spanning}, the bound vanishes in the dense-strike limit and uniqueness of the risk-neutral measure is recovered. The proposition below states the bound.

\begin{prop}[Finite-sample error]\label{prop:proj_errorBound}
Let Assumption \ref{asmp:compact} hold. Suppose further that
\begin{enumerate}[({A}1)]
    \item \label{item:a1} The risk-neutral measure admits a density $f_{\ttt}^Q$ with respect to Lebesgue measure on $\R_{++}$. 

    \item \label{item:a2} The chi-squared divergence between $f_{\ttt}^Q$ and $\omega$ is finite:
    \begin{equation*}
    \chi^2(f_{\ttt}^Q \,\|\, \omega)  \;:=\; \int_0^\infty 
        \frac{\bigl(f_{\ttt}^Q(s) - \omega(s) \bigr)^2}{\omega(s)} \diff s \;<\; \infty.
    \end{equation*}
\end{enumerate}
Then,
\begin{equation}\label{eq:finite_bound}
\abs{\Expect_t^Q g(S_T) - \Expect_t^Q \hat{g}(S_T)} \le \underbrace{\norm{g - \hat{g}}_{L^2(\R_{++},\omega)}}_{\text{hedging error}} \underbrace{\sqrt{\chi^2(f_{\ttt}^Q \,\|\, \omega)}}_{\text{prior misspecification}}.
\end{equation}

\end{prop}

Assumption \ref{item:a1} is a mild regularity condition: it rules out point masses but allows the distribution to assign zero mass to regions of the state space. Virtually all option-pricing models satisfy this property. Assumption \ref{item:a2} requires
the chi-squared divergence between $f_{\ttt}^Q$ and the weighting density $\omega$ to be finite. This is precisely the square-integrability condition on the likelihood ratio $f_{\ttt}^Q/\omega$ that underlies the density-approximation theory of \citet{filipovic2013density}, which uses an $L^2$ framework closely related to ours. A sufficient condition is that the tails of $\omega$ decay no faster than those of $f_{\ttt}^Q$, both as $s \to \infty$ and as $s \to 0^+$. For theoretical consistency, we therefore set $\omega$ to the Variance Gamma (VG) density in our applications.\footnote{However, in unreported simulations we find that the specific choice of weighting density has little impact on the resulting risk-neutral moment estimates, as long as it has unbounded support.}  Both tails decay polynomially, which ensures a finite chi-squared divergence with many risk-neutral measures. The polynomial tails of the VG density are also consistent with empirical evidence that risk-neutral distributions feature heavier tails than a lognormal benchmark \citep{figlewski2010estimating}.

The bound in \eqref{eq:finite_bound} decomposes the pricing error into two interpretable factors. The chi-squared divergence in the second factor measures how well the weighting density approximates the risk-neutral density: it is small when $\omega$ is close to $f_{\ttt}^Q$, and vanishes when $\omega = f_{\ttt}^Q$, so that the projection estimator is exact for any finite number of strikes.\footnote{Chi-squared belongs to a broader family of $f$-divergences. The bound in \eqref{eq:finite_bound} extends to this family by replacing the 
$L^2$ projection with an appropriate $L^p$ projection.} The first factor is the weighted $L^2$ projection error and is directly computable: it measures how well the option basis $\cF_{2+n_k}$ replicates the target payoff $g$ under the chosen weighting. In Internet Appendix \ref{app:conv_rate}, we show that the projection error converges to zero quadratically with the number of strike prices. Thus, as more options are included, the weighting density matters less. This parallels the empirical finding of \citet{almeida2023nonparametric} that, in the related problem of estimating the risk-neutral density, the choice of divergence in the Cressie-Read family becomes less important as more options are used.

A useful way to read \eqref{eq:finite_bound} is as a generalization of put-call parity. If $g(S_T) = \pmax{K - S_T}$ is the put payoff with strike $K$, and the traded basis includes a call with the same strike, the underlying, and the risk-free asset, then $\norm{g - \hat{g}}_{L^2(\R_{++}, \omega)} = 0$ and the bound collapses to put-call parity exactly. More generally, \eqref{eq:finite_bound} delivers a near-replication statement for any payoff $g \in L^2(\R_{++}, \omega)$, including discontinuous functions such as the indicator $\ind{S_T \le s}$ used to estimate the risk-neutral CDF. No analogous finite-sample bound is available for the CM approximation, because the CM portfolio is determined by the second derivative of $g$ rather than by minimization of an approximation criterion, and is therefore not a projection in any $L^p$ space.

Finally, the bound in \eqref{eq:finite_bound} extends to any number of underlyings with no change in the proof. This is important for Section \ref{sec:multivariate}, where we estimate joint risk-neutral moments from options on multiple assets. In that setting the option basis typically cannot span all continuous payoffs (Theorem \ref{thm:dense_multi}), so a pointwise replication argument breaks down. The bound in \eqref{eq:finite_bound} nevertheless quantifies the approximation error directly, which is particularly useful in our FX application, where triangular no-arbitrage enriches but does not complete the basis.

\subsection{Spanning and the second Fundamental Theorem of Asset Pricing}
We now establish the spanning property. To this end, we use the following assumption.
\begin{asmp}[Dense-strike limit]\label{asmp:dense} 
The sequence of strikes $\{K_1^{(n_k)}, \dots, K_{n_k}^{(n_k)}\}$ is such that, for every compact $A \subset \R_{++}$,
\begin{equation*} 
\max_{x \in A} \min_{j=1,\dots,n_k} \abs{x - K_j^{(n_k)}} \to 0 \qquad \text{as } n_k \to \infty.
\end{equation*}                             
\end{asmp} 
The assumption requires the observed strikes to become dense in every bounded subset of the positive real line. Intuitively, this condition allows us to synthesize, for any state $s$, a portfolio of options that pays out only at $S_T = s$. The next proposition shows that any continuous payoff can therefore be hedged (or spanned) arbitrarily well.

\begin{prop}[Spanning]\label{prop:spanning}
Let $A\subset\R_{++}$ be compact and let $C(A)$ denote the space of continuous functions on $A$ equipped with the sup norm $\norm{g} = \sup_{x \in A} |g(x)|$. Let Assumption \ref{asmp:dense} hold. Then, for every $g \in C(A)$ there exists $f_{n_k} \in \spn(\cF_{2+n_k})$ such that $\norm{g-f_{n_k}}_\infty \to 0$ as $n_k \to \infty$.
\end{prop}

Finally, we can combine the spanning property and the finite-sample error bound to obtain a new version of the second FTAP, tailored to our finite-strike setting. Let 
\begin{multline*}
\mathcal{Q}_{\omega} \;:=\; \big\{ Q_t : Q_t \text{ admits a continuous,  strictly positive density}\\ 
f_{\ttt}^Q
\text{ on } \R_{++}  \text{ with } \chi^2(f_{\ttt}^Q \,\|\, \omega) < \infty \big\}
\end{multline*}
denote the class of risk-neutral probability measures with finite chi-squared
divergence with respect to $\omega$, and assume that $\mathcal{Q}_{\omega}$ is non-empty. The only substantive condition this imposes beyond no-arbitrage is the existence of a 
risk-neutral measure with strictly positive density on $\R_{++}$, a property satisfied by virtually all standard option-pricing models. The remaining requirement, finiteness of the chi-squared divergence, is not a restriction on $Q_t$ itself but on the pairing of $f^Q_{\ttt}$ with the weighting density $\omega$: for any $Q_t$ with strictly positive density, one can always choose $\omega$ with sufficiently heavy tails to ensure $\chi^2(Q_t \,\|\ \omega) < \infty$. Of course, the tails of $Q_t$ are unknown in practice, which is why we choose the VG density in applications, whose polynomial tails can accommodate many risk-neutral measures.

\begin{thm}[FTAP II in the option-pricing setting]\label{thm:ftap_II}
Let $Q_t \in \mathcal{Q}_{\omega}$ and let Assumption \ref{asmp:compact} hold. The following are equivalent:
\begin{enumerate}
    \item[(i)] \emph{Spanning}:
    $\bigcup_{n_k \ge 1} \spn(\cF_{2+n_k})$ is dense in $L^2(\R_{++}, \omega)$.
  \item[(ii)] \emph{Uniqueness of $Q_t$}: For any $Q_t' \in \mathcal{Q}_{\omega}$ that satisfies $\Expect_t^{Q'} \phi(S_T) = \Expect_t^{Q} \phi(S_T)$ for every $\phi \in \bigcup_{n_k \ge 1} \cF_{2+n_k}$, we have $Q_t' = Q_t$. 
\end{enumerate}
\end{thm}

\begin{cor}[Quantitative refinement]\label{cor:refine}
Let $Q_t, Q_t' \in \mathcal{Q}_{\omega}$ agree on the prices of all basis functions in $\cF_{2+n_k}$. Then for any $g \in L^2(\R_{++}, \omega)$,   
\begin{equation*}
\abs{\Expect_t^Q g(S_T) - \Expect_t^{Q'} g(S_T)}
\;\le\; \norm{g - \hat g}_{L^2(\R_{++}, \omega)}
 \Bigl( \sqrt{\chi^2(Q_t \|\omega)} + \sqrt{\chi^2(Q_t'\|\omega)} \Bigr).
\end{equation*}
\end{cor}

Theorem \ref{thm:ftap_II} is closely related to \citet{breeden1978prices, nachman1988spanning, bakshi2000spanning,Carr_Madan_2001}. These papers show that with a continuum of strikes, the risk-neutral density is recovered from option prices, thereby implicitly establishing asymptotic uniqueness of $Q_t$. Furthermore, the CM formula \eqref{eq:cm_taylor} establishes asymptotic spanning of any sufficiently smooth payoff. In each case the result is the forward direction of FTAP II (spanning $\Rightarrow$ uniqueness) and is asymptotic in the strike grid;
the converse direction is not addressed.

Theorem~\ref{thm:ftap_II} and Corollary~\ref{cor:refine} differ from this prior literature in three substantive ways. First, the forward direction (spanning $\Rightarrow$ uniqueness) is established with a finite-sample bound at
every $n_k$ (Corollary~\ref{cor:refine}), rather than using a continuum of strikes. Second, all of the option-pricing precedents address only the forward direction, while the converse direction (uniqueness $\Rightarrow$ 
spanning) is, to our knowledge, novel in this setting.  Economically, the converse direction equates pricing
uniqueness with replication completeness: if option prices admit only one consistent risk-neutral measure, then every payoff in $L^2(\R_{++}, \omega)$ can be replicated by a static portfolio of options.  Third, the framework uses observed option prices directly: no density recovery or characteristic-function machinery is needed.

A separate strand of literature \citep{harrison1981martingales, delbaen1994general} establishes the FTAP II in a dynamic trading setting, where payoffs are replicated by self-financing strategies. Closer to our setup, \citet{jarrow1999second} prove a static FTAP II  but require a lattice condition which implies that options on options are traded, as well as options on portfolios of options, options on options on options, and so on \textit{ad infinitum}; these instruments are not traded in practice. By contrast, our setup is static and uses only simple tradable options on the underlying: the investor purchases a portfolio at time $t$ and holds it to maturity. The proof of Theorem~\ref{thm:ftap_II} accordingly requires only Hilbert-space theory in $L^2(\R_{++}, \omega)$, without recourse to stochastic calculus and semimartingales.

\subsection{Projection and the CM portfolio}
The finite-sample bound established in Section~\ref{sec:proj_errorBound} has no counterpart for the CM approximation. A natural question is when the two approaches yield similar replicating portfolios. Because the analysis is not tractable in finite samples, we turn to an asymptotic perspective where the number of options grows to infinity. When the strike prices are equally spaced over a compact interval, the following proposition shows that the projection and CM portfolio weights are asymptotically equal.

\begin{prop}[Projection and CM portfolio weights]\label{prop:cm_proj_weights}
	Let $A=[\amin,\amax]$ and let
	\begin{equation*}
	\amin=K_0<K_1<\cdots<K_{n_k}<K_{n_k+1}=\amax
	\end{equation*}
	be uniformly spaced with mesh size $h:=K_i-K_{i-1}$. Assume
	$g\in C^4(A)$ and $\omega \in C^4(A)$ with $\omega>0$ on $A$. Let $\hat g$ be the
	$L^2(A,\omega)$-projection of $g$ onto $\spn(\cF_{2+n_k})$, written in the option-payoff basis as
	\begin{equation*}
		\hat g(x)=\hat\beta_1+\hat\beta_2 x+\sum_{i=1}^{n_k}\hat\gamma_i\pmax{x-K_i}.
	\end{equation*}
	Then there exist constants $C<\infty$ and $\rho\in(0,1)$, independent of
	$h$ and $i$, such that for every $i=1,\ldots,n_k$,
	\begin{equation}\label{eq:gamma_boundary_layer}
		\left|\hat\gamma_i-hg''(K_i)\right|
		\le C h^3+C h^2\left(\rho^i+\rho^{n_k+1-i}\right).
	\end{equation}
	Consequently,
	\begin{equation*}
		\hat\gamma_i= \underbrace{h g''(K_i)}_{\text{CM weight}}+O(h^2)
		\qquad\text{uniformly for }i=1,\ldots,n_k,
	\end{equation*}
	and, for every fixed $\delta>0$,
	\begin{equation*}
		\hat\gamma_i=h g''(K_i)+O(h^3)
		\qquad\text{uniformly over all } i \text{ such that }
		K_i\in[\amin+\delta,\amax-\delta].
	\end{equation*}
\end{prop}

This result may appear surprising at first because the projection method seems global, in the sense that each coefficient estimate depends on the full set of strikes. However, results from the series regression literature suggest that it depends on the number of basis functions: when the number of strikes is small the estimator is effectively global, whereas as the strike grid becomes dense the projection behaves increasingly like a local method (see, e.g., \citet[Section 20.7]{hansen2022econometrics}).\footnote{In Internet Appendix \ref{app:conv_rate} we also prove that projection and the CM approximation converge with the same speed to the true risk-neutral expectation, under the same assumptions of Proposition \ref{prop:cm_proj_weights}.}

Why, then, prefer the projection method? First, the results above are asymptotic and may not accurately describe the finite-sample behavior that is relevant in practice. Second, Proposition \ref{prop:cm_proj_weights} relies on idealized assumptions, such as a uniformly spaced strike grid and a mesh that becomes dense all the way to the endpoints of $A$. Furthermore, in most applications the risk-neutral measure is supported on $\R_{++}$ and not on a bounded interval. Hence, when any of these assumptions fails, as is typical in option data, the asymptotic approximation in Proposition \ref{prop:cm_proj_weights} need not hold, and the implied portfolio weights can differ substantially from those obtained by the CM  approximation. Indeed, in simulation (Section~\ref{sec:simulate}) the projection portfolio is more accurate than the CM portfolio by orders of magnitude, and in the empirical application (Section~\ref{sec:target_zone}) it significantly reduces out-of-sample hedging error.

One notable exception where the assumptions of Proposition~\ref{prop:cm_proj_weights} are close to being satisfied, is the market for 30-day options on the S\&P 500, one of the most liquid options markets in the world. In Internet Appendix~\ref{app:svix_vix} we find little difference between the projection and CM portfolios when estimating the SVIX of \citet{martin2017expected} at this maturity. For longer maturities, such as 90 or 180 days, the strike grid is sparser and the two methods start to diverge. Using projection, we cannot reject the null hypothesis that the SVIX is a tight lower bound on the equity premium, while using the CM approximation we reject this null at the 180-day maturity. Because the CM portfolio is downward biased relative to projection (Figure~\ref{fig:approx_svix}), the rejection under CM reflects an understated bound rather than genuine evidence against tightness.

\subsection{Estimation of the risk-neutral distribution}
Because options complete the market and projection can approximate any continuous payoff, it is natural to ask which risk-neutral distribution is implied by projection. Consider $g(s) = \ind{s \le x}$, which is used to compute the risk-neutral CDF: $F_{\ttt}^Q(x) = \Expect_t^Q \ind{S_T \le x}$. In this case, the projection estimates obtained in \eqref{eq:ols_cont} will also depend on $x$, because
\begin{equation*}
\seq{\phi_j,\ind{\cdot \le x}} = \int_0^\infty \phi_j(s) \ind{s \le x} \omega(s) \diff s = \int_0^x \phi_j(s) \omega(s) \diff s. 
\end{equation*}
Let $\hat{\beta}(x)$ denote the coefficient vector obtained by projecting the target payoff $g_x(s) = \ind{s \le x}$ onto the basis $\set{\phi}_j$. The risk-neutral CDF is then estimated by
\begin{equation}\label{eq:rn_cdf}
\hat{F}_{\ttt}^Q(x) = \hat{\beta}_1(x) + \hat{\beta}_2(x) F_{\ttt} +  R_{f,\ttt} \lro{\sum_{j=1}^{n_k^P} \hat{\beta}_j^P(x) P_{\ttt}(K_j) + \sum_{j=1}^{n_k^C} \hat{\beta}_j^C(x) C_{\ttt}(K_j)}.
\end{equation}
The following proposition shows that $\hat{F}_{\ttt}^Q(x)$ obtained in this way satisfies many of the natural CDF requirements.

\begin{prop}[Risk-neutral distribution]\label{prop:rn_expectation}
Let $\omega \in C^d(\R_{++})$ satisfy Assumption \ref{asmp:compact}. Then:
\begin{enumerate}[(i)]
    \item The estimated CDF satisfies the natural boundary limits  \label{item:rn_1}
\begin{equation*}
\lim_{x \to 0^+} \hat{F}_{\ttt}^Q(x) = 0, \quad \text{and} \quad \lim_{x \to \infty} \hat{F}_{\ttt}^Q(x) = 1.
\end{equation*}
\item $\hat{F}_{\ttt}^Q(x)$ is continuously differentiable on $(0,\infty)$, with density estimate $\hat{f}_{\ttt}^Q = (\hat{F}_{\ttt}^Q)'$; moreover, $\hat{f}_{\ttt}^Q$ is piecewise $C^d(\R_{++})$. \label{item:rn_2}

\item (Pricing consistency) The estimated value of a nonlinear contract in \eqref{eq:E_estimate} equals the moment implied by the estimated distribution: \label{item:rn_3}
\begin{equation*}
\Expect_t^Q \lr{\hat{g}(S_T)} = \int_0^\infty g(x) \diff \hat{F}_{\ttt}^Q(x).
\end{equation*}
\end{enumerate}
\end{prop}

Property~\ref{item:rn_3} is the most important. For any \emph{finite} set of strikes, the price obtained by integrating the payoff $g$ against the projection-implied distribution is exactly the same as the price obtained by projecting $g$ directly, that is, by using \eqref{eq:E_estimate}. In this sense, the implied distribution is pricing consistent with the projection estimator in finite samples. This pricing consistency is typically not guaranteed by existing risk-neutral density estimators. In particular, the value of a nonlinear contract computed from a density estimate will generally not coincide with the estimate given by the CM approximation. This discrepancy can be undesirable in applications, because the density estimate and the direct pricing formula will not agree even when they are applied to the same option data. Therefore, when the entire risk-neutral distribution is not of primary interest, the CM approximation is often preferred since it is more robust (see, e.g., \citet{martin2017expected}). By construction, the projection approach avoids this issue and yields a density that is consistent with any moment obtained by direct projection. Consequently, the risk-free asset, the stock, and all observed options are priced with zero error under the projection-implied distribution, because each is itself a basis function in the regression. Especially in sparse-strike settings, this distribution can serve as a better-behaved input for applications that require the full risk-neutral distribution, such as \citet{MartinShi2025crashes} or \citet{haddad2025whatever}. We illustrate this use in Section~\ref{sec:target_zone}.

Despite these desirable properties, the projection-based CDF estimate need not be monotone. Violations can arise when strikes are sparse relative to the evaluation grid, since the option basis cannot tightly constrain the implied CDF between strikes. In practice, the problem typically disappears when the weighting density $\omega$ is chosen to have heavier tails, since this improves the conditioning of the Gram matrix in \eqref{eq:ols_cont}. A general-purpose remedy is to apply the rearrangement approach of \citet{chernozhukov2013inference}, which sorts the estimated CDF values on the grid to enforce monotonicity. Moreover, \citet{chernozhukov2009improving} show that, unless the original estimate is already monotone, the rearranged CDF has strictly smaller estimation error. However, the rearranged CDF will generally no longer satisfy the price consistency property \ref{item:rn_3}.

\section{Spanning and joint dependence in multi-asset markets}\label{sec:multivariate}

It is of great interest to generalize the projection approach to higher dimensions. For example, the risk premium of an individual return can often be related to its risk-neutral covariance with the market return (see Example~\ref{exmp:correlation}). The key challenge is that the claim paying $S_{1,T} S_{2,T}$ is not traded, so $\Expect_t^Q (S_{1,T} S_{2,T})$ must be estimated from tradable options.

A naive approach is to take options on the two assets separately and project the target function, such as $g(s_1,s_2) = s_1 s_2$ (needed for the covariance) or $g(s_1,s_2) = \ind{s_1 \le x_1} \ind{s_2 \le x_2}$ (the joint CDF), onto the basis generated by the individual-asset options, the stock prices, and the risk-free payoff. However, the next proposition shows that this approach is uninformative about dependence: any estimated dependence is inherited from the weighting density rather than identified from individual options.

\begin{prop}[Dependence inherited from the weighting density]\label{prop:additive_weights}
Let $f_{1,\ttt}^{Q}, f_{2,\ttt}^{Q}$ denote the marginal risk-neutral densities of $S_{1,T}$ and $S_{2,T}$ respectively. Let $\omega$ be any joint weighting density whose marginals coincide with $f_{1,\ttt}^{Q}$ and $f_{2,\ttt}^{Q}$, and let $\hat{g}$ be the $L^2(\R_{++}^2, \omega)$ projection of $g \in L^2(\R_{++}^2, \omega)$ onto the span of the risk-free payoff, the individual stock payoffs, and individual-asset option payoffs. Then,
\begin{equation*}
\Expect_t^Q\lr{\hat{g} (S_{1,T}, S_{2,T})} = \Expect^{\omega}[g(S_{1,T}, S_{2,T})],
\end{equation*}
where $\Expect^{\omega}$ denotes the expectation under $\omega$. Thus, when the basis contains only individual-asset options, any dependence estimate is inherited from the weighting density $\omega$ rather than identified from option prices.
\end{prop}

In the dense-strike limit for each asset (Assumption~\ref{asmp:dense}), the marginal distributions are identified (Proposition~\ref{prop:spanning}). Their product weighting density $\omega(s_1,s_2) = f_{1,\ttt}^{Q}(s_1) f_{2,\ttt}^{Q}(s_2)$ then gives a natural two-dimensional weighting density for the projection. In this special case, for any $g(s_1,s_2)=g_1(s_1)g_2(s_2)$, it follows from Proposition \ref{prop:additive_weights} that

\begin{equation*}
\Expect_t^Q\lr{\hat{g} (S_{1,T}, S_{2,T})} = 
\Expect_t^Q\lr{g_1(S_{1,T})}
\Expect_t^Q\lr{g_2(S_{2,T})}.
\end{equation*}
Setting $g_1=\ind{\cdot\le x}$ and $g_2=\ind{\cdot\le y}$ shows that the projection-implied joint CDF factors under product-of-marginals weighting, thus imposing independence.


Intuitively, options on the individual assets are sufficient to identify the marginal distributions, but not the joint distribution. With product weights, the projection-implied joint distribution is independent. With non-product weights, the projection can generate nonzero dependence, but this dependence comes from the weighting density rather than from option prices. To estimate a nonzero correlation, or more generally any joint dependence from options alone, the basis must include payoffs that depend jointly on both assets, either through nonlinear functions of the individual returns or through multi-asset instruments such as basket options. A natural possibility is to use options on a market index, since the index payoff depends jointly on the performance of its constituents. For example, the return on the SPDR index ETF can be written as a weighted average of the 11 sector returns:

\begin{equation*}
\sum_{i=1}^{11} w_{i,t} R_{i,\ttt} = R_{\ttt},
\end{equation*}
where $w_{i,t}$ and $R_{i,\ttt}$ denote the weight and realized return on sector $i$, and $R_{\ttt}$ represents the return on the index portfolio. Thus, options on the SPDR ETF reveal information about the joint distribution of returns. In combination with options on the individual sectors, they allow more precise inference about correlations. Nevertheless, the information conveyed by options on the market index and on the sectors is limited: with three or more sectors, correlations cannot be identified from these derivatives alone. We establish this non-identification result below.

\subsection{Multivariate spanning by simple options}
The finite-sample bound of Proposition~\ref{prop:proj_errorBound} holds in any dimension and ensures that the projection error vanishes when the basis becomes dense in $L^2(\R_{++}^d, \omega)$, where $d \ge 1$ denotes the number of assets. Density follows from a multivariate analogue of the spanning property in Proposition~\ref{prop:spanning}, namely that options uniformly approximate any continuous payoff. We therefore ask when simple option payoffs span multivariate contingent claims in this sense. Suppose, as in practice, that there are $d$ sectors (or stocks) that aggregate to the index return:\footnote{For sector returns, we have $d=11$, corresponding to the 11 sector ETFs whose value-weighted average is the SPDR S\&P 500 ETF (SPY) return. For individual stocks, we have $d=500$, corresponding to the constituents of the S\&P 500.}
\begin{equation*}
\sum_{i=1}^d w_{i,t} R_{i,\ttt} = R_{\ttt}.
\end{equation*}

Assume now that the assumptions of Proposition~\ref{prop:spanning} hold for each sector and for the index return, so that any continuous function of any individual sector return, or of the index return, can be uniformly approximated by options. By combining the options on each of the sectors and on the index return in a portfolio, we conclude that the set of option payoffs spans the space\footnote{Although option payoffs depend only on nonnegative prices, we state the definition with $g \in C(\R)$ to match the statement of Theorem~\ref{thm:dense_multi} and to accommodate the FX setting in Section~\ref{sec:exchange}, where the cross-rate direction after a change of num{\'e}raire has mixed signs.}
\begin{equation*}
\cM(\Omega) \coloneqq \spn\Bigl\{\, x\mapsto g(a' x)\ :\ a\in\Omega,\ g\in C(\R)\Bigr\},
\end{equation*}
where $\Omega\subset\R^d$ is the set of available portfolio directions. In our baseline setting,
\begin{equation}\label{eq:a}
\Omega = \{e_1,\dots,e_d,\ w_t\}, \qquad w_t = (w_{1,t},\dots,w_{d,t})',
\end{equation}
where $e_i$ is the $i$th basis vector in $\R^d$ (i.e., it gives full weight to sector $i$). Functions of the form $g(a' x)$ are known as \emph{ridge functions} in the approximation theory literature \citep{pinkus2015ridge}. The question of multivariate spanning by simple options can therefore be phrased as a question about when ridge functions with directions in $\Omega$ are dense in $C(\R^d)$. The following result by \citet{vostrecov1961approximation} provides necessary and sufficient conditions (see also \citet{lin1993fundamentality}):

\begin{thm}\label{thm:dense_multi}
Any continuous payoff on $\R^d$ can be approximated arbitrarily well on compact sets by a static portfolio of options on the linear combinations $a' x$ with $a \in \Omega$, if and only if no nontrivial homogeneous polynomial vanishes on $\Omega$.\footnote{A polynomial in several variables is homogeneous if all monomials have the same total degree. Formally, the theorem states that $\cM(\Omega)$ is dense in $C(\R^d)$ in the topology of uniform convergence on compacta.}
\end{thm}

In the special case $d = 2$, the set of option payoffs is dense if $\Omega$ contains an infinite number of pairwise linearly independent vectors. This connects to two classical results. \citet{ross1976options} establishes spanning in $d = 2$ on a finite state space, while \citet[Result~2]{martin2018options} shows that the joint risk-neutral distribution can be recovered in any dimension by inverting a Radon transform, provided options on a continuum of linear combinations are observable, which is difficult to meet in practice. Theorem~\ref{thm:dense_multi} complements these results by characterizing density as an algebraic property of the direction set $\Omega$, with a condition that is both necessary and sufficient and does not require continuum observability. In applications, we still cannot hope to approximate the price of \emph{every} multivariate contingent claim arbitrarily well, since we only observe the finite set of twelve direction vectors in~\eqref{eq:a} associated with the $d=11$ sector portfolios. Nevertheless, projecting an arbitrary claim onto the corresponding sector and market option payoffs, which are actively traded, still delivers a meaningful approximation that is fully implementable from observed derivatives. Theorem~\ref{thm:dense_multi} further suggests that adding options on a sector portfolio whose weights differ from the market portfolio can sharpen the approximation. Recently introduced options on an equally weighted sector portfolio (EQL) provide exactly this additional variation, and could improve estimates of sector correlations (see Internet Appendix \ref{sec:proj_equi}).

\begin{rem}                                               
Theorem~\ref{thm:dense_multi} is closely related to the Cram{\'e}r--Wold theorem from probability theory. If ridge payoffs $g(a'x)$ are available for all directions $a \in \mathbb{R}^d$, then equality of prices for all such payoffs pins down the (risk-neutral) measure, since ridge functions are dense on compact sets. In our context, the benefit of Theorem \ref{thm:dense_multi} is that it shows the same conclusion holds even when only a restricted set of directions is traded. From a hedging perspective, the density result is also stronger than uniqueness of the measure alone, because it implies that continuous payoffs can be approximated arbitrarily well by tradable static portfolios.
\end{rem} 

\begin{rem}
\citet{ross1976options,nachman1988spanning} noted that payoffs formed by products of call options on multiple assets are sufficient to complete the market. In particular, derivatives with payoffs of the form
\begin{equation*}
\pmax{S_{1,T}-K_1} \pmax{S_{2,T}-K_2},
\end{equation*}
for arbitrary strikes $K_1$ and $K_2$, together with standard call options on $S_{1,T}$ and $S_{2,T}$, span all contingent claims in dimension two.\footnote{\citet{bakshi2000spanning} refer to such securities as correlation options.} In our framework, this market completeness result follows directly from Proposition~\ref{prop:spanning} and the fact that tensor products of spline basis functions are dense in the space of continuous functions on compact sets.  This argument immediately generalizes to arbitrary dimension $d$ by considering tensor products of spline basis functions in $\R_{++}^d$. In practice, however, derivatives involving products of options on more than one underlying are not traded on major exchanges and are typically only available over the counter. For this reason, we focus on market completeness achieved using simple options, which are widely traded and liquid.

\end{rem}

\subsection{Identification and approximation of risk-neutral covariances and correlations}
Theorem~\ref{thm:dense_multi} implies that it is impossible to identify the price of an \emph{arbitrary} claim using options, unless options on infinitely many distinct portfolios are observed. In several cases of empirical interest, however, identification or accurate approximation is still possible. A leading example is the risk-neutral covariance between an individual stock and the market: \citet{martin2019expected} show that this covariance forecasts individual stock returns, so its identification has direct empirical content. Focusing on the two-dimensional case first and letting $R_{\ttt} = w_{1,t} R_{1,\ttt} + w_{2,t} R_{2,\ttt}$, the following identity obtains:
\begin{equation}\label{eq:cov_id}
R_{1,\ttt} R_{\ttt} = \frac{1}{2w_{1,t}} R_{\ttt}^2 + \frac{w_{1,t}}{2} R_{1,\ttt}^2 - \frac{w_{2,t}^2}{2 w_{1,t}} R_{2,\ttt}^2.   
\end{equation}
The prices of each of the payoffs on the right-hand side can be inferred from options on the market index, sector~1, and sector~2, respectively. Hence, in $d=2$, the covariance between any pair of these returns can be identified from option prices.\footnote{This is unsurprising, since $\Var_t^Q R_{\ttt} = w_{1,t}^2 \Var_t^Q R_{1,\ttt} + w_{2,t}^2 \Var_t^Q R_{2,\ttt} + 2 w_{1,t} w_{2,t}\Cov_t^Q\lro{R_{1,\ttt},R_{2,\ttt}}$ and because each individual variance is identified from option prices, the covariance must also be identifiable.} A benefit of the projection approach is that it automatically finds the best replicating portfolio using traded options, regardless of whether an explicit identity such 
as~\eqref{eq:cov_id} is available, which becomes important when the dimension increases.

Generally, the question of identifying the price of a payoff thus depends on whether there is an exact algebraic identity linking the payoff function and a linear combination of ridge functions. It is useful to have a simple algebraic condition that determines whether such a separable identity holds. \citet{diaconis1984nonlinear} derive a simple necessary and sufficient condition for this to hold in $\R^2$.


When $d \ge 3$, the situation becomes more involved. Necessary and sufficient conditions were derived by \citet{lin1993fundamentality}, although they are not straightforward to verify in practice. For completeness, we state their result in Internet Appendix~\ref{app:sufficient} and provide a more elementary argument showing why correlations in dimensions $d \geq 3$ cannot be identified solely from options on the individual sectors and the market portfolio. The following Proposition summarizes this result.

\begin{prop}[Non-replication of the covariance payoff]\label{prop:impossible}
Let $d \ge 3$, and let $R_{i,\ttt}$ denote the return on sector $i$ and $R_{\ttt}$ the return on the market portfolio with weight vector $w \in \R^d$. Suppose that at least two sectors other than $i$ carry nonzero weight. Then no static portfolio of European options on the individual sector returns and on the market return can replicate the payoff $R_{i,\ttt}  R_{\ttt}$. Consequently, the risk-neutral covariance between sector $i$ and the market cannot be identified from options on the individual sectors and on the market portfolio alone.
\end{prop}

To forecast individual stock returns, \citet{martin2019expected} need precisely this risk-neutral covariance between an individual stock and the market. They proceed via a linearization under the assumption that risk-neutral betas are close to one, which makes it possible to estimate the covariance from option prices but imposes a substantive restriction. Proposition~\ref{prop:impossible} clarifies why such a workaround is needed: the covariance is not identified from options on individual stocks and the market alone. 

Projection offers an alternative that imposes no assumption on betas, computes its estimate directly from observed option prices, and accepts the resulting approximation error. The cost is that the weighting density $\omega$ becomes more consequential than in the univariate case. With a single underlying, the choice of $\omega$ becomes asymptotically irrelevant as the strike grid becomes dense, because options span the payoff space. In contrast, with multiple underlyings and only a single market portfolio, options do not span the covariance payoff. The covariance is therefore not point-identified from the observed option prices, so the weighting density continues to affect the limiting projection estimate. This is analogous to the phenomenon emphasized by \citet{baumeister2015sign}: under set identification, prior assumptions continue to matter asymptotically, whereas under point identification they wash out in the limit. For example, product weighting tilts the projection toward independence. Internet Appendix~\ref{sec:proj_equi} formalizes this interpretation and shows that projection nests the CBOE equicorrelation estimator, which corresponds to a Gaussian weighting density whose shrinkage target is equal pairwise correlation.

\subsection{Joint dependence in FX markets}\label{sec:exchange}
The results above show that joint dependence can be estimated reliably from
options only when options are traded on a sufficiently rich set of portfolio
directions. In practice, however, the set of traded portfolio indexes is
limited. For example, alongside the value-weighted SPDR ETF, there is also an
equal-weighted ETF, EQL, but options on the latter began trading only recently.

There is, however, a more precise way of estimating dependence from options in FX markets. The key idea, which has been exploited before
\citep[e.g.,][]{mueller2017international,martin2018options}, is that exchange rates satisfy a
triangular parity relation that embeds information about dependence. We show
what information about dependence is revealed by this relation, and how
projection can be used to estimate it. Our approach extends \citet{martin2018options} in three ways. First, options on different bilateral rates are quoted in different currencies, so combining them in a single-currency portfolio requires a state-dependent change of measure; the projection approach handles this conversion explicitly. Second, our basis consists of vanilla options that are actually quoted, yielding a replicating portfolio that is directly investable from the perspective of a domestic investor. Third, Proposition~\ref{prop:proj_errorBound} provides an explicit finite-sample error bound for the projection-based portfolio, whereas \citet[Result 2]{martin2018options} inverts a Radon transform under the assumption that options on a continuum of linear combinations are observable, with no analogous error guarantee.

For concreteness and in anticipation of the empirical application, let $S_{1,T}$ denote the EUR/CHF exchange rate, $S_{2,T}$ the USD/CHF rate, and $S_{3,T}$ the EUR/USD rate at maturity $T$. By triangular no-arbitrage, $S_{3,T}=S_{1,T}/S_{2,T}$. Hence, options on EUR/USD reveal joint information not captured by options on EUR/CHF and USD/CHF, which only reveal the marginal distributions. Incorporating this additional source of variation should yield a better estimate of joint dependence. Throughout we use the convention that $S_1$ and $S_2$ are quoted in CHF, while $S_3$ is in USD units.\footnote{This convention is the same as for the Bloomberg options data that we use in the empirical application in Section \ref{sec:target_zone}.} 

With $R_{f,\ttt}$ and $R_{f,\ttt}^\$$ denoting the Swiss and U.S.~gross risk-free rates, the call option prices are
\begin{align*}
C_{i,\ttt}(K) &= \frac{1}{R_{f,\ttt}} \Expect_t^{Q}\pmax{S_{i,T} - K}, \quad i = 1,2,\\
C_{\ttt}^{\$}(K) &= \frac{1}{R_{f,\ttt}^\$} \Expect_t^{\Qdollar}\pmax{S_{3,T} - K} ,
\end{align*}
where $Q$ and $\Qdollar$ are the risk-neutral measures using the Swiss and U.S.~money-market accounts as num{\'e}raires, respectively. This distinction is needed because EUR/USD options are USD-quoted. Our convention is that the risk-neutral measure of the domestic investor, which is Swiss in this case, is denoted without superscript. 

Using the change of num{\'e}raire result \citep[Chapter 9]{shreve2004stochastic}, it follows that the Radon-Nikodym derivative between the domestic and foreign risk-neutral measures is given by
\begin{equation*}
\frac{\diff Q_T}{\diff \Qdollar_T} \bigg/ \frac{\diff Q_t}{\diff \Qdollar_t}  =  \frac{R_{f,\ttt}}{R_{f,\ttt}^\$} \frac{S_{2,t}}{S_{2,T}}.
\end{equation*}
Using this result, we obtain the following expression for a judicious choice of payoff function under $Q$
\begin{align*}
\Expect_t^{Q} \lr{ S_{2,T} \pmax{\frac{S_{1,T}}{S_{2,T}} - K} } &= \frac{R_{f,\ttt}}{R_{f,\ttt}^\$} S_{2,t} \Expect_t^{\Qdollar}\lr{\pmax{S_{3,T} - K}}\\
&= R_{f,\ttt} S_{2,t} C_{\ttt}^{\$}(K).
\end{align*}

We consider this payoff because its risk-neutral expectation is directly linked to observed market prices. In particular, the change of num\'eraire converts the USD-quoted EUR/USD option price into a CHF-valued payoff through the factor $S_{2,t}$. A key advantage of the projection approach is that it accommodates this state-dependent num\'eraire adjustment when combining options quoted in different currencies, thereby yielding a theoretically consistent CHF-denominated replicating portfolio. By contrast, much of the existing FX literature effectively ignores this state dependence, or treats the conversion kernel as approximately constant when extracting dependence measures from option prices (e.g., \citet{mueller2017international}). Further, it is possible to obtain the expected value of EUR/CHF and USD/CHF under the Swiss risk-neutral measure because
\begin{equation*}
\Expect_t^{Q} S_{1,T} = \frac{R_{f,\ttt}}{R_{f,\ttt}^{\text{\euro}}} S_{1,t} = F_{1,\ttt}, \quad \Expect_t^{Q} S_{2,T} = \frac{R_{f,\ttt}}{R_{f,\ttt}^{\$}} S_{2,t} = F_{2,\ttt},
\end{equation*}
where $F_{i,\ttt}$ denotes the $T$-maturity forward FX rate for pair $i = 1,2$.

The foregoing discussion suggests a way to estimate the covariance and correlation between EUR/CHF and USD/CHF. Specifically, we project the payoff
\begin{equation*}
\lro{S_{1,T} - F_{1,\ttt}}\lro{S_{2,T} - F_{2,\ttt}}
\end{equation*}
onto basis functions of the form
\begin{equation*}
1,\ S_{1,T},\ \pmax{S_{1,T} - K},\ S_{2,T},\ \pmax{S_{2,T} - K},\ S_{2,T}\pmax{\frac{S_{1,T}}{S_{2,T}} - K}.
\end{equation*}
Taking risk-neutral expectations under the Swiss money-market num\'eraire, each basis payoff reduces to an observable market object: a constant, a forward rate, a CHF-denominated option price, or a USD-denominated option price multiplied by known discounting and conversion terms. 


If options on all three bilateral rates are available and, for each rate, the assumptions of Proposition \ref{prop:spanning} hold, then static portfolios in these options can uniformly approximate any payoff of the form
\begin{equation}\label{eq:gen_ridge}
g(S_{1,T},S_{2,T}) \;=\; g_1(S_{1,T}) \;+\; g_2(S_{2,T}) \;+\; S_{2,T}\, g_3\!\left(\frac{S_{1,T}}{S_{2,T}}\right),
\end{equation}
where each $g_i$ is continuous. This class is not dense in the space of continuous functions on compact sets. For example, the function $g(x,y)=xy$ cannot be represented in the form \eqref{eq:gen_ridge}. Hence, vanillas on the three bilateral exchange rates do not in general identify the covariance, nor do they characterize the full joint risk-neutral distribution. Nevertheless, our simulations show that projection onto \eqref{eq:gen_ridge} yields highly accurate approximations for the covariance and for economically relevant measures of joint tail risk. In the empirical application, we use this approach to study the effect of central bank announcements on the joint distribution of EUR/CHF and USD/CHF.

\section{Simulation}\label{sec:simulate}

\subsection{Univariate projection}\label{sec:uni_proj}
To illustrate the benefits of the projection-based approach, we consider the problem of approximating the value of the SVIX and VIX discussed in Examples \ref{exmp:var}--\ref{exmp:vix}. Across Monte Carlo iterations, the number of strikes varies over $\set{5, 10, , \dots, 130}$; in each iteration, strikes are drawn uniformly from a fixed range where the endpoints cover 90\% of the return probability distribution.  This allows us to study the approximation error as a function of the number of strikes available in the market. In addition, we  consider a design where the number of strikes is fixed, but the range of the strike prices is increasing to cover a larger portion of the support. Throughout, the weighting density for projection is based on the VG process.

Based on the strikes, we obtain the corresponding call and put option prices from the two-factor stochastic-volatility model with jumps of \citet{andersen2015risk}. More details on the simulation and calibration of this model are given in Internet Appendix \ref{app:simulation}.\footnote{In this appendix, we also show that the projection approach is robust to alternative option pricing models.} The accuracy of the approximation for each number of strikes is measured by the relative error,
\begin{equation*}
\text{Relative error} = \frac{\abs{\widehat{\text{SVIX}} - \text{SVIX}}}{\text{SVIX}},
\end{equation*}
where $\widehat{\text{SVIX}}$ is the SVIX estimate obtained by either the CM approximation \eqref{eq:cm_approx}, or the projection method. The relative error for VIX is defined analogously.

\begin{figure}[htb!]
\centering
\begin{subfigure}{0.48\linewidth}
\includegraphics[width=\linewidth]{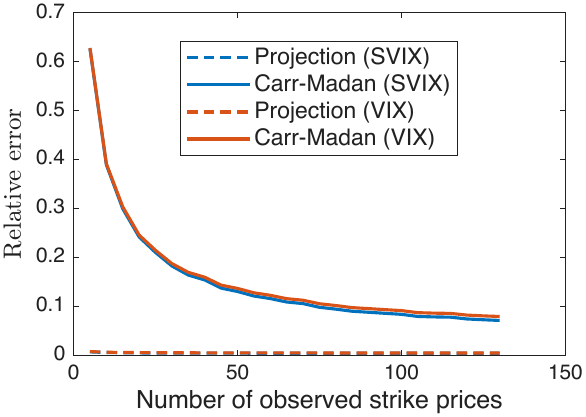}
\caption{}
\label{fig:aft_fixedRange}
\end{subfigure}
\begin{subfigure}{0.48\linewidth}
\includegraphics[width=\linewidth]{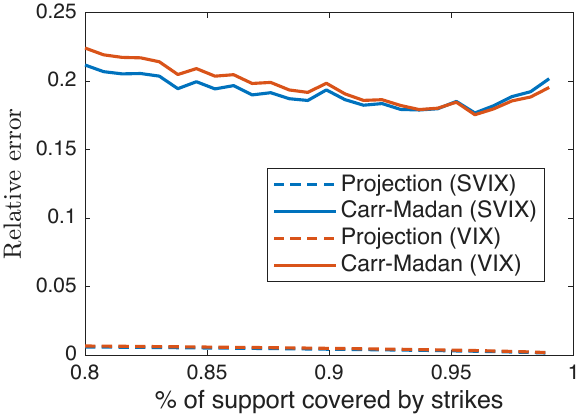}
\caption{}
\label{fig:aft_varRange}
\end{subfigure}
\caption{\textbf{Relative approximation error}. The left panel shows the convergence rate as a function of the number of strikes and as a function of the strike range (right panel).}\label{fig:mse}
\end{figure}

Figure~\ref{fig:aft_fixedRange} shows convergence as the number of strikes increases. Even with as few as 5 strikes, the projection error is almost zero. In contrast, the CM portfolio achieves a relative error of 60\% in the sparse strike case. The convergence of the CM method is gradual and with 130 strikes the relative error is between 5\% and 10\%, depending on the regime. Moreover, at every strike count, the projection estimate is pointwise closer to SVIX/VIX than the CM estimate. For the SVIX, we know that the CM estimate is downward biased (see Figure~\ref{fig:approx_svix}), the projection estimate therefore lies above the CM estimate in every simulation. This is a useful benchmark when applying the projection approach to actual data, where we expect the same relation. In Internet Appendix~\ref{app:svix_vix}, we find that the projection SVIX estimate is almost always larger than the CM estimate.\footnote{There are a handful of cases when this relation is violated, due to extrapolation: on these days no listed expiration falls below 30 days, so the 30-day estimate must be backward-extrapolated from two longer-dated maturities, one of which has a thin option chain that biases the CM estimate downward. See Internet Appendix~\ref{app:svix_vix} for more explanation.}

Panel~\ref{fig:aft_varRange} shows the convergence of the relative error when the strike range expands, holding the number of strikes fixed at $n_k = 30$. The projection error declines steadily as the strike range widens and is already close to zero when the strikes cover only 80\% of the support of the return distribution. In contrast, the CM error stays in the 20-27\% range across both models and improves only modestly as the strike range increases. The improvement is even non-monotone: as the range widens, the average strike spacing becomes larger, which eventually offsets the benefit of better tail coverage because the accuracy of the integral approximation deteriorates as the spacing increases.\footnote{Replacing the trapezoidal rule by Simpson's rule yields essentially identical results, because the CM error in this regime is driven by tail truncation and finite grid spacing rather than the choice of quadrature; projection continues to dominate by an order of magnitude.} Overall, the projection estimate is much more accurate than the CM approximation in every panel of Figure~\ref{fig:mse}.

\subsection{Multivariate projection for exchange rates}
We simulate exchange-rate outcomes under the risk-neutral measure from a bivariate normal distribution:
\begin{equation*}
\begin{bmatrix}
    S_{1,T}\\
    S_{2,T}
\end{bmatrix}
\sim \mathsf{N}\lro{
\begin{bmatrix}
1\\
1
\end{bmatrix},
\begin{bmatrix}
0.1^2 & 0.1 \cdot 0.05 \cdot \rho\\
0.1 \cdot 0.05 \cdot \rho & 0.05^2
\end{bmatrix}
}.
\end{equation*}
Across Monte-Carlo iterations, we let $\rho$ cycle over an equally spaced grid on  $[-0.99,0.99]$.  For the option inputs, we take five strikes each on $S_{1,T}$, $S_{2,T}$, and $S_{1,T}/S_{2,T}$. The strikes are evenly spaced between the 5th and 95th percentiles of the respective marginal distributions. This choice mirrors OTC FX practice: quotes out to the 5-delta put and 95-delta call (under forward-delta conventions) roughly correspond to the 5th and 95th percentiles for 1-month tenors. 
For the weighting density, we use the product of two VG densities, each calibrated to most closely match the observed option prices. These marginals are misspecified by design. We do so intentionally, since in practice the VG proxy will likewise be a misspecified approximation of the true risk-neutral marginals. As the strike grid becomes dense, this misspecification does not matter, but for the sparse-strike setting considered here it can play a role.

We then project the payoff $(S_{1,T}-1)(S_{2,T}-1)$ onto the span of the payoffs
\begin{equation*}
1, \quad
S_{1,T}, \quad 
\pmax{S_{1,T}-K_{1j}},\quad
S_{2,T}, \quad 
\pmax{S_{2,T}-K_{2j}},\quad
S_{2,T} \pmax{S_{1,T}/S_{2,T}-K_{3j}},
\end{equation*}
with strikes $\{K_{1j},K_{2j},K_{3j}\}_{j=1}^5$ generated as above. To recover the correlation, we also estimate the standard deviations by projecting $(S_{1,T}-1)^2$ onto the constant function, $S_{1,T}$, and options on $S_{1,T}$ (and analogously for $S_{2,T}$).

In addition, we consider a setting where $S_{2,T}$ is generated as above and then perturbed to $\tilde{S}_{2,T} = S_{2,T} + 0.1 S_{1,T}^3$. $\tilde{S}_{2,T}$ is further normalized so that the mean is 1.  We estimate the correlation between $S_{1,T}$ and $\tilde{S}_{2,T}$ to introduce nonlinear dependence and verify that our results are not driven by the normality assumption. 

\begin{figure}[!htb]
    \centering
    \begin{subfigure}{0.49\linewidth}
    \includegraphics[width=\linewidth]{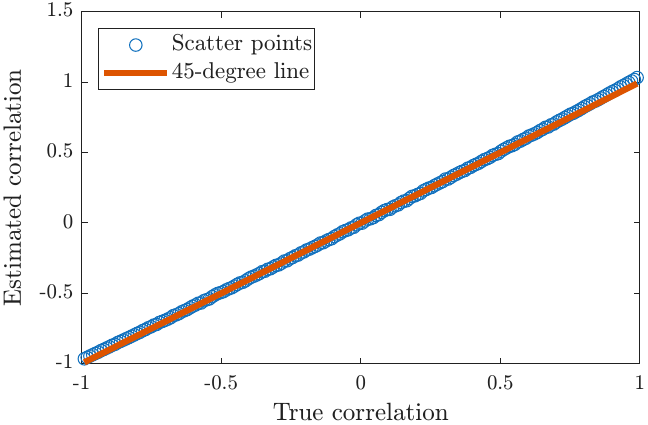}
    \caption{Multivariate normal}
    \end{subfigure}
    \begin{subfigure}{0.49\linewidth}
    \includegraphics[width=\linewidth]{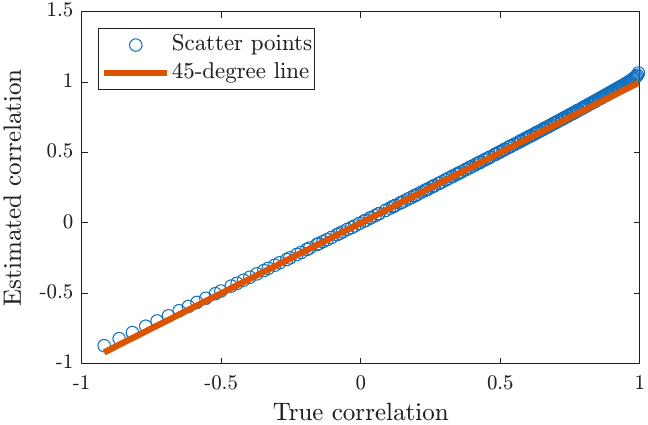}
    \caption{Nonlinear dependence}
    \end{subfigure}
    \begin{subfigure}{0.49\linewidth}
    \includegraphics[width=\linewidth]{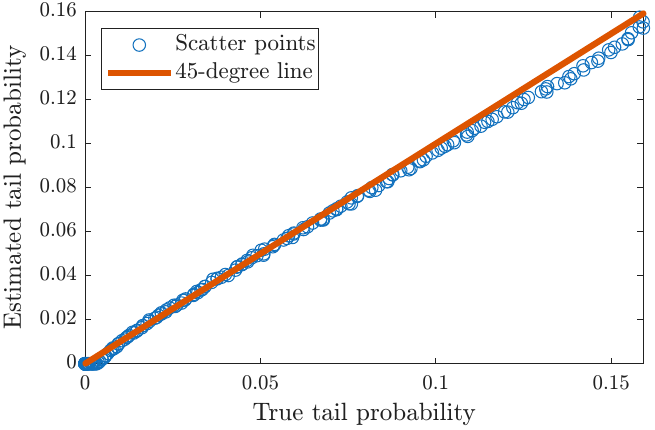}
    \caption{Multivariate normal}
    \end{subfigure}
    \begin{subfigure}{0.49\linewidth}
    \includegraphics[width=\linewidth]{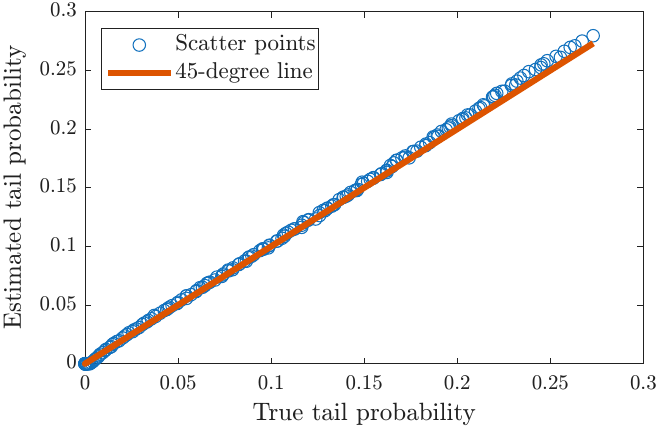}
    \caption{Nonlinear dependence}
    \end{subfigure}
    \caption{\textbf{Estimated correlation and joint tail risk in exchange-rate markets.}
    Each point is one of 200 grid points on $[-0.99,0.99]$. Top: true correlation versus its projection-based estimate. Bottom: true joint left-tail probability $Q_t(S_{1,T} \le 0.95, S_{2,T} \le 0.95)$ versus its projection-based estimate.}
    \label{fig:exchange}
\end{figure}

The upper panels in Figure~\ref{fig:exchange} report the results for different values of the correlation parameter. In both panels, the projection approach recovers the true correlation with high accuracy: the scatter points lie nearly on the $45^\circ$ line. This is encouraging because the correlation is not exactly identifiable within the restricted function class (see Section~\ref{sec:exchange}). We conclude that projection delivers an excellent approximation to the true correlation in the FX setting, regardless of whether the dependence is linear or nonlinear in our simulated data.

In the bottom panels, we use the same generated data to estimate the joint probability that both returns are below a certain threshold, which can be interpreted as a measure of joint tail risk. Specifically, we estimate $Q_t(S_{1,T} \le 0.95, S_{2,T} \le 0.95)$ by projection, using the target payoff 
\begin{equation*}
\ind{S_{1,T} \le 0.95} \ind{S_{2,T} \le 0.95}
\end{equation*}
The bottom panels of Figure \ref{fig:exchange} report fitted versus true probabilities. The estimates line up closely with the $45^\circ$ line, albeit slightly less tightly than for the correlation results, indicating that the projection method recovers joint tail probabilities with high accuracy. This motivates the use of projection to recover joint crash probabilities in the SNB event study of Section~\ref{sec:target_zone}, where the joint crash probability is the central object.

\section{Empirical application: Cross-currency contagion under a policy floor}\label{sec:target_zone}

Our empirical application uses an event-study design based on FX options. This setting is well suited to the projection approach because the method delivers accurate estimates of joint moments from option prices. It also has two features that make the projection approach particularly attractive relative to the CM approximation. First, for each currency we observe only five option quotes, so finite-sample performance is especially important. Second, FX option strikes provide limited tail coverage, making it valuable to use a method that approximates the replicating payoff well over the entire domain, including states beyond the outermost strike.

We focus on two unexpected SNB announcements that define the empirical setting. On September 6, 2011, the SNB unexpectedly introduced a floor of 1.20 on the EUR/CHF exchange rate. Because the exchange rate moved from approximately 1.11 to above 1.20 within the same trading day, our end-of-day option quotes already reflect the post-announcement regime. On January 15, 2015, the SNB unexpectedly removed the floor while EUR/CHF was at the boundary value of 1.20, causing an immediate drop to 0.97. The floor had remained in place throughout the intervening three-year period. Our objective is to study how these announcements altered dependence across exchange rates and, in particular, how they changed the price of an Arrow security that pays one Swiss Franc when both EUR/CHF and USD/CHF returns crash.

Our analysis connects to the framework of \citet{haddad2025whatever}, who show that conditional policy promises redistribute Arrow prices across states, creating asymmetric ``policy puts'' that provide greater price support in bad states. The SNB floor is a natural application of this logic: the commitment to maintain EUR/CHF above 1.20 generates a policy put that compresses state prices in the left tail. Closely related, \citet{krugman1991target} argues that currency target zones induce option-like payoffs, as the exchange rate becomes a nonlinear function of fundamentals (see also \citet{bertola1992target,bertola1993stochastic}). These models predict that state prices in \emph{both} tails compress when the exchange rate is near the boundary (the ``honeymoon effect''), rather than only in the left tail. However, because the SNB implements the floor through interventions that expand the Swiss monetary base, the policy affects all CHF-denominated exchange rates, not just EUR/CHF. A univariate framework, whether in \citet{haddad2025whatever} or in the target-zone literature, can capture these marginal effects, but it cannot determine whether the policy also changes dependence across exchange rates.

\subsection{Data}\label{sec:tz_data}

Before turning to the empirical results, we describe the EUR/CHF, USD/CHF, and EUR/USD options data used in the analysis.

From Bloomberg we obtain daily end-of-day OTC quotes using the BGN (Bloomberg Generic) composite source. For each of the three currency pairs we retrieve: (i)~spot exchange rates; (ii)~1-month money-market deposit rates for each currency\footnote{EURIBOR for EUR, USD~LIBOR for USD, CHF~LIBOR for CHF.}; and (iii)~1-month constant-maturity implied volatilities at the standard OTC smile pillars: ATM delta-neutral, 10- and 25-delta risk reversals and butterflies. All series are sampled at the same daily frequency using a single pricing source (BGN) to ensure temporal alignment across instruments.\footnote{The BGN composite aggregates end-of-day dealer contributions. Because FX spot, FX option volatility, and interbank rate fixings are produced by different contributor panels, their snapshot times can in principle differ. We retrieve the \texttt{TRADING\_DAY\_END\_TIME\_EOD} field from Bloomberg for each spot and implied volatility ticker and verify that they share the same end-of-day snapshot time (22:59:59~UTC). This ensures that the spot and volatility quotes are aligned to the same Bloomberg end-of-day timestamp.}

We construct 1-month forward rates via covered interest parity. Wing implied volatilities are constructed from the ATM, risk-reversal, and butterfly quotes via the standard market identities $\sigma_{\text{call}} = \sigma_{\text{ATM}} + \text{BF} + \tfrac{1}{2}\text{RR}$ and $\sigma_{\text{put}} = \sigma_{\text{ATM}} + \text{BF} - \tfrac{1}{2}\text{RR}$. We then invert the spot-delta formula to obtain strike prices and compute option prices using the Garman--Kohlhagen model.\footnote{The Garman--Kohlhagen formula serves only as the quoting convention that maps implied volatilities to prices; it is not imposed as the true pricing model.} This yields five option prices per currency pair per day (10-delta put, 25-delta put, ATM, 25-delta call, 10-delta call). Our sample spans July~2008 to December~2022. For each day and currency, we use VG weighting for projection, with the density calibrated to that day's option prices.\footnote{When the VG calibration is degenerate, we fall back to a Black--Scholes log-normal density.}

The projection method evaluates cross-rate option payoffs at $S_{\text{EUR/USD}} = S_{\text{EUR/CHF}}/S_{\text{USD/CHF}}$, imposing the no-arbitrage triangle on the state-space grid by construction. The market data, however, are obtained independently for each pair, so the triangular relation $F_{\text{EUR/USD}} = F_{\text{EUR/CHF}}/F_{\text{USD/CHF}}$ is not enforced on the observed forwards. We verify that violations are negligible. Over the full sample (3,407 trading days), the forward triangular deviation, as measured by 
\begin{equation*}
 \frac{F_{\text{EUR/USD}} - F_{\text{EUR/CHF}} / F_{\text{USD/CHF}}}{F_{\text{EUR/USD}}},
\end{equation*}
has a median below 0.1~basis points and a standard deviation of 2.4~basis points, which is economically small. Crucially, the deviations on the two SNB event days are small: $-0.7$~bps on September~6, 2011 (floor introduction) and $7.2$~bps on January~15, 2015 (floor removal). Figure \ref{fig:triangle_deviation} in the Internet Appendix shows that the largest outliers (up to 24~bps) occur during the 2008--2009 financial crisis, when FX market microstructure was strained, and none fall within our event windows.

\subsection{Contagion decomposition and the joint honeymoon}\label{sec:tz_decomp}

\begin{figure}[htb!]
\centering
\begin{subfigure}{0.49\linewidth}
\includegraphics[width=\linewidth]{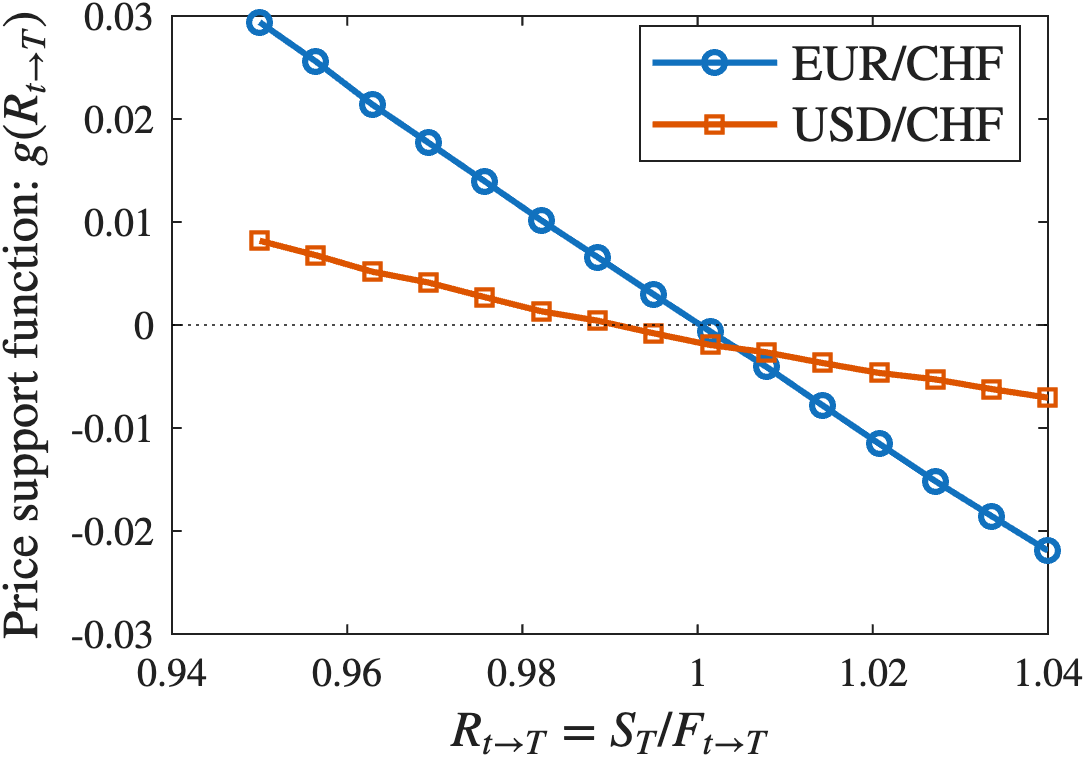}
\caption{Floor introduction }
\end{subfigure}
\begin{subfigure}{0.49\linewidth}
\includegraphics[width=\linewidth]{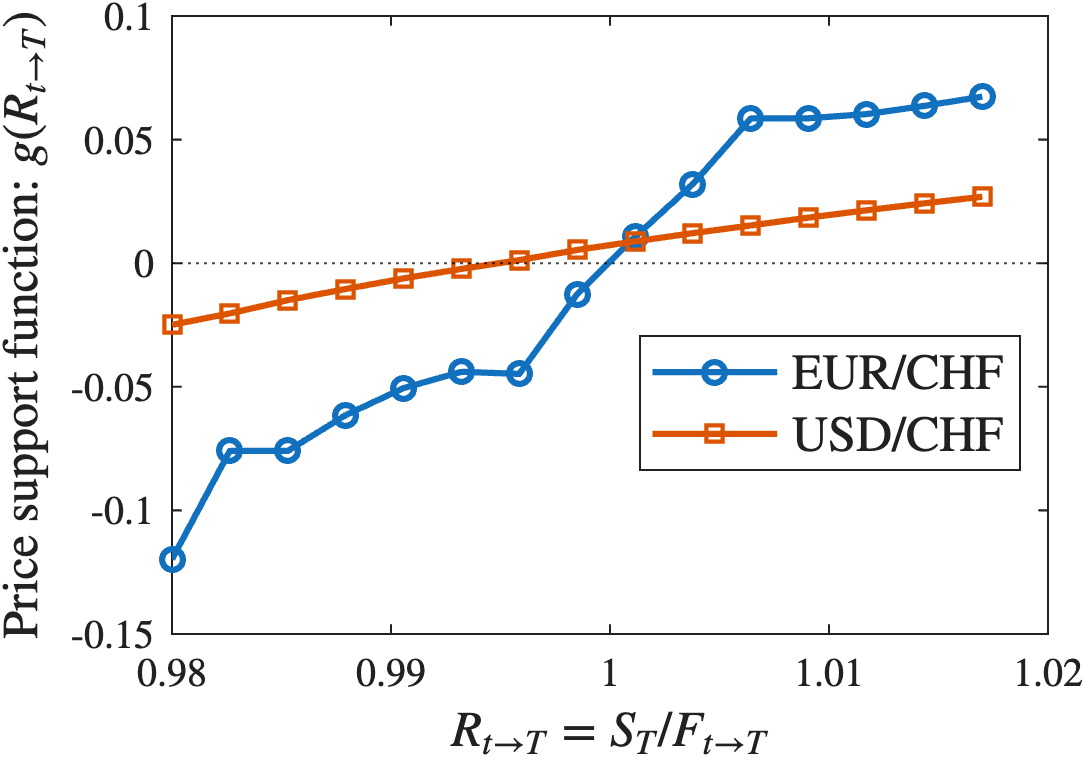}
\caption{Floor removal}
\end{subfigure}
\caption{\textbf{Price support function.} Each panel plots the price support function \eqref{eq:p_support} for the SNB floor introduction and removal respectively.}
\label{fig:price_support}
\end{figure}

Let $R_{i, \ttt} = S_{i,T}/F_{i,\ttt}$ be the forward normalized return on currency $i = 1, 2$, where we continue to identify currency~1 with EUR/CHF and currency~2 with USD/CHF. By construction $\Expect_t^Q R_{i, \ttt} = 1$, where $Q$ refers to the risk-neutral measure from the perspective of a Swiss investor. We first analyze the impact of the SNB policy on the currencies separately. \citet{haddad2025whatever} define a price support function by 
\begin{equation}\label{eq:p_support}
g(R) = \frac{F_{t+1 \to T+1}^{Q -1} \lro{F_{\ttt}^Q(R)} - R }{R},
\end{equation}
where $F_{t+1 \to T+1}^{Q -1}$ is the quantile function of the first post-announcement trading day, and $F_{\ttt}^Q$ is the usual risk-neutral CDF right before the announcement. This function measures how much each return increases/decreases due to the policy, relative to when there was no policy.\footnote{See \citet{haddad2025whatever} for the conditions necessary to interpret $g$ as measuring the state-contingent price impact.} The price support functions for both policy announcements are shown in Figure \ref{fig:price_support}. In contrast to the asymmetric ``policy put'' documented by \citet{haddad2025whatever} for Federal Reserve asset purchases, the estimated price support functions are approximately symmetric around zero.  This is consistent with the target-zone framework of \citet{krugman1991target}, who argues that when the exchange rate is close to the boundary it becomes insensitive to fundamentals, which compresses \emph{both} tails of the return distribution (the ``honeymoon effect''). The removal reverses the pattern with a larger magnitude: the function shifts from compressing both tails to expanding them.

The price support function captures the marginal effect of the policy on each currency, but is silent on whether the policy also changes the dependence between them. We therefore turn to the analysis of the joint distribution. Specifically, define the joint Arrow security that pays out in a crash state by
\begin{equation*}
\texttt{Crash}_{\ttt} =\ind{R_{1,\ttt} \le 0.97} \ind{R_{2,\ttt} \le 0.97}.
\end{equation*}
The price of this joint claim, which equals the joint risk-neutral probability up to discounting, is estimated using projection. We also estimate the risk-neutral correlation.

\begin{figure}[htb!]
\centering
\begin{subfigure}{0.49\linewidth}
\includegraphics[width=\linewidth]{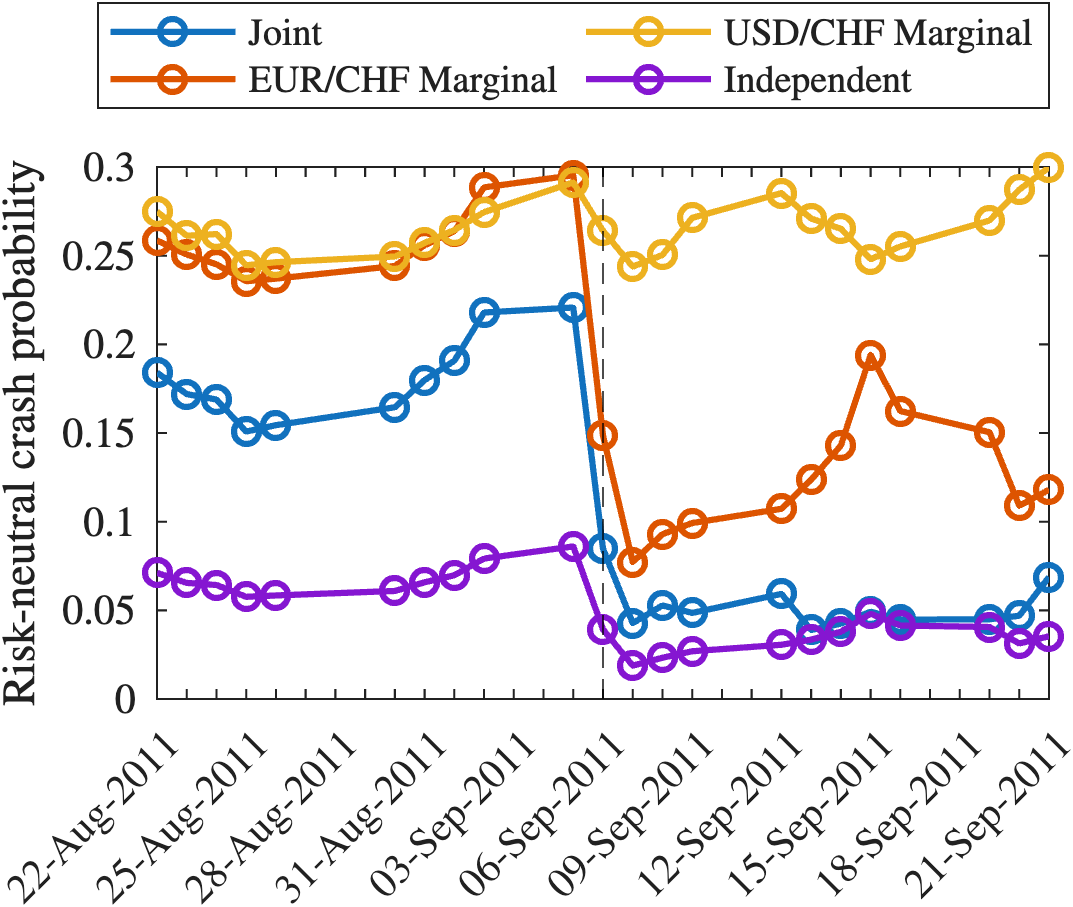}
\caption{Floor introduction (Sep 2011)}
\end{subfigure}
\begin{subfigure}{0.49\linewidth}
\includegraphics[width=\linewidth]{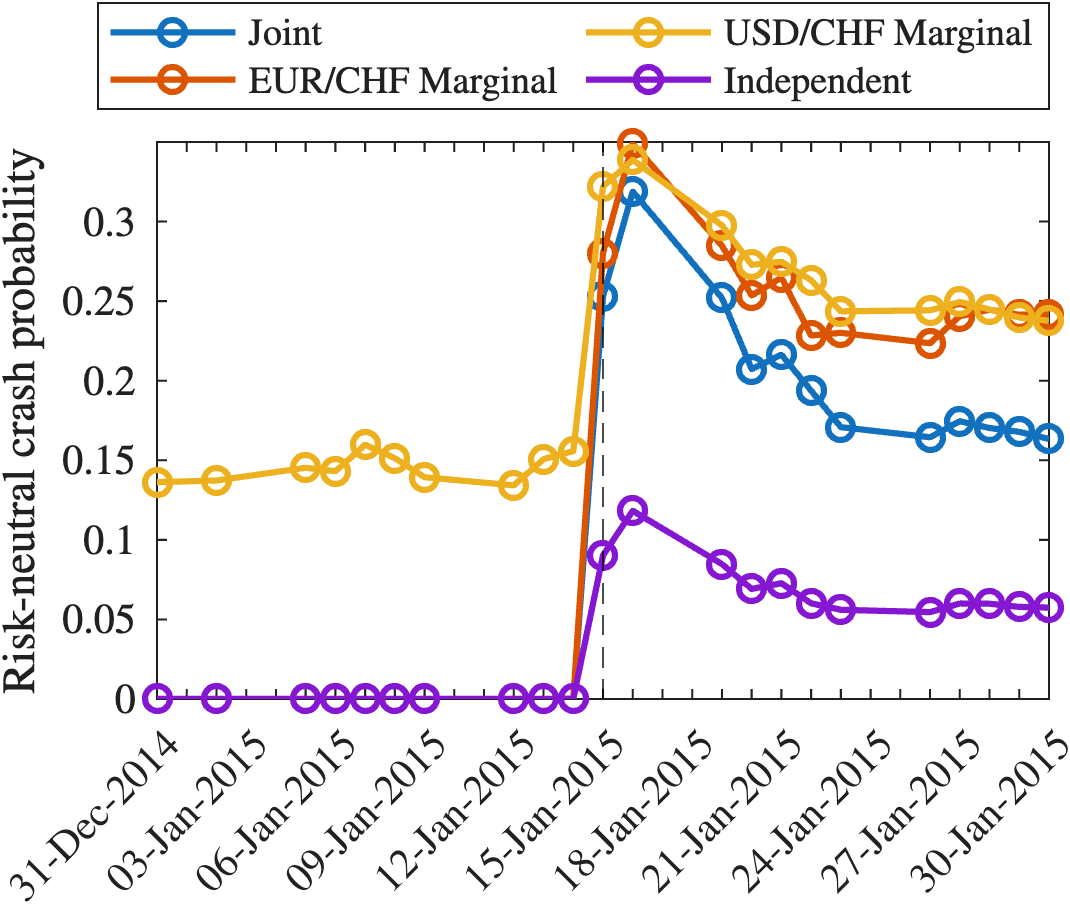}
\caption{Floor removal (Jan 2015)}
\end{subfigure}
\caption{\textbf{Joint crash probability around the SNB floor.} Each panel plots the risk-neutral probability that both EUR/CHF and USD/CHF monthly returns fall below $-3\%$ (blue), together with the independence benchmark (purple). The red and yellow lines denote the EUR/CHF and USD/CHF marginal crash probabilities.}
\label{fig:snb_joint}
\end{figure}

Figure~\ref{fig:snb_joint} plots the 30-day joint crash probability around the floor introduction and removal.\footnote{Figure \ref{fig:placebo} in the Internet Appendix shows that the changes in joint crash probability at both announcements are the two largest in absolute value in the empirical distribution.} For comparison, we also plot the independence benchmark, $Q_t\lro{R_{1,\ttt} \le 0.97} \cdot Q_t\lro{R_{2,\ttt} \le 0.97}$, which uses the updated marginals but ignores dependence. The gap between the two series (the dependence wedge) represents the excess joint crash probability relative to the independence benchmark, capturing how the market prices dependent tail risk.

We decompose the total change in the joint crash probability as
\begin{align}\label{eq:decomp}
\underbrace{\Delta Q_t(R_{1,\ttt} \le 0.97,\, R_{2,\ttt} \le 0.97)}_{\text{total}} &= \underbrace{\Delta \lr{Q_t(R_{1,\ttt} \le 0.97) \cdot Q_t(R_{2,\ttt} \le 0.97)}}_{\text{marginal channel}} \nonumber \\
&+ \underbrace{\Delta\, \text{Wedge}}_{\text{dependence channel}}.
\end{align}
The marginal channel captures changes in the individual distributions and corresponds to the standard prediction of univariate target-zone models. For instance, \citet{kremens2019theory} document a compression of EUR/CHF risk-neutral volatility under the floor and a sharp rise at its removal. However, such marginal compression alone does not pin down whether the joint distribution changed beyond what marginal moves would imply. The dependence channel captures changes in joint tail pricing beyond those implied by the marginal distributions. In the terminology of \citet{forbes2002no}, the marginal channel corresponds to “interdependence,” mechanical co-movement induced by changes in marginal volatility, whereas the dependence channel corresponds to “contagion,” that is, a change in the joint dependence structure. Crucially, unlike sample correlations, which \citet{forbes2002no} show can be mechanically biased upward during high-volatility episodes, the dependence channel is less exposed to this critique because it benchmarks the joint crash probability against the product of the prevailing marginals. Visually, in both cases the SNB announcement caused a large shift in both marginal and joint crash probabilities. At the same time, the effects are asymmetric. When the floor was introduced, the marginal and joint risk-neutral crash probabilities did not immediately fall to zero, suggesting that markets initially had some doubts about the credibility of the announcement. By contrast, when the floor was abandoned, both the EUR/CHF crash probability and the joint crash probability jumped immediately from near zero to above 0.25. Relatedly, \citet{jermann2017financial} finds that the floor's credibility increased over time using EUR/CHF options data.

Table~\ref{tab:snb_summary} reports the decomposition~\eqref{eq:decomp} together with the levels of the dependence wedge, joint crash probability, and risk-neutral correlation in tight windows around the two events. For both events, approximately two-thirds of the total change in the joint crash probability is attributable to the dependence channel, showing that the SNB policy altered the cross-currency joint tail beyond what would be implied by marginal distributions alone.

After the floor introduction, the dependence wedge falls from 0.131 to 0.030, while the risk-neutral correlation declines from 0.82 to 0.38. The floor therefore generates a \emph{joint} honeymoon effect: cross-currency crash risk declines by more than would be implied by compression of the EUR/CHF marginal distribution alone. The economic magnitude is also substantial: the floor reduced the price of the joint crash claim from 0.210 to 0.057, a 73\% decline in the cost of hedging joint currency tail risk. The removal reverses this pattern sharply: the wedge rises from $0$ to 0.167, and the correlation increases from 0.24 to 0.91. In this sense, the removal undoes the joint honeymoon almost immediately. Taken together, these results show that a policy targeting a single exchange rate can fundamentally reshape the joint risk structure of the broader currency market.

\begin{table}[htb!]
\centering
\small
\begin{tabular}{l d{2.3} d{2.3} d{2.3} d{1.2} d{1.2}}
\toprule
 & \multicolumn{1}{c}{Wedge} & \multicolumn{1}{c}{Joint crash} & \multicolumn{1}{c}{$\rho^Q$} & \multicolumn{1}{c}{$\Delta$Marg.} & \multicolumn{1}{c}{Frac.} \\
\midrule
\multicolumn{6}{l}{\textit{Panel A: Floor introduction}} \\
Pre  [$-5$, $-1$]   & 0.131  & 0.210  & 0.82  &        &      \\
Post [$0$, $+5$]    & 0.030  & 0.057  & 0.38  &        &      \\
Change              & -0.101 & -0.153 & -0.44 & -0.051 & 0.66 \\
Placement $p$       & {<}0.001 & {<}0.001 & {<}0.001 &  &  \\
\addlinespace
\multicolumn{6}{l}{\textit{Panel B: Floor removal}} \\
Pre  [$-5$, $-1$]   & 0.000  & 0.000  & 0.24  &        &      \\
Post [$0$, $+5$]    & 0.167  & 0.258  & 0.91  &        &      \\
Change              & 0.167  & 0.258  & 0.66  & 0.091  & 0.65 \\
Placement $p$       & {<}0.001 & {<}0.001 & {<}0.001 &  &  \\
\bottomrule
\end{tabular}
\caption{\textbf{Contagion decomposition and joint honeymoon effect.} The table reports the dependence wedge, joint crash probability, and risk-neutral correlation in tight windows around both SNB events. The pre-event window covers days $[-5, -1]$ and the post-event window covers days $[0, +5]$ relative to the announcement. The Change row decomposes the total change in the joint crash probability into a marginal channel ($\Delta$Marg.) and a dependence channel (the change in the Wedge) via \eqref{eq:decomp}; ``Frac.'' is the share attributable to the dependence channel. ``Placement $p$'' is a rank-based $p$-value: for each non-event date $t$ in the sample, we compute the placebo statistic $\Delta_t := \mathrm{mean}(\,\cdot\,;[t,t+5]) - \mathrm{mean}(\,\cdot\,;[t-5,t-1])$ for the column variable, and report $\hat p = (1 + \#\{|\Delta_t| \ge |\mathrm{Change}|\})/(N+1)$, where $N = 3{,}313$ is the number of placebo dates.}
\label{tab:snb_summary}
\end{table}

\subsection{Validation of the replicating portfolio}
\begin{figure}[!htb]                                                            \centering  
\begin{subfigure}{0.49\linewidth}                                               \includegraphics[width=\linewidth]{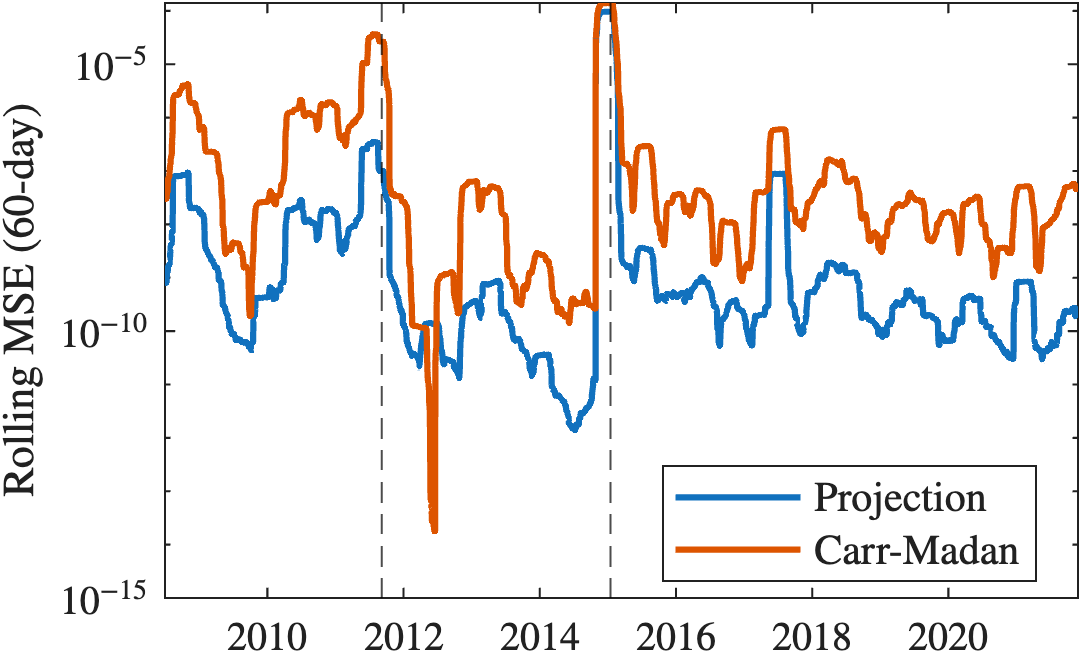}
\caption{OOS variance error}
\label{fig:oos_var}
\end{subfigure}   
\begin{subfigure}{0.49\linewidth}                                               \includegraphics[width=\linewidth]{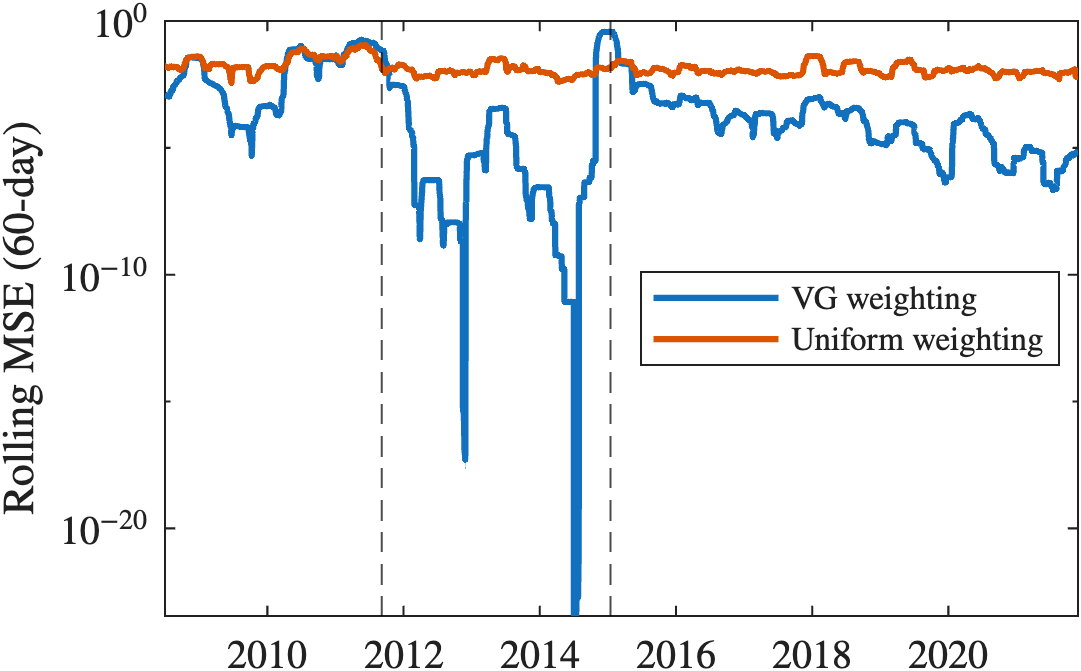}      \caption{OOS tail-risk error}     
\label{fig:oos_joint_error}
\end{subfigure}                                           
\caption{\textbf{Out-of-sample hedging error.}
The left panel plots the squared difference between the value of the replicating
portfolio formed at time~$t$ and the realized variance payoff $(R_{1,t\to T}-1)^2$
at maturity, comparing the projection portfolio with the CM
approximation. The right panel plots the analogous squared difference for the joint-crash claim $\texttt{Crash}_{\ttt}$, comparing the projection portfolio under VG
weighting (baseline) with one under uniform weighting on
$[0.9\,\underline{K},\,1.1\,\overline{K}]$ for each currency. In both panels, errors
are averaged within a 60-day rolling window to obtain the MSE; the $y$-axis is on a
log scale and dashed vertical lines mark the two SNB announcement dates.}
\label{fig:oos_hedging_error}
\end{figure}

The estimates in this section rely on the accuracy of the projection-implied replicating portfolio. We validate its accuracy by comparing the realized payoff at
expiry to the payoff of the projection-based replicating portfolio formed
at time~$t$, separately for the EUR/CHF variance claim,
$(R_{1,t\to T} - 1)^2$, and the joint crash indicator.\footnote{The realized return comes from the physical measure and is therefore not the pricing object of interest. Nevertheless, this exercise is informative because it shows the state-by-state replication accuracy of the portfolio, which is crucial for estimating risk-neutral moments.} We focus on the
variance claim first. Because this claim is univariate, the CM approximation provides a natural benchmark. With a continuum of observed
strikes both methods recover the variance payoff exactly; with only five
out-of-the-money strikes per day, the projection's weights and the CM weights differ meaningfully and produce
distinct hedging errors. Figure~\ref{fig:oos_var}  plots the 60-day rolling mean squared
hedging error for the two portfolios. The projection portfolio achieves
a smaller error than CM on 85\% of the 3{,}199 trading days in the
sample, with the gap most pronounced around the SNB event days and other
periods of elevated volatility. In aggregate, projection reduces the
out-of-sample root mean squared hedging error by~28\% relative to CM.

Figure~\ref{fig:oos_joint_error} turns to the joint-crash claim and compares the
projection portfolio formed under VG weighting (the baseline) with one formed under
uniform weighting on $[0.9\,\underline{K},\, 1.1\,\overline{K}]$, where
$\underline{K}$ and $\overline{K}$ denote the lowest and highest observed strikes
for each currency on a given day. Relative to the VG density, the uniform density places
substantially more mass in the tails. Across most of the sample, the VG-weighted
portfolio achieves a noticeably lower 60-day rolling mean squared hedging error.
The pattern reverses around the SNB floor removal in January~2015, when the
realized exchange-rate jump falls in the deep left tail and the uniform-weighted
portfolio, by placing mass there ex~ante, hedges the realized joint crash
more accurately. The trade-off is clear: VG weighting prioritizes accurate
state-by-state replication where the risk-neutral measure concentrates its mass,
while uniform weighting trades that off for accurate replication when the
realized return lands in the deep tails.\footnote{The uniform density is a poor approximation to the true risk-neutral density. In unreported figures, more plausible parametric alternatives such as a calibrated log-normal yield OOS hedging performance very close to the VG baseline.}

\FloatBarrier

\section{Conclusion}\label{sec:conclusion}
Recovering the joint risk-neutral distribution from option prices has remained an open problem since \citet{ross1976options}. Options on individual assets reveal each asset's marginal distribution but not how the assets move together. Existing approaches for estimating dependence require contracts that are not traded in practice, such as basket options or options on a continuum of portfolios of the underlying assets.

We propose a projection method to address this problem. The idea is to construct a replicating portfolio of traded options whose weights minimize the approximation error to the target payoff. The portfolio is accurate across the entire state space, works in any number of dimensions, and admits a finite-sample error bound that can be interpreted as a quantitative measure of market incompleteness. No analogous bound is available for the widely used CM approach. Using approximation theory, we derive conditions for risk-neutral moments to be identified from option prices, and show that in practice covariances and other joint moments generally cannot be pinned down from simple options alone.

FX markets provide a particularly clean environment for approximating joint dependence with traded options. Here, projection delivers highly accurate joint moments using readily available option quotes. In an event study around two unexpected SNB announcements regarding the EUR/CHF floor, we find that the intervention fundamentally altered the pricing of cross-currency tail risk, rather than merely shifting the marginal distributions. Even after accounting for contemporaneous changes in those marginals, about two-thirds of the change in the joint crash probability of EUR/CHF and USD/CHF comes from the dependence channel. The removal of the floor reverses this pattern almost immediately, consistent with markets treating the removal as credible and permanent. 

Finally, even in the univariate case the framework delivers new results. In simulations, projection's approximation error is an order of magnitude smaller than CM's, particularly in sparse-strike settings. The framework also yields a new and quantitative version of the second Fundamental Theorem of Asset Pricing: spanning by simple options and uniqueness of the risk-neutral measure are equivalent. More broadly, projection provides a unified framework for extracting risk-neutral information from option prices in finite samples, ranging from univariate moments and risk-neutral distributions to joint dependence in settings where the option basis is sufficiently rich.

\begin{refcontext}[sorting=nyt]
  \printbibliography

@Article{BlackScholes1973,
  author  = {Black, Fischer and Scholes, Myron},
  journal = {Journal of Political Economy},
  title   = {The Pricing of Options and Corporate Liabilities},
  year    = {1973},
  number  = {3},
  pages   = {637-654},
  volume  = {81},
  doi     = {10.1086/260062},
}

@Book{Trefethen2018,
  author    = {Trefethen, Lloyd N.},
  publisher = {Society for Industrial and Applied Mathematics},
  title     = {Approximation Theory and Approximation Practice},
  year      = {2018},
  address   = {Philadelphia, PA},
  edition   = {Extended},
  doi       = {10.1137/1.9781611975949},
}

@article{chernozhukov2013inference,
  title={Inference on counterfactual distributions},
  author={Chernozhukov, Victor and Fern{\'a}ndez-Val, Iv{\'a}n and Melly, Blaise},
  journal={Econometrica},
  volume={81},
  number={6},
  pages={2205--2268},
  year={2013},
}

@article{breeden1978prices,
  title={Prices of state-contingent claims implicit in option prices},
  author={Breeden, Douglas T. and Litzenberger, Robert H.},
  journal={Journal of Business},
  pages={621--651},
  year={1978},
}

@inbook{Carr_Madan_2001, 
title={Towards a Theory of Volatility Trading}, 
booktitle={Handbooks in Mathematical Finance: Option Pricing, Interest Rates and Risk Management}, publisher={Cambridge University Press}, 
author={Carr, Peter and Madan, Dilip B.}, 
year={2001}, 
pages={458–476}
}

@article{martin2017expected,
  title={What is the Expected Return on the Market?},
  author={Martin, Ian W. R.},
  journal={Quarterly Journal of Economics},
  volume={132},
  number={1},
  pages={367--433},
  year={2017},
  publisher={MIT Press}
}

@article{ross1976options,
  title={Options and Efficiency},
  author={Ross, Stephen A.},
  journal={Quarterly Journal of Economics},
  volume={90},
  number={1},
  pages={75--89},
  year={1976},
}

@article{filipovic2013density,
  title={Density approximations for multivariate affine jump-diffusion processes},
  author={Filipovi{\'c}, Damir and Mayerhofer, Eberhard and Schneider, Paul},
  journal={Journal of Econometrics},
  volume={176},
  number={2},
  pages={93--111},
  year={2013},
  doi = {j.jeconom.2012.12.003},
}

@incollection{figlewski2010estimating,
    author = {Figlewski, Stephen},
    title = {Estimating the Implied Risk‐Neutral Density for the {US} Market Portfolio},
    booktitle = {Volatility and Time Series Econometrics: Essays in Honor of Robert Engle},
    publisher = {Oxford University Press},
    year = {2010},
    month = {03},
    doi = {10.1093/acprof:oso/9780199549498.003.0015},
}

@article{lebesgue1898approximation,
  title={Sur l’approximation des fonctions},
  author={Lebesgue, Henri},
  journal={Bulletin des Sciences Mathématiques},
  volume={22},
  number={10},
  pages={278--287},
  year={1898}
}

@article{kremens2019theory,
  title={The quanto theory of exchange rates},
  author={Kremens, Lukas and Martin, Ian W. R.},
  journal={American Economic Review},
  volume={109},
  number={3},
  pages={810--843},
  year={2019},
  doi={10.1257/aer.20180019}
}

@article{martin2018options,
  title={Options and the gamma knife},
  author={Martin, Ian W. R.},
  journal={Journal of Portfolio Management},
  volume={44},
  number={6},
  pages={47--55},
  year={2018},
  doi={10.3905/jpm.2018.44.6.047}
}

@misc{embree2010numerical,
  author       = {Mark Embree},
  title        = {Numerical Analysis I},
  year         = {2010},
  howpublished = {Lecture notes, Rice University},
  note         = {Pages 1--207}
}

@article{britten2000option,
author = {Britten-Jones, Mark and Neuberger, Anthony},
title = {Option Prices, Implied Price Processes, and Stochastic Volatility},
journal = {Journal of Finance},
volume = {55},
number = {2},
pages = {839-866},
doi = {10.1111/0022-1082.00228},
year = {2000}
}

@article{martin2025information,
   author = "Martin, Ian W. R.",
   title = "Information in Derivatives Markets: Forecasting Prices with Prices",
   journal = "Annual Review of Financial Economics",
   issn = "1941-1367",
   year = "2025",
   doi = "10.1146/annurev-financial-082123-105811",
}

@article{chernozhukov2009improving,
  title={Improving point and interval estimators of monotone functions by rearrangement},
  author={Chernozhukov, Victor and Fern{\'a}ndez-Val, Iv{\'a}n and Galichon, Alfred},
  journal={Biometrika},
  volume={96},
  number={3},
  pages={559--575},
  year={2009},
  doi={10.1093/biomet/asp030}
}

@article{schneider2019almost,
  title={(Almost) model-free recovery},
  author={Schneider, Paul and Trojani, Fabio},
  journal={Journal of Finance},
  volume={74},
  number={1},
  pages={323--370},
  year={2019},
  doi={10.1111/jofi.12737}
}

@article{bates1991crash,
  author  = {Bates, David S.},
  title   = {The crash of '87: Was it expected? Evidence from options markets},
  journal = {Journal of Finance},
  year    = {1991},
  volume  = {46},
  number  = {3},
  pages   = {1009--1044},
  doi     = {10.1111/j.1540-6261.1991.tb03775.x}
}

@article{bakshi2003stock,
  title={Stock return characteristics, skew laws, and the differential pricing of individual equity options},
  author={Bakshi, Gurdip and Kapadia, Nikunj and Madan, Dilip B.},
  journal={Review of Financial Studies},
  volume={16},
  number={1},
  pages={101--143},
  year={2003},
  doi={10.1093/rfs/16.1.0101}
}

@article{chabi2020conditional,
  title={The conditional expected market return},
  author={Chabi-Yo, Fousseni and Loudis, Johnathan},
  journal={Journal of Financial Economics},
  volume={137},
  number={3},
  pages={752--786},
  year={2020},
  doi={10.1016/j.jfineco.2020.03.009}
}

@article{kozhan2013skew,
  title={The skew risk premium in the equity index market},
  author={Kozhan, Roman and Neuberger, Anthony and Schneider, Paul},
  journal={Review of Financial Studies},
  volume={26},
  number={9},
  pages={2174--2203},
  year={2013},
  doi={10.1093/rfs/hht039}
}

@article{carr2009variance,
  title={Variance risk premiums},
  author={Carr, Peter and Wu, Liuren},
  journal={Review of Financial Studies},
  volume={22},
  number={3},
  pages={1311--1341},
  year={2009},
  doi={10.1093/rfs/hhn038}
}

@article{bollerslev2009expected,
  title={Expected stock returns and variance risk premia},
  author={Bollerslev, Tim and Tauchen, George and Zhou, Hao},
  journal={Review of Financial Studies},
  volume={22},
  number={11},
  pages={4463--4492},
  year={2009},
  doi={10.1093/rfs/hhp008}
}

@inproceedings{vostrecov1961approximation,
  title        = {Approximation of continuous functions by superpositions of plane waves},
  author       = {Vostrecov, B. A. and Kreines, M. A.},
  booktitle    = {Doklady Akademii Nauk SSSR},
  volume       = {140},
  pages        = {1237--1240},
  year         = {1961}
}

@article{lin1993fundamentality,
title = {Fundamentality of Ridge Functions},
journal = {Journal of Approximation Theory},
volume = {75},
number = {3},
pages = {295-311},
year = {1993},
doi = {10.1006/jath.1993.1104},
author = {V.Y. Lin and A. Pinkus},
}

@article{diaconis1984nonlinear,
  title={On nonlinear functions of linear combinations},
  author={Diaconis, Persi and Shahshahani, Mehrdad},
  journal={SIAM Journal on Scientific and Statistical Computing},
  volume={5},
  number={1},
  pages={175--191},
  year={1984},
}

@book{pinkus2015ridge,
  title={Ridge Functions},
  author={Pinkus, Allan},
  volume={205},
  year={2015},
  publisher={Cambridge University Press}
}

@article{engle2012dynamic,
  title={Dynamic equicorrelation},
  author={Engle, Robert F. and Kelly, Bryan},
  journal={Journal of Business \& Economic Statistics},
  volume={30},
  number={2},
  pages={212--228},
  year={2012},
}

@book{shreve2004stochastic,
  title={Stochastic Calculus for Finance II: Continuous-Time Models},
  author={Shreve, Steven E.},
  volume={11},
  year={2004},
  publisher={Springer}
}

@article{bondarenko2024option,
title={Option-Implied Dependence and Correlation Risk Premium},
volume={59},
doi={10.1017/S0022109023000960}, 
number={7}, 
journal={Journal of Financial and Quantitative Analysis}, 
author={Bondarenko, Oleg and Bernard, Carole}, 
year={2024}, 
pages={3139–3189}
}

@article{lustig2014countercyclical,
  title={Countercyclical currency risk premia},
  author={Lustig, Hanno and Roussanov, Nikolai and Verdelhan, Adrien},
  journal={Journal of Financial Economics},
  volume={111},
  number={3},
  pages={527--553},
  year={2014},
}

@article{jiang2005model,
  title={The model-free implied volatility and its information content},
  author={Jiang, George J. and Tian, Yisong S.},
  journal={Review of Financial Studies},
  volume={18},
  number={4},
  pages={1305--1342},
  year={2005},
}

@article{andersen2017short,
author = {Andersen, Torben G. and Fusari, Nicola and Todorov, Viktor},
title = {Short-Term Market Risks Implied by Weekly Options},
journal = {Journal of Finance},
volume = {72},
number = {3},
pages = {1335-1386},
doi = {10.1111/jofi.12486},
year = {2017}
}

@article{mueller2017international,
title = {International correlation risk},
journal = {Journal of Financial Economics},
volume = {126},
number = {2},
pages = {270-299},
year = {2017},
doi = {10.1016/j.jfineco.2016.09.012},
author = {Philippe Mueller and Andreas Stathopoulos and Andrea Vedolin},
}

@book{hansen2022econometrics,
  title={Econometrics},
  author={Hansen, Bruce E.},
  year={2022},
  publisher={Princeton University Press}
}

@article{nachman1988spanning,
  title={Spanning and completeness with options},
  author={Nachman, David C.},
  journal={Review of Financial Studies},
  volume={1},
  number={3},
  pages={311--328},
  year={1988},
  doi={10.1093/rfs/1.3.311}
}

@article{bakshi2000spanning,
  title={Spanning and derivative-security valuation},
  author={Bakshi, Gurdip and Madan, Dilip B.},
  journal={Journal of Financial Economics},
  volume={55},
  number={2},
  pages={205--238},
  year={2000},
  doi={10.1016/S0304-405X(99)00050-1}
}

@Article{krugman1991target,
  author  = {Krugman, Paul R.},
  journal = {Quarterly Journal of Economics},
  title   = {Target Zones and Exchange Rate Dynamics},
  year    = {1991},
  volume  = {106},
  number  = {3},
  pages   = {669--682},
  doi     = {10.2307/2937922}
}

@Article{bertola1992target,
  author  = {Bertola, Giuseppe and Caballero, Ricardo J.},
  journal = {American Economic Review},
  title   = {Target Zones and Realignments},
  year    = {1992},
  volume  = {82},
  number  = {3},
  pages   = {520--536},
}

@Article{bertola1993stochastic,
  author  = {Bertola, Giuseppe and Svensson, Lars E. O.},
  journal = {Review of Economic Studies},
  title   = {Stochastic Devaluation Risk and the Empirical Fit of Target-Zone Models},
  year    = {1993},
  volume  = {60},
  number  = {3},
  pages   = {689--712},
  doi     = {10.2307/2298131}
}

@Article{forbes2002no,
  author  = {Forbes, Kristin J. and Rigobon, Roberto},
  journal = {Journal of Finance},
  title   = {No Contagion, Only Interdependence: Measuring Stock Market Comovements},
  year    = {2002},
  volume  = {57},
  number  = {5},
  pages   = {2223--2261},
  doi     = {10.1111/0022-1082.00494}
}

@Article{jermann2017financial,
  author  = {Jermann, Urban J.},
  journal = {Journal of Money, Credit and Banking},
  title   = {Financial Markets' Views about the Euro--Swiss Franc Floor},
  year    = {2017},
  volume  = {49},
  number  = {2--3},
  pages   = {553--565},
}

@Article{haddad2025whatever,
  author  = {Haddad, Valentin and Moreira, Alan and Muir, Tyler},
  journal = {American Economic Review},
  title   = {Whatever It Takes? {T}he Impact of Conditional Policy Promises},
  year    = {2025},
  volume  = {115},
  number  = {1},
  pages   = {295--329},
}

@article{harrison1981martingales,
title = {Martingales and stochastic integrals in the theory of continuous trading},
journal = {Stochastic Processes and their Applications},
volume = {11},
number = {3},
pages = {215-260},
year = {1981},
doi = {https://doi.org/10.1016/0304-4149(81)90026-0},
author = {J.Michael Harrison and Stanley R. Pliska},
}

@article{delbaen1994general,  
author  = {Delbaen, Freddy and Schachermayer, Walter},     
title   = {A general version of the fundamental theorem of asset pricing}, 
journal = {Mathematische Annalen},  
volume  = {300},                                                                    number  = {1},                                                                      pages   = {463--520},                                                               year    = {1994},                                                                   doi     = {10.1007/BF01450498}                                                      }

@article{jarrow1999second,                                                 
author  = {Jarrow, Robert A. and Jin, Xing and Madan, Dilip B.},
title   = {The Second Fundamental Theorem of Asset Pricing},
journal = {Mathematical Finance},     
volume  = {9},                            
number  = {3},                                                               
pages   = {255--273},
year    = {1999},                                                             
doi     = {10.1111/1467-9965.00069}                     
}

@article{madan1998variance,
  author  = {Madan, Dilip B. and Carr, Peter P. and Chang, Eric C.},
  title   = {The Variance Gamma Process and Option Pricing},
  journal = {Review of Finance},
  year    = {1998},
  volume  = {2},
  number  = {1},
  pages   = {79--105},
  doi     = {10.1023/A:1009703431535}
}

@unpublished{MartinShi2025crashes,
  author = {Martin, Ian W. R. and Shi, Ran},
  title  = {Forecasting Crashes with a Smile},
  year   = {2025},
  month  = sep,
  note   = {Unpublished manuscript, version dated September 2025},
  url    = {https://personal.lse.ac.uk/martiniw/oib_martin_shi_latest.pdf}
}

@article{martin2019expected,
  author  = {Martin, Ian W. R. and Wagner, Christian},
  title   = {What Is the Expected Return on a Stock?},
  journal = {Journal of Finance},
  year    = {2019},
  volume  = {74},
  number  = {4},
  pages   = {1887--1929},
  doi     = {10.1111/jofi.12778}
}

@article{baumeister2015sign,
  title={Sign restrictions, structural vector autoregressions, and useful prior information},
  author={Baumeister, Christiane and Hamilton, James D.},
  journal={Econometrica},
  volume={83},
  number={5},
  pages={1963--1999},
  year={2015},
  doi={10.3982/ECTA12356}
}

@article{andersen2015risk,
  title={The risk premia embedded in index options},
  author={Andersen, Torben G. and Fusari, Nicola and Todorov, Viktor},
  journal={Journal of Financial Economics},
  volume={117},
  number={3},
  pages={558--584},
  year={2015},
  doi={10.1016/j.jfineco.2015.06.005}
}

@article{beason2022,
  title={Dissecting the equity premium},
  author={Beason, Tyler and Schreindorfer, David},
  journal={Journal of Political Economy},
  volume={130},
  number={8},
  pages={2203--2222},
  year={2022},
  doi={10.1086/720394}
}

@misc{gatheral2004,
  title={A parsimonious arbitrage-free implied volatility parameterization with application to the valuation of volatility derivatives},
  author={Gatheral, Jim},
  year={2004},
  note={Presentation at Global Derivatives \& Risk Management, Madrid}
}

@article{almeida2023nonparametric,
  title={Nonparametric option pricing with generalized entropic estimators},
  author={Almeida, Caio and Freire, Gustavo and Azevedo, Rafael and Ardison, Kym},
  journal={Journal of Business \& Economic Statistics},
  volume={41},
  number={4},
  pages={1173--1187},
  year={2023},
  doi={10.1080/07350015.2022.2115499}
}
\end{refcontext}

\clearpage
\phantomsection
\addcontentsline{toc}{section}{Internet Appendix}
\begin{center}
  {\huge\bfseries Internet Appendix}\\[0.75em]
\end{center}

\appendix
\section{Proofs}\label{app:proofs}

\subsection{Proof of Proposition \ref{prop:ols_cont}}
\begin{proof}                               
The weighted normal equations yield $X'\Omega X \hat{\beta}_{n_s} = X'\Omega Y$, where $\Omega = \diag(\omega(s_1),\dots,\omega(s_{n_s}))$. The $(i,j)$-element of $X'\Omega X$ and the $i$th element of $X'\Omega Y$ are
\begin{equation*}                                           
(X'\Omega X)_{ij} =  \sum_{z = 1}^{n_s} \omega(s_z) \phi_i(s_z) \phi_j(s_z), \quad (X'\Omega Y)_{i} =  \sum_{z=1}^{n_s} \omega(s_z) \phi_{i}(s_z) g(s_z).  
\end{equation*}                                                                
Assuming that the grid is equally spaced on $[s_1^{(n)}, s_{n_s}^{(n)}]$ with length $m(n_s) = (s_{n_s}^{(n)} - s_1^{(n)})/n_s$, the conditions $s_1^{(n)} \to 0$, $s_{n_s}^{(n)} \to \infty$, and $m(n_s) \to 0$ together yield the 
Riemann-sum approximation                                                   
\begin{align*}   
m(n_s) (X'\Omega X)_{ij} &\to \int_0^\infty \phi_i(s) \phi_j(s) \omega(s) \diff s, \\
m(n_s) (X'\Omega Y)_{i} &\to  \int_0^\infty \phi_{i}(s) g(s) \omega(s) \diff s,
\end{align*}                                                                where finiteness of both limits is ensured by Assumption \ref{asmp:compact}. The factor $m(n_s)$ cancels in $(X'\Omega X)^{-1}X'\Omega Y$, yielding the limit in \eqref{eq:ols_cont}. The argument continues to hold if the grid is not equally spaced but the mesh goes to zero.
The Gram matrix in \eqref{eq:ols_cont} is invertible because $\omega$ is strictly positive on $\R_{++}$, so the basis functions are linearly independent in $L^2(\R_{++}, \omega)$, and the limit $\hat{\beta}$ is therefore well defined. The proof that $\hat{\beta}$ also solves the minimization problem in \eqref{eq:beta_continuous} follows immediately from the first-order conditions
\begin{equation*}                                                
\lro{\int_0^\infty  \phi(s)  \phi(s)' \omega(s) \diff s } \hat{\beta} = \int_0^\infty \phi(s) g(s) \omega(s) \diff s,
\end{equation*}                                                                     
where $\phi(s) = [\phi_1(s),\dots,\phi_{2+n_k}(s)]'$. These coincide with the equations determining $\hat{\beta}$ in \eqref{eq:ols_cont}.
\end{proof}

\subsection{Proof of Proposition \ref{prop:proj_errorBound}}
\begin{proof}
Write the difference of risk-neutral expectations as an integral against $\omega$,
using the Radon-Nikodym derivative $Z := f_{\ttt}^Q / \omega$:
\begin{align*}
  \Expect_t^Q\!\left[g(S_T) - \hat g(S_T)\right]
  &= \int_0^\infty \bigl(g(s) - \hat g(s)\bigr) f_{\ttt}^Q(s) \diff s\\
  &= \int_0^\infty \bigl(g(s) - \hat g(s)\bigr) Z(s)\, \omega(s) \diff s\\
  &= \int_0^\infty \bigl(g(s) - \hat g(s)\bigr) (Z(s) - 1)\, \omega(s) \diff s,
\end{align*}
the last line follows because the constant function (risk-free asset) is a basis function and projection forces each basis function to be orthogonal to the residual.  By the Cauchy-Schwarz inequality,
\begin{equation*}
  \abs{\int_0^\infty \bigl(g - \hat g\bigr) (Z - 1)\, \omega \diff s}
  \;\le\;
  \norm{g - \hat g}_{L^2(\R_{++}, \omega)} \norm{Z - 1}_{L^2(\R_{++}, \omega)}.
\end{equation*}
It remains to identify $\norm{Z - 1}_{L^2(\R_{++}, \omega)}$ with the chi-squared factor:
\begin{equation*}
  \norm{Z - 1}_{L^2(\R_{++}, \omega)}^2
  \;=\; \int_0^\infty \lro{\frac{f_{\ttt}^Q(s) - \omega(s)}{\omega(s)}}^2\, \omega(s) \diff s
  \;=\; \int_0^\infty \frac{(f_{\ttt}^Q(s) - \omega(s))^2}{\omega(s)} \diff s.
\end{equation*}
\end{proof}

\subsection{Proof of Proposition \ref{prop:spanning}}
The following proof is well known (see \citet{lebesgue1898approximation}), but we include it for completeness and because Assumption \ref{asmp:dense} on the strikes results in some slight modifications of the original proof. The proof below is presented for call options, but applies verbatim to put options as well. 

\begin{proof}
Let  $g \in C(A)$. Because $g$ is continuous on a compact set it is uniformly continuous: for every $\varepsilon > 0$ there exists a $\delta > 0$ (independent of $x$), such that $\sup_{\abs{x - y}< \delta} \abs{g(x) - g(y)} < \varepsilon/4$. Let $a_{\min} = x_1 < x_2 < \dots < x_n = a_{\max}$ be a partition of an arbitrary compact set $A$, such that $x_{j+1} - x_{j} < \delta \ \forall j$, where $a_{\min} = \min(A)$ and  $a_{\max} = \max(A)$.  On each interval $[x_j,x_{j+1}]$ construct a linear function $\tilde{g}_j(x) = a_j x + b_j$ such that $\tilde{g}_j(x_j) = g(x_j)$ and $\tilde{g}_j(x_{j+1}) = g(x_{j+1})$.  For every $x_c \in (x_j,x_{j+1})$ it follows that 
\begin{equation*}
\abs{g(x_c) - \tilde{g}_j(x_c)} \le \abs{g(x_c) - g(x_j)} + \abs{g(x_j) - \tilde{g}_j(x_c)} < \varepsilon/2,
\end{equation*}
because
\begin{equation*}
\abs{g(x_j) - \tilde{g}_j(x_c)} = \abs{\frac{x_c - x_j}{x_{j+1} - x_j}} \abs{g(x_{j+1}) - g(x_j)}.
\end{equation*}
Since $x_c$ is arbitrary, it follows that  $\sup_{x \in [x_j,x_{j+1}]} |g(x) - \tilde{g}_j(x)| < \varepsilon/2$. Now, define the polygonal function
\begin{equation}\label{eq:g_tilde}
\tilde{g}(x) = \sum_{j=1}^{n-2} \tilde{g}_j(x) \ind{x \in [x_j, x_{j+1})} + \tilde{g}_{n-1}(x) \ind{x \in [x_{n-1},x_n]}.
\end{equation}
From the construction above it follows that $\tilde{g}$ is continuous and $\sup_{x \in A} |g(x) - \tilde{g}(x)| <  \varepsilon$. We claim that the polygonal function constructed in this way can be written as
\begin{equation}\label{eq:polygon}
\tilde{g}(x) = \beta_1 + \sum_{j=1}^{n-1} \beta_{j+1} \pmax{x - x_j}.
\end{equation}
To see this, proceed inductively. On $[x_1,x_2]$, \eqref{eq:g_tilde} can be written as
\begin{equation*}
\tilde{g}(x) = a_1 x + b_1 = a_1 \pmax{x - x_1} + \tilde{b}_1,
\end{equation*}
where $\tilde{b}_1 = b_1 + a_1 x_1$. On $[x_1,x_3]$, we can write 
\begin{equation*}
\tilde{g}(x) = a_1 \pmax{x - x_1} + \tilde{b}_1 + \tilde{a}_2\pmax{x - x_2},
\end{equation*}
where 
\begin{equation*}
a_1 + \tilde{a}_2 = a_2
\end{equation*}
which can be solved for to obtain $\tilde{a}_2$. Continuing inductively, we obtain \eqref{eq:polygon}. It remains to show that $\tilde{g}$ can be uniformly approximated by a function of the form
\begin{equation*}
\tilde{g}_{n_k}(x) = \beta_1 + \sum_{j=1}^{n_k-1} \beta_{j+1} \pmax{x - K_j},
\end{equation*}
where $K_j$ is among the observed call option strike prices. But this can be achieved if $n_k$ is large enough. Specifically, let $n_k$ be large enough such that $\max_{j = 1,\dots, n_k-1}\abs{x_j - K_j} < \varepsilon/(2M)$, where $M = \sum_{j = 1}^{n_k-1} \abs{\beta_{j+1}}$. By assumption such $n_k$ can always be found since $\set{K_j}_{j=1}^{n_k}$ is dense in $A$ as $n_k \to \infty$. Considering that 
\begin{equation*}
\sup_{x \in A} \abs{\pmax{x-x_j} - \pmax{x - K_j}}  < \varepsilon/(2M),
\end{equation*}
it follows by another application of the triangle inequality that 
\begin{equation*}
\sup_{x \in A} \abs{g(x) - \tilde{g}_{n_k}(x)} < \varepsilon.
\end{equation*}
\end{proof}

\subsection{Proof of Theorem \ref{thm:ftap_II}}
\begin{proof}
\emph{(i) $\Rightarrow$ (ii).} Let $Q_t' \in \mathcal{Q}_{\omega}$ satisfy $\Expect_t^{Q'} \phi = \Expect_t^{Q} \phi$ for every $\phi \in \bigcup_{n_k \ge 1} \cF_{2+n_k}$. Fix $g \in L^2(\R_{++}, \omega)$. By (i), there exist $f_{n_k} \in \spn(\cF_{2+n_k})$ with $\norm{g - f_{n_k}}_{L^2(\R_{++}, \omega)} \to 0$. Applying Proposition~\ref{prop:proj_errorBound} separately to $Q_t$ and $Q_t'$ (the crude Cauchy–Schwarz bound without centering, valid for any element of the span), \begin{equation*}
\left| \Expect_t^{Q} g - \Expect_t^{Q} f_{n_k} \right| \le \norm{g - f_{n_k}}_{L^2(\R_{++}, \omega)}  \sqrt{1 + \chi^2(f^Q_{\ttt} \,\|\ \omega)} \to 0,
\end{equation*}     
and analogously for $Q_t'$. Since $\Expect_t^{Q'} f_{n_k} = \Expect_t^{Q} f_{n_k}$ for every $n_k$ by hypothesis, taking limits gives $\Expect_t^{Q'} g = \Expect_t^{Q} g$ for every $g \in L^2(\R_{++}, \omega)$. Since $L^2(\R_{++}, \omega)$ contains every bounded measurable function on $\R_{++}$, and bounded measurable functions form a measure-determining class, $Q_t' = Q_t$.

\medskip
\emph{(ii) $\Rightarrow$ (i).} We prove the contrapositive. Suppose
\begin{equation*}
V \;:=\; \overline{\bigcup_{n_k \ge 1} \spn(\cF_{2+n_k})}
\end{equation*}
is a proper closed subspace of $L^2(\R_{++}, \omega)$. We construct a measure
$Q_t' \in \mathcal{Q}_{\omega}$ with $Q_t' \ne Q_t$ that produces identical option prices. For future reference, $L^\infty_c(\R_{++})$ denotes the space of bounded measurable functions on $\R_{++}$ with compact support contained in $\R_{++}$. 

\smallskip
\emph{Step 1: $V^\perp \cap L^\infty_c(\R_{++}) \neq \{0\}$.}

Let
\begin{equation*}
K^{*} \;:=\; \bigcup_{n_k \ge 1}\set{K_{1}^{(n_k)}, \dots, K_{n_k}^{(n_k)}}
\end{equation*}
denote the set of all strikes appearing in the basis. We argue first that $K^{*}$ cannot be dense in $\R_{++}$ when $V$ is a proper subspace.

Suppose, for contradiction, that $K^{*}$ is dense in $\R_{++}$. We will show $V = L^2(\R_{++}, \omega)$. Since $C_c(\R_{++})$ is dense in $L^2(\R_{++}, \omega)$ and $V$ is closed, it
suffices to show $C_c(\R_{++}) \subseteq V$. Fix $f \in C_c(\R_{++})$ with $\mathrm{supp}(f) \subseteq [a, b] \subset \R_{++}$ and choose any $[a', b'] \subset \R_{++}$ with $0 < a' < a < b
< b'$. By density of $K^{*}$, pick strikes $K_{i_0} \in K^{*} \cap [a', a)$ and $K_{i_m} \in K^{*} \cap (b, b']$, together with $m - 1$ further strikes $K_{i_j} \in K^{*} \cap (K_{i_0},
K_{i_m})$ for $1 \le j \le m - 1$. Because $K_{i_0}, K_{i_m} \notin \mathrm{supp}(f)$, the continuous piecewise linear interpolant $g$ of $f$ at $K_{i_0}, \dots, K_{i_m}$ vanishes at the
endpoints, and its extension by zero outside $[K_{i_0}, K_{i_m}]$ is continuous on $\R_{++}$. A direct calculation shows $g$ can be written as
\begin{equation*}
g(s) \;=\; \sum_{j=0}^{m} \alpha_j \pmax{s - K_{i_j}}
\quad\text{with}\quad
\sum_{j=0}^{m} \alpha_j = 0, \;\; \sum_{j=0}^{m} \alpha_j K_{i_j} = 0,
\end{equation*}
the two constraints ensuring that $g \equiv 0$ on both $(0, K_{i_0}]$ and $[K_{i_m}, \infty)$. Thus $g \in \spn(\cF_{2+n_k})$ for any $n_k$ large enough to include all of $K_{i_0}, \dots,
K_{i_m}$, hence $g \in V$. As the strike grid becomes dense in $[K_{i_0}, K_{i_m}]$, uniform continuity of $f$ gives $\norm{f - g}_\infty \to 0$, and therefore $\norm{f - g}_{L^2(\R_{++},
\omega)} \to 0$. Since $V$ is closed, $f \in V$, so $V = L^2(\R_{++}, \omega)$, contradicting $V \subsetneq L^2(\R_{++}, \omega)$.

Hence $K^{*}$ is not dense in $\R_{++}$ and there exists an open interval $(c, d) \subset \R_{++}$ with $K^{*} \cap (c, d) = \emptyset$. Pick any $c < c' < d' < d$. We construct a continuous, bounded, nonzero function $h$, supported on $[c',d']$, that is orthogonal to $1$ and $s$ in $L^2(\R_{++},\omega)$. Partition $[c',d']$ into three consecutive closed subintervals $I_1, I_2, I_3$ with pairwise disjoint interiors, and let $\varphi_i$ be a tent function on $I_i$: continuous, positive on the interior of $I_i$, and zero at the endpoints of $I_i$ (hence zero outside $I_i$). The two conditions
\begin{equation*}
\sum_{i=1}^{3} a_i \int_{c'}^{d'} \varphi_i(s)\,\omega(s)\,\mathrm{d}s = 0,
\qquad
\sum_{i=1}^{3} a_i \int_{c'}^{d'} s\,\varphi_i(s)\,\omega(s)\,\mathrm{d}s = 0,
\end{equation*}
form a linear system of two equations in the three unknowns $(a_1,a_2,a_3)$, so a nontrivial solution exists. Define $h_0 := \sum_{i=1}^{3} a_i \varphi_i$. Because the supports of the $\varphi_i$ have disjoint interiors, $h_0$ is not identically zero, and by construction $\int h_0\,\omega\,\mathrm{d}s = \int s\,h_0\,\omega\,\mathrm{d}s = 0$. Moreover, $h_0(c') = h_0(d') = 0$, so extending $h_0$ by zero outside $[c',d']$ defines a function $h$ that is continuous on all of $\R_{++}$, bounded, with $\operatorname{supp}(h) \subseteq [c',d']$ and $\|h\|_{L^2(\R_{++},\omega)} > 0$.


It remains to verify $h \in V^\perp$. Every basis function $\phi \in \bigcup_{n_k \ge 1} \cF_{2+n_k}$ is either $1$, $s$, or an option with strike $K \in K^{*}$. Since $K^{*} \cap (c, d) =
\emptyset$, every such $K$ satisfies $K \le c$ or $K \ge d$, and the corresponding option payoff is therefore affine on $[c', d'] \subset (c, d)$: $\pmax{K - s} = K - s$ when $K \ge d$ and
$\pmax{K - s} = 0$ when $K \le c$, and analogously for $\pmax{s - K}$. Writing $\phi(s) = a_\phi + b_\phi s$ on $[c', d']$,
\begin{equation*}
\int_0^\infty \phi(s)\, h(s)\, \omega(s) \diff s
\;=\; a_\phi \int_{c'}^{d'} h(s)\, \omega(s) \diff s
\;+\; b_\phi \int_{c'}^{d'} s\, h(s)\, \omega(s) \diff s
\;=\; 0.
\end{equation*} 
By linearity and continuity of the inner product, $h \perp V$.

\smallskip
\emph{Step 2: Construct a perturbed measure.} Define
\begin{equation*}
  f_{\ttt}^{Q'}(s) \;:=\; f_{\ttt}^Q(s) \;+\; \varepsilon\, h(s)\, \omega(s),
\end{equation*}
where $\varepsilon \in (0, \varepsilon_0)$ with
\begin{equation*}
M := \max_{s \in [c',d']} |h(s)|, \qquad
\varepsilon_0 := \min_{s \in [c',d']} \frac{f_{\ttt}^{Q}(s)}{M\,\omega(s)} \; > \; 0,
\end{equation*}
finite and strictly positive by continuity and strict positivity of $f_{\ttt}^{Q}$ and $\omega$ on the compact set $[c',d']$. We verify the five required properties.


\textbf{Normalization.} Since the constant function $1$ belongs to
$\cF_{2+n_k} \subset V$ and $h \perp V$,
\begin{equation*}
  \int_0^\infty h(s)\, \omega(s) \diff s \;=\; 0.
\end{equation*}
Therefore $\int_0^\infty f_{\ttt}^{Q'}(s) \diff s = \int_0^\infty f_{\ttt}^Q(s)
\diff s + \varepsilon \int_0^\infty h(s) \omega(s) \diff s = 1$.


\textbf{Strict positivity.} On the support of $h$, $\varepsilon\,|h(s)|\,\omega(s) \le \varepsilon M \omega(s) < \varepsilon_0 M \omega(s) \le f_{\ttt}^{Q}(s)$, so $f_{\ttt}^{Q'}(s) > 0$. Outside the support of $h$, $f_{\ttt}^{Q'}(s) = f_{\ttt}^{Q}(s) > 0$.

\textbf{Continuity.} $f_{\ttt}^{Q'} = f_{\ttt}^{Q} + \varepsilon h \omega$ is continuous on $\R_{++}$: $f_{\ttt}^{Q}$ is continuous because $Q_t \in \mathcal{Q}_\omega$, $\omega$ is continuous by Assumption \ref{asmp:compact}, and $h$ is continuous by construction.

\textbf{Finite chi-squared divergence.} Expanding,
\begin{equation*}
  \chi^2(Q_t' \,\|\, \omega)
  = \chi^2(f_{\ttt}^Q \,\|\, \omega)
    + 2\varepsilon \int (f_{\ttt}^Q - \omega) h \diff s
    + \varepsilon^2 \int h^2 \omega \diff s.
\end{equation*}
The first term is finite by $Q_t \in \mathcal{Q}_{\omega}$. The middle term is
finite by Cauchy-Schwarz and $\int h^2 \omega \diff s < \infty$. The third term
is finite since $h \in L^\infty_c$. Therefore $Q_t' \in \mathcal{Q}_{\omega}$.

\textbf{Identical option prices.} For every $\phi \in
\bigcup_{n_k \ge 1} \cF_{2+n_k} \subset V$, the orthogonality $h \perp V$
gives $\int \phi h \omega \diff s = 0$, hence
\begin{equation*}
  \Expect_t^{Q'} \phi(S_T)
  = \int \phi f_{\ttt}^Q \diff s + \varepsilon \int \phi h \omega \diff s
  = \Expect_t^Q \phi(S_T).
\end{equation*}

Since $h \ne 0$ and $\varepsilon > 0$, the densities
$f_{\ttt}^{Q'}$ and $f_{\ttt}^Q$ differ on a set of positive Lebesgue measure. Therefore $Q_t' \in \mathcal{Q}_{\omega}$ produces identical option prices to $Q_t$
but is distinct from $Q_t$, contradicting (ii).
\end{proof}

\subsection{Proof of Corollary \ref{cor:refine}}
By assumption $\Expect_t^Q \hat{g} = \Expect_t^{Q'} \hat{g}$. Hence, 
\begin{align*}
\abs{\Expect_t^Q g(S_T) - \Expect_t^{Q'} g(S_T)} &= \abs{\Expect_t^Q [g(S_T) - \hat{g}(S_T)] - \Expect_t^{Q'} [g(S_T) - \hat{g}(S_T)]}\\
&\le \norm{g - \hat g}_{L^2(\R_{++}, \omega)}
 \Bigl( \sqrt{\chi^2(Q_t \|\omega)} + \sqrt{\chi^2(Q_t'\|\omega)} \Bigr),
\end{align*}
using the triangle inequality and \eqref{eq:finite_bound} in the second line.

\subsection{Proof of Proposition \ref{prop:cm_proj_weights}}
\begin{proof}
Let $N:=n_k+1$. The points $K_0,\ldots,K_N$ are the nodes of the uniform grid. The space
\begin{equation*}
\PL_h:=\spn\{1,x,(x-K_1)^+,\ldots,(x-K_{n_k})^+\}
\end{equation*}
is exactly the space of continuous piecewise-linear functions on this grid. Let
$B_0,\ldots,B_N$ denote the nodal hat functions, so that
\begin{equation*}
B_i(K_j)=\delta_{ij},\qquad \supp(B_i)\subseteq [K_{i-1},K_{i+1}],
\end{equation*}
with the obvious one-sided modification at the boundary. Write
\begin{equation*}
\hat g(x)=\sum_{i=0}^N \alpha_i B_i(x).
\end{equation*}
The projection normal equations are
\begin{equation}\label{eq:mass_normal}
M_h^\omega \alpha=b_h^\omega,
\end{equation}
where
\begin{equation*}
(M_h^\omega)_{ij}=\int_A B_i(x)B_j(x)\omega(x)\diff x,
\qquad
(b_h^\omega)_i=\int_A B_i(x)g(x)\omega(x)\diff x.
\end{equation*}
Set
\begin{equation*}
A_h^\omega:=h^{-1}M_h^\omega,
\qquad
 y_i:=h^{-1}(b_h^\omega)_i.
\end{equation*}
Then \eqref{eq:mass_normal} is equivalent to
\begin{equation}\label{eq:scaled_mass_normal}
A_h^\omega\alpha=y.
\end{equation}
The advantage of the hat basis is that $A_h^\omega$ is tridiagonal: two hat functions overlap only when their indices differ by at most one. Moreover, since $\omega$ is continuous and strictly positive on the compact set $A$, the eigenvalues of $A_h^\omega$ are bounded above and below uniformly in $h$.

We use the following expansion.

\begin{lem}[Nodal expansion of the weighted spline projection]\label{lem:nodal_expansion}
Under the assumptions of Proposition~\ref{prop:cm_proj_weights}, there exist constants $C<\infty$ and $\rho\in(0,1)$, independent of $h$, such that
\begin{equation}\label{eq:alpha_expansion}
\alpha_i=g(K_i)-\frac{h^2}{12}g''(K_i)+e_i,
\qquad i=0,\ldots,N,
\end{equation}
where
\begin{equation}\label{eq:e_bound_new}
|e_i|\le C h^4+C h^3\left(\rho^i+\rho^{N-i}\right),
\qquad i=0,\ldots,N.
\end{equation}
\end{lem}

For now, take Lemma~\ref{lem:nodal_expansion} as given. Since $\hat g$ is linear on each interval $[K_i,K_{i+1}]$, its slope on that interval is
\begin{equation*}
p_i=\frac{\alpha_{i+1}-\alpha_i}{h}.
\end{equation*}
The coefficient $\hat\gamma_i$ on $(x-K_i)^+$ is the jump in the slope of the fitted spline at $K_i$. Hence, for $i=1,\ldots,n_k=N-1$,
\begin{equation}\label{eq:gamma_jump_new}
\hat\gamma_i=p_i-p_{i-1}
=\frac{\alpha_{i+1}-2\alpha_i+\alpha_{i-1}}{h}.
\end{equation}
Using \eqref{eq:alpha_expansion},
\begin{align}\label{eq:gamma_decomposition_new}
\hat\gamma_i
&=\frac{g(K_{i+1})-2g(K_i)+g(K_{i-1})}{h} \notag\\
&\quad -\frac{h}{12}\left[g''(K_{i+1})-2g''(K_i)+g''(K_{i-1})\right]
+\frac{e_{i+1}-2e_i+e_{i-1}}{h}.
\end{align}
By Taylor expansion and $g\in C^4(A)$,
\begin{equation*}
\frac{g(K_{i+1})-2g(K_i)+g(K_{i-1})}{h}=h g''(K_i)+O(h^3)
\end{equation*}
uniformly in $i$, while
\begin{equation*}
\frac{h}{12}\left[g''(K_{i+1})-2g''(K_i)+g''(K_{i-1})\right]=O(h^3).
\end{equation*}
Finally, \eqref{eq:e_bound_new} gives
\begin{equation*}
\left|\frac{e_{i+1}-2e_i+e_{i-1}}{h}\right|
\le C h^3+C h^2\left(\rho^i+\rho^{N-i}\right).
\end{equation*}
Combining these bounds and using $N=n_k+1$ yields \eqref{eq:gamma_boundary_layer}. The two displayed consequences follow immediately: the boundary term is uniformly $O(h^2)$, and on compact subsets of the interior it is exponentially small in $1/h$, hence $O(h^3)$.
\end{proof}

\begin{proof}[Proof of Lemma~\ref{lem:nodal_expansion}]
Define the local approximate nodal vector
\begin{equation}\label{eq:q_def_new}
q_i:=g(K_i)-\frac{h^2}{12}g''(K_i),
\qquad i=0,\ldots,N.
\end{equation}
We show that $q$ solves the scaled normal equations \eqref{eq:scaled_mass_normal} up to a small residual. Let
\begin{equation*}
r:=A_h^\omega q-y.
\end{equation*}
For an interior index $i=1,\ldots,N-1$, the local expansions of the scaled mass matrix are obtained by setting $x=K_i+hu$:
\begin{align*}
(A_h^\omega)_{ii}
&=\int_{-1}^{1}(1-|u|)^2\omega(K_i+hu)\diff u
=\frac{2}{3}\omega(K_i)+\frac{h^2}{30}\omega''(K_i)+O(h^4),\\
(A_h^\omega)_{i,i+1}
&=\int_0^1 u(1-u)\omega(K_i+hu)\diff u
=\frac16\omega(K_i)+\frac{h}{12}\omega'(K_i)+\frac{h^2}{40}\omega''(K_i)+O(h^3),\\
(A_h^\omega)_{i,i-1}
&=\int_0^1 u(1-u)\omega(K_i-hu)\diff u
=\frac16\omega(K_i)-\frac{h}{12}\omega'(K_i)+\frac{h^2}{40}\omega''(K_i)+O(h^3).
\end{align*}
Writing $G:=g\omega$, the right-hand side satisfies
\begin{equation}\label{eq:y_expansion_new}
y_i=\int_{-1}^{1}(1-|u|)G(K_i+hu)\diff u
=G(K_i)+\frac{h^2}{12}G''(K_i)+O(h^4).
\end{equation}
The odd terms vanish because the interior hat function is symmetric.

A Taylor expansion of \eqref{eq:q_def_new} around $K_i$, together with the preceding formulas, gives
\begin{align*}
(A_h^\omega q)_i
&=G(K_i)+\frac{h^2}{12}G''(K_i)+O(h^4).
\end{align*}
Comparing with \eqref{eq:y_expansion_new},
\begin{equation}\label{eq:interior_residual_new}
r_i=O(h^4),
\qquad i=1,\ldots,N-1.
\end{equation}
At the two boundary nodes the hat functions are one-sided. The same calculation, using one-sided Taylor expansions, gives
\begin{equation}\label{eq:boundary_residual_new}
r_0=O(h^3),
\qquad r_N=O(h^3).
\end{equation}
Thus
\begin{equation}\label{eq:residual_all_new}
|r_j|\le C h^4+C h^3\left(\ind{j=0}+\ind{j=N}\right),
\qquad j=0,\ldots,N.
\end{equation}

Now write $\alpha=q+e$. Since $A_h^\omega\alpha=y$ and $r=A_h^\omega q-y$, we have
\begin{equation}\label{eq:e_eq_new}
A_h^\omega e=-r.
\end{equation}
The matrix $A_h^\omega$ is tridiagonal and uniformly well-conditioned. Therefore its inverse has geometric off-diagonal decay: there exist $C<\infty$ and $\rho\in(0,1)$, independent of $h$, such that
\begin{equation}\label{eq:demko_decay_new}
|(A_h^\omega)^{-1}_{ij}|\le C\rho^{|i-j|},
\qquad i,j=0,\ldots,N.
\end{equation}
This is the inverse-decay property for uniformly well-conditioned banded matrices. Combining \eqref{eq:e_eq_new}, \eqref{eq:residual_all_new}, and \eqref{eq:demko_decay_new},
\begin{align*}
|e_i|
&\le \sum_{j=0}^N |(A_h^\omega)^{-1}_{ij}| |r_j|\\
&\le C h^4\sum_{j=0}^N \rho^{|i-j|}+C h^3\rho^i+C h^3\rho^{N-i}\\
&\le C h^4+C h^3\left(\rho^i+\rho^{N-i}\right).
\end{align*}
This proves the lemma.
\end{proof}

\subsection{Proof of Proposition \ref{prop:rn_expectation}}
\begin{proof}
Part \ref{item:rn_1} follows immediately from the continuous-state problem \eqref{eq:beta_continuous}, because when $x \to 0^+$, $\ind{S_T \le x} \to 0$ in $L^2(\R_{++}, \omega)$, so $\hat\beta(x) \to 0$ and hence $\hat{F}_{\ttt}^Q(x) \to 0$. As $x \to \infty$, $\ind{S_T \le x} \to 1$, and since the constant function $1$ lies in the basis, $\hat\beta(x) \to (1, 0, \ldots, 0)$, giving $\hat{F}_{\ttt}^Q(x) \to 1$.

Part \ref{item:rn_2}: We need to establish differentiability of $\hat{\beta}(x)$. The risk-neutral distribution can easily be derived from \eqref{eq:ols_cont} and \eqref{eq:rn_cdf}. In particular, from \eqref{eq:ols_cont} we deduce that
\begin{equation*}
\frac{\partial}{\partial x} \hat{\beta}(x) = \begin{bmatrix}
\seq{\phi_1,\phi_1} & \dots & \seq{\phi_{1},\phi_{2 + n_k}} \\
\vdots & \ddots &\vdots \\
\seq{\phi_{2 + n_k},\phi_1} & \dots & \seq{\phi_{2 + n_k},\phi_{2 + n_k}}
\end{bmatrix}^{-1} 
\begin{bmatrix}
\omega(x)\\
x \omega(x) \\
\vdots\\
\phi_j(x) \omega(x)\\
\vdots \\
\phi_{2 + n_k}(x) \omega(x)
\end{bmatrix}.
\end{equation*}
Continuity of $\hat\beta'(x)$ on $\R_{++}$ follows because each basis function $\phi_j$ is continuous, hence so is $\phi(x) \omega(x)$. On each open subinterval $(K_j, K_{j+1})$ the entries are products of $C^d$ functions, so $\hat\beta'$ is $C^d$ between strikes.

Part \ref{item:rn_3}: By the Gram-Schmidt process, we can assume that $\set{\phi_i}_{i=1}^{2+n_k}$ is an orthonormal basis w.r.t.\ the inner product $\seq{\phi_i,\phi_j} = \int_0^\infty \phi_i(s) \phi_j(s) \omega(s) \diff s$. Hence, the risk-neutral CDF and PDF can be expressed as
\begin{align}
\hat{F}_{\ttt}^Q(x) &= \sum_{j = 1}^{2 + n_k} \seq{\ind{S_T \le x}, \phi_j(S_T)} \Expect_t^Q \phi_j(S_T) \nonumber\\
\hat{f}_{\ttt}^Q(x) &= \frac{\partial}{\partial x} \hat{F}_{\ttt}^Q(x) = \sum_{j = 1}^{2 + n_k} \phi_j(x) \omega(x) \Expect_t^Q \phi_j(S_T). \label{eq:pdf}
\end{align}
Notice that $\Expect_t^Q \phi_j(S_T)$ is now a linear combination of put and call option prices due to the Gram-Schmidt process. It follows from \eqref{eq:pdf} that
\begin{align*}
\int_{0}^\infty g(x) \diff \hat{F}_{\ttt}^Q(x)  &= \sum_{j=1}^{2+n_k} \Expect_t^Q \lr{\phi_j(S_T)} \int_0^\infty g(x) \phi_j (x) \omega(x) \diff x\\
&= \sum_{j=1}^{2+n_k} \Expect_t^Q \lr{\phi_j(S_T)} \seq{g,\phi_j}\\
&= \Expect_t^Q \hat{g}(S_T).
\end{align*}
The last line follows because, under the Gram-Schmidt process,  $\hat{\beta}_j$ from \eqref{eq:ols_cont} equals $\seq{g,\phi_j}$ since $\seq{\phi_i,\phi_j} = \delta_{ij}$ by orthonormality. 
\end{proof}

\subsection{Proof of Proposition \ref{prop:additive_weights}}
\begin{proof}
Because the marginals of $\omega$ coincide with the risk-neutral marginals, and because $\hat g$ is a linear combination of payoffs that depend on $S_{1,T}$ or $S_{2,T}$ separately, its expectation under $Q$ equals its expectation under $\omega$:
\begin{equation*}
\Expect_t^Q [\hat{g}(S_{1,T}, S_{2,T})] = \Expect^{\omega} [\hat{g}(S_{1,T}, S_{2,T})].
\end{equation*}
Moreover, because the risk-free payoff is contained in the projection space, the projection residual is orthogonal to constants:
\begin{equation*}
\Expect^{\omega}[g-\hat g]=0.
\end{equation*}
Therefore,
\begin{equation*}
\Expect_t^Q [\hat{g}(S_{1,T}, S_{2,T})] = \Expect^{\omega}[\hat{g} (S_{1,T}, S_{2,T}) ] = \Expect^{\omega}[g(S_{1,T}, S_{2,T})].
\end{equation*}
\end{proof}

\subsection{Sufficient conditions for ridge representation and the proof of Proposition \ref{prop:impossible}}\label{app:sufficient}
For completeness, we state the result of \citet{lin1993fundamentality}, giving necessary and sufficient conditions for ridge representation to hold. To state the result, some additional terminology is necessary. A polynomial $p(x_1,\dots, x_d)$ can be associated to the differential operator $p(\frac{\partial}{\partial x_1},\dots , \frac{\partial }{\partial x_d})$. Let $P(a^1,\dots,a^r)$ be the set of polynomials which vanish on all lines $\set{\lambda a^i, \lambda \in \R}$. Let $Q$ be the set of polynomials $q(x_1,\dots,x_d)$ such that $p(\frac{\partial}{\partial x_1},\dots , \frac{\partial }{\partial x_d}) q = 0$ for all $p \in P(a^1,\dots,a^r)$.

\begin{prop*}[\citet{lin1993fundamentality}]
Let $a^1,\dots,a^r$ be pairwise linearly independent vectors in $\R^d$. A function $g \in C(\R^d)$ can be expressed in the form
\begin{equation*}
g(x) = \sum_{i=1}^r g_i(a^i \cdot x)
\end{equation*}
if and only if $g$ belongs to the closure of the linear span of Q.
\end{prop*}

In many practical situations, a more elementary argument suffices to show that a function cannot be written as a ridge combination with given directions $a^i$. For example, in the case $d = 3$, the following reasoning shows that $g(x) = x_1 (w' x)$ cannot be expressed as 
\begin{equation*}
g(x) = g_1(x_1) + g_2(x_2) + g_3(x_3) + g_4(w'x). 
\end{equation*}
Suppose, by contradiction, that such a representation exists. Then, by differentiating twice, we have $\frac{\partial^2 g}{\partial x_2 \partial  x_3}  = 0$. However, $\frac{\partial^2 g_i}{\partial x_2 \partial  x_3} = 0$ for $i = 1,\dots,3$, while $\frac{\partial^2 g_4}{\partial x_2 \partial  x_3} = w_2 w_3 g_4''(w'x)$. This implies that $g_4$ must be affine, but this cannot possibly hold since $g(x)$ contains the cross terms $x_1 x_2$ and $x_1 x_3$. This proves Proposition \ref{prop:impossible} in case $d=3$. Notice that we tacitly assume the most favorable scenario where options complete the market for each asset (e.g. using the same assumptions as in Proposition \ref{prop:spanning}), so that each $g_i$ can be estimated with arbitrary accuracy. 

The argument generalizes directly to $d \ge 3$, thus showing that in higher dimensions it is not possible to perfectly estimate the risk-neutral covariance or correlation of sector $i$ with the market portfolio.

\subsection{Proof of Proposition \ref{prop:conv_ols}}
\begin{proof}
Without loss of generality, we assume that all strike prices
correspond to call options. We start by deriving an error bound on
the piecewise linear polynomial, denoted by $\tilde g$, that
interpolates the points
\begin{equation*}
  \set{(\amin, g(\amin)),\, (K_j, g(K_j))_{j=1}^{n_k},\, (\amax, g(\amax))}.
\end{equation*}
By the proof of Proposition~\ref{prop:spanning}, $\tilde g$ restricted
to $A$ belongs to $\spn(\cF_{2+n_k})$. Standard results in
approximation theory (see, e.g., \citealp[Lecture~11]{embree2010numerical})
yield
\begin{align}
  \max_{s \in [K_j, K_{j+1}]} \abs{g(s) - \tilde g(s)}
  &\;\le\; \lro{\max_{\xi \in [K_j, K_{j+1}]} \frac{\abs{g''(\xi)}}{2}}
           \lro{\max_{s \in [K_j, K_{j+1}]} (s - K_j)(K_{j+1} - s)} \nonumber\\
  &\;\le\; \frac{\norm{g''}_\infty}{8} \lro{K_{j+1} - K_j}^2,
  \label{eq:interp_bound}
\end{align}
where $\norm{g''}_\infty := \max_{\xi \in [\amin, \amax]} \abs{g''(\xi)}$,
which is finite by $g \in C^2[\amin, \amax]$. Squaring and integrating
against $\omega$, and using that Assumption~\ref{asmp:compact} guarantees
$\norm{\omega}_\infty := \max_{s \in A} \omega(s) < \infty$,
\begin{equation*}
  \int_{K_j}^{K_{j+1}} \lro{g(s) - \tilde g(s)}^2 \omega(s) \diff s
  \;\le\; \frac{\norm{g''}_\infty^2 \norm{\omega}_\infty}{64}
          \lro{K_{j+1} - K_j}^5.
\end{equation*}
By additivity of the integral,
\begin{align}
  \int_{K_1}^{K_{n_k}} \lro{g(s) - \tilde g(s)}^2 \omega(s) \diff s
  &\;=\; \sum_{j=1}^{n_k - 1}
         \int_{K_j}^{K_{j+1}} \lro{g(s) - \tilde g(s)}^2 \omega(s) \diff s
         \nonumber \\
  &\;\le\; \frac{n_k \norm{g''}_\infty^2 \norm{\omega}_\infty}{64}\, \Delta^5
         \nonumber \\
  &\;=\; O\!\lro{\frac{1}{n_k^4}},
  \label{eq:delta_4}
\end{align}
where in the last line we used $\Delta = O(1/n_k)$. Applying
\eqref{eq:interp_bound} on the boundary interval $[K_{n_k}, \amax]$
gives
\begin{equation}\label{eq:amax}
  \max_{s \in [K_{n_k}, \amax]} \abs{g(s) - \tilde g(s)}
  \;\le\; \frac{\norm{g''}_\infty}{8} \lro{\amax - K_{n_k}}^2,
\end{equation}
and therefore
\begin{equation*}
  \int_{K_{n_k}}^{\amax} \lro{g(s) - \tilde g(s)}^2 \omega(s) \diff s
  \;\le\; \frac{\norm{g''}_\infty^2 \norm{\omega}_\infty}{64}
          \lro{\amax - K_{n_k}}^5
  \;=\; O\!\lro{\frac{1}{n_k^4}},
\end{equation*}
where the last equality uses $\amax - K_{n_k} = O(1/n_k^{4/5})$. A
symmetric bound holds on $[\amin, K_1]$. Putting the three estimates
together,
\begin{equation*}
  \int_{\amin}^{\amax} \lro{g(s) - \tilde g(s)}^2 \omega(s) \diff s
  \;=\; O\!\lro{\frac{1}{n_k^4}}.
\end{equation*}
Since $\hat g$ is the $L^2(A, \omega)$-projection of $g$ onto
$\spn(\cF_{2+n_k})$ and $\tilde g$ restricted to $A$ lies in this
span, the minimization property yields
\begin{align*}
  \norm{(g - \hat g)\, \ind{s \in A}}_{L^2(\R_{++}, \omega)}^2
  &\;=\; \int_{\amin}^{\amax} \lro{g(s) - \hat g(s)}^2 \omega(s) \diff s\\
  &\;\le\; \int_{\amin}^{\amax} \lro{g(s) - \tilde g(s)}^2 \omega(s) \diff s\\
  &\;=\; O\!\lro{\frac{1}{n_k^4}}.
\end{align*}
The finite-sample bound in Proposition~\ref{prop:proj_errorBound},
applied to $(g - \hat g)\, \ind{s \in A}$, then immediately gives
\begin{equation*}
  \abs{\Expect_t^Q\!\lr{g(S_T)\, \ind{S_T \in A}}
       - \Expect_t^Q\!\lr{\hat g(S_T)\, \ind{S_T \in A}}}
  \;=\; O\!\lro{\frac{1}{n_k^2}}.
\end{equation*}
\end{proof}

\subsection{Proof of Proposition \ref{prop:conv_cm}}
\begin{proof}
Over $A$, the CM Taylor expansion in \eqref{eq:cm_taylor} is given by 
\begin{align*}
g(x) &= g(F_{\ttt}) + g'(F_{\ttt}) \lro{x - F_{\ttt}}\\
&+ \int_{\amin}^{F_{\ttt}} g''(K) \pmax{K - x} \diff K + \int_{F_{\ttt}}^{\amax} g''(K) \pmax{x - K} \diff K.
\end{align*}
We will focus on the case $x \le F_{\ttt}$ (the case $x > F_{\ttt}$ is identical). The integral is discretized using the trapezoidal rule, which is known to satisfy
\begin{equation*}
\sum_{j:K_j\le F_{\ttt}} \Delta K_j\, g''(K_j) \pmax{K_j - x} = \int_{K_1}^{F_{\ttt}} g''(K) \pmax{K - x} \diff K + O\lro{\frac{1}{n_k^2}},
\end{equation*}
uniformly in $x$. Hence, for $x \in [K_1,F_{\ttt}]$, we obtain 
\begin{equation*}
\max_{x \in [K_1,F_{\ttt}]}  \abs{g(x) - \hat{g}_{\mathrm{CM}}(x)} = O\lro{\frac{1}{n_k^2}}.
\end{equation*}
For $x \in [\amin,K_1]$, we get 
\begin{align*}
\abs{g(x) - \hat{g}_{\mathrm{CM}}(x)} &= \abs{  \int_{x}^{K_1} g''(K) (K - x)  \diff K}  \\
&\le \norm{g''}_{\infty} \frac{1}{2} \lro{K_1 - x}^2.
\end{align*} 
Analogous reasoning yields a similar bound for $x > F_{\ttt}$. The same reasoning at the end of Proposition \ref{prop:conv_ols} then finally gives
\begin{align*}
&\abs{\int_{\amin}^{\amax} \lro{g(x) - \hat{g}_{\mathrm{CM}}(x)}f_{\ttt}^Q(x) \diff x }\\
&\le \lro{\int_{\amin}^{\amax} f_{\ttt}^Q(x)^2 \diff x}^{1/2} \bigg(\int_{\amin}^{K_1}  \lro{g(x) - \hat{g}_{\mathrm{CM}}(x)}^2 \diff x \\
&+ \int_{K_1}^{K_{n_k}}  \lro{g(x) - \hat{g}_{\mathrm{CM}}(x)}^2 \diff x + \int_{K_{n_k}}^{\amax}  \lro{g(x) - \hat{g}_{\mathrm{CM}}(x)}^2 \diff x \bigg)^{1/2}\\
&= \lro{O\lro{ K_1 - \amin }^5 + O\lro{\frac{1}{n_k^4}} +  O\lro{ \amax - K_{n_k} }^5 }^{1/2}\\
&= O\lro{\frac{1}{n_k^2}}.
\end{align*}

\end{proof}

\subsection{Proof of Proposition \ref{prop:cov_addition}}
\begin{proof}
We start from the identity
\begin{equation*}
x_M^2 = \sum_{k=1}^d w_k^2 x_k^2 + 2 \sum_{1 \le i < j \le d} w_i w_j x_i x_j.
\end{equation*}
Apply the linear projection operator $\widehat{\Pi}_{\cF}$ to both sides. Since $x_1^2,\dots,x_d^2,x_M^2$ belong to $\cF$, the projection leaves these terms unchanged. Hence
\begin{equation*}
x_M^2 = \sum_{k=1}^d w_k^2 x_k^2 + 2 \sum_{1 \le i < j \le d} w_i w_j \widehat{\Pi}_{\cF}[x_i x_j].
\end{equation*}
Taking risk-neutral expectations on both sides gives
\begin{equation*}
\Var_t^Q R_{\ttt} = \sum_{k=1}^d w_k^2 \Var_t^Q R_{k,\ttt} + 2 \sum_{1 \le i < j \le d} w_i w_j \widehat{\Cov}_{ij,t}^Q,
\end{equation*}
where we used $\Expect_t^Q[x_M]=\Expect_t^Q[x_k]=0$ under the no-dividend assumption. This is \eqref{eq:cov_sum}.
\end{proof}

\subsection{Proof of Proposition \ref{prop:gauss_equicor_collapse}}
\begin{proof}
Let $\omega_\rho$ be the Gaussian weighting density specified in the proposition. The proof will use the multivariate Hermite decomposition of \(L^2(\omega_\rho)\). For $q\ge 0$, let $\mathcal P_q$ denote the space of all polynomials in $x$ of total degree at most $q$.

We first show that basis elements of degree larger than two do not affect the projection of $x_i x_j$. Consider any basis element $z$ in $\cF_p$ of degree $m\ge 3$, such as $z=x_k^m$ or $z=x_M^m$. Let $\tilde z$ be its residual after projecting onto the span of lower-degree elements in $\cF_p$. Under a Gaussian weighting density, $L^2(\omega_\rho)$ admits the Hermite orthogonal grading by total polynomial degree, and the degree-$m$ block is orthogonal to every polynomial of degree strictly smaller than $m$. Because $\cF_p$ contains every univariate polynomial in $x_k$ and in $x_M$ of degree at most $p$---with the degree-one term $x_M=w'x$ already contained in the span of $x_1,\dots,x_d$---the residualization of $z=x_k^m$ or $z=x_M^m$ against the lower-degree elements of $\cF_p$ removes exactly the lower-degree Hermite components of $z$. Hence $\tilde z$ lies in the degree-$m$ Hermite block. Therefore, since $x_i x_j\in \mathcal P_2$ and $m\ge 3$,
\begin{equation*}
    \seq{x_i x_j,\tilde z}=0.
\end{equation*}
Thus the coefficient on every residualized basis element of degree larger than two is zero. Consequently, for the purpose of projecting $x_i x_j$, the space $\cF_p$ is equivalent to its quadratic part,
\begin{equation*}
    \cF_2
    =
    \operatorname{span}
    \set{1,x_1,\dots,x_d,x_1^2,\dots,x_d^2,x_M^2}.
\end{equation*}

It remains to evaluate the price of the quadratic projection. By construction, $\omega_\rho$ matches the option-implied sector means, sector variances, and market variance. Hence, for every $g\in\cF_2$,
\begin{equation*}
    \Expect_t^Q\lr{g(X)}
    =
    \Expect^{\omega_\rho}\lr{g(X)}.
\end{equation*}
Since $\widehat{\Pi}_{\cF_2}[x_i x_j]\in\cF_2$, this gives
\begin{equation*}
    \Expect_t^Q\lr{\widehat{\Pi}_{\cF_2}[x_i x_j]}
    =
    \Expect^{\omega_\rho}\lr{\widehat{\Pi}_{\cF_2}[x_i x_j]}.
\end{equation*}
Moreover, because the constant belongs to $\cF_2$, the projection residual is orthogonal to constants:
\begin{equation*}
    \Expect^{\omega_\rho}\lr{x_i x_j-\widehat{\Pi}_{\cF_2}[x_i x_j]}=0.
\end{equation*}
Combining the previous displays yields
\begin{equation*}
    \Expect_t^Q\lr{\widehat{\Pi}_{\cF_p}[x_i x_j]}
    =
    \Expect_t^Q\lr{\widehat{\Pi}_{\cF_2}[x_i x_j]}
    =
    \Expect^{\omega_\rho}\lr{x_i x_j}
    =
    \rho_\omega\sigma_i\sigma_j,
\end{equation*}
where the first equality follows from the preceding reduction of $\cF_p$ to its quadratic part. 
\end{proof}


\section{Convergence rate}\label{app:conv_rate}
In this section, we establish the rate at which the estimated risk-neutral expectation converges as a function of the number of strikes. From approximation theory, we expect the convergence rate to depend on the smoothness of the underlying function (see, e.g., \citet[Chapter 10]{Trefethen2018}). To facilitate the comparison with the CM formula, we assume that the underlying function is twice continuously differentiable.  The following proposition derives the convergence rate of the projection approach under this assumption.

\begin{prop}\label{prop:conv_ols}
Let everything be as in Proposition~\ref{prop:proj_errorBound}, and
let Assumption~\ref{asmp:compact} hold. Assume that
$g \in C^2(A)$, where $A = [\amin, \amax]$. Let $\Delta := \max_j (K_{j+1} - K_j)$,
where the strikes are ordered
$\amin < K_1 < K_2 < \cdots < K_{n_k} < \amax$, and assume that
$\Delta = O(1/n_k)$, $K_1 - \amin = O(1/n_k^{4/5})$, and
$\amax - K_{n_k} = O(1/n_k^{4/5})$. Then as $n_k \to \infty$,
\begin{equation*}
  \Expect_t^Q\!\lr{g(S_T)\, \ind{S_T \in A}}
  \;=\; \Expect_t^Q\!\lr{\hat g(S_T)\, \ind{S_T \in A}}
       + O\!\lro{\frac{1}{n_k^2}}.
\end{equation*}
\end{prop}

Proposition~\ref{prop:conv_ols} can be viewed as yielding the speed with which options complete the market. For the CM formula, the integral representation can be approximated using the composite trapezoidal rule, which is the method employed by the CBOE to calculate the VIX. Under the same assumptions, the CM approximation with the trapezoidal rule attains the same convergence rate.

\begin{prop}\label{prop:conv_cm}
Let everything be as in Proposition \ref{prop:conv_ols}, and denote the CM replicating portfolio by 
\begin{align*}
\hat g_{\mathrm{CM}}(S_T)
&= g(F_{\ttt}) + g'(F_{\ttt})(S_T-F_{\ttt})\\
&+ \sum_{j:K_j\le F_{\ttt}} \Delta K_j\, g''(K_j) \pmax{K_j - S_T}
+ \sum_{j:K_j>F_{\ttt}} \Delta K_j\, g''(K_j) \pmax{S_T - K_j} .
\end{align*}
where 
\begin{equation*}
\Delta K_j = 
\begin{cases}
\frac{K_{j+1} - K_{j-1}}{2},  & j = 2,\dots,n_k - 1   \\
K_2 - K_1, & j = 1\\
K_{n_k} - K_{n_{k} - 1}, & j = n_k.
\end{cases}
\end{equation*}
Then, as $n_k \to \infty$
\begin{align*}
\Expect_t^Q \lr{g(S_T) \ind{S_T \in A }} &=  \Expect_t^Q \lr{ \hat{g}_{\mathrm{CM}}(S_T)\ind{S_T \in A}}  + O\lro{\frac{1}{n_k^2}}.  
\end{align*}
\end{prop}

\section{Projection and equicorrelation}\label{sec:proj_equi}
Vanilla options on the individual sectors and on the market portfolio do not, in general, identify the full matrix of pairwise risk-neutral correlations (Proposition \ref{prop:impossible}). Closing this identification gap requires additional structure, and a common choice is to impose equicorrelation. We will show that the equicorrelation estimator admits a portfolio-replication interpretation, and we then generalize it via projection. For ease of exposition we maintain the no-dividend assumption; incorporating dividends is straightforward at the cost of slightly heavier notation.\footnote{Under this assumption, $\Expect_t^Q R_{i,\ttt} = R_{f,\ttt}$. If dividends are included, $\Expect_t^Q R_{i,\ttt} = F_{i,\ttt}/S_{i, t}$.}

The equicorrelation estimator of \citet{engle2012dynamic} assumes that the pairwise correlation is the same for every pair of assets. In that case, the implied common correlation can be written as
\begin{equation}\label{eq:rho_equi}
\hat{\rho}_t = \frac{\Var_t^Q(R_{\ttt}) - \sum_{j=1}^d w_{j,t}^2 \Var_t^Q(R_{j,\ttt}) }{2 \sum_{1 \le i < j \le d} w_{i,t} w_{j,t} \sqrt{\Var_t^Q(R_{i,\ttt}) \Var_t^Q(R_{j,\ttt})} },
\end{equation}
which is the formula the CBOE uses to construct its implied correlation index. It is useful to reinterpret \eqref{eq:rho_equi} as a portfolio-replication problem. The target payoff is
\begin{equation*}
\frac{(R_{i,\ttt} - R_{f,\ttt})(R_{j,\ttt} - R_{f,\ttt})}{\sqrt{\Var_t^Q(R_{i,\ttt}) \Var_t^Q(R_{j,\ttt})}},
\end{equation*}
and the basis functions are the quadratic payoffs
\begin{equation*}
\lro{R_{\ttt} - R_{f,\ttt}}^2 \quad \text{and} \quad  \lro{R_{j,\ttt} - R_{f,\ttt}}^2, \quad \text{for } j = 1, \dots,d.
\end{equation*}
Viewed this way, the replicating portfolio combines the market-variance payoff and sector-variance payoffs, with coefficients determined by $w_j$ and sector volatilities $\sigma_j$. The estimator is identified from option prices because each variance is identified from individual options, but the cross-sectional restriction is strong: a single common $\rho$ is imposed on $d(d-1)/2$ distinct pairs.

The projection approach optimizes and relaxes these features. For shorthand, let $x_k := R_{k,\ttt} - R_{f,\ttt}$ and $x_M := R_{\ttt} - R_{f,\ttt}$ denote the excess returns on asset $k$ and on the market, respectively. Let $x = [x_1,\dots,x_d]'$, so that $x_M = w\cdot x$, where $w$ is the vector of market weights. To generalize the equicorrelation estimator, we seek the optimal replicating portfolio for $x_i x_j$, which directly targets the risk-neutral covariance between returns $i$ and $j$. 

The projection approach would search over univariate functions $g_1,\dots,g_d,g_M$ and solve
\begin{equation*}
\min_{g_1,\dots,g_d,g_M}
\int_{\R^d} \Bigl(x_i x_j - \sum_{k=1}^d g_k(x_k) - g_M(x_M)\Bigr)^2 \omega(x)\diff x,
\end{equation*}
with $x_M=w'x$. Solving this infinite-dimensional problem directly is computationally demanding. We approximate the additive space by low-degree polynomial payoffs because they can be reliably estimated from option prices.

For a fixed degree $p$, define
\begin{equation}\label{eq:F_p}
\cF_p = \operatorname{span}\Bigl(\set{1} \cup \set{x_k^m: k=1,\dots,d,\; m=1,\dots,p} \cup \set{x_M^m: m=2,\dots,p}\Bigr).
\end{equation}
We omit $x_M$ itself because $x_M=w'x$ already lies in the span of the sector-level linear terms. In the application below we set $p=4$, which allows the replicating portfolio to use variance, skewness, and kurtosis-type payoffs of every sector and of the market portfolio. The specification has two virtues over the equicorrelation benchmark: it exploits higher moments of the individual assets and of the index, and it permits the projection coefficients to differ across pairs $(i,j)$, so heterogeneous covariances can be estimated rather than collapsed onto a single common correlation.

Let $\widehat{\Pi}_{\cF_p}[x_i x_j]$ denote the $L^2(\omega)$-projection of $x_i x_j$ onto $\cF_p$. We write this projection as
\begin{equation}\label{eq:poly_projection}
\widehat{\Pi}_{\cF_p}[x_i x_j]
=
\hat{\beta}_{0,ij}
+
\sum_{m=1}^p \sum_{k=1}^d \hat{\beta}_{k,m,ij} x_k^m
+
\sum_{m=2}^p \hat{\beta}_{M,m,ij}x_M^m.
\end{equation}
The polynomial projection separates the multivariate and univariate parts of the problem. First, we compute the pair-specific coefficients in \eqref{eq:poly_projection} under the weighting density $\omega$. Second, we price each univariate power payoff $x_k^m$ and $x_M^m$ using the one-dimensional projection method of Section~\ref{sec:gen_proj}.  Define
\begin{equation*}
m_{k,m,t}^Q := \Expect_t^Q\lr{x_k^m}, \qquad m_{M,m,t}^Q := \Expect_t^Q\lr{x_M^m}.
\end{equation*}
Based on \eqref{eq:poly_projection}, the covariance estimator is
\begin{align}\label{eq:cov_hat}
\widehat{\Cov}_{ij,t}^Q
\coloneqq
\Expect_t^Q\lr{\widehat{\Pi}_{\cF_p}[x_i x_j]}
&=
\hat{\beta}_{0,ij}
+
\sum_{m=1}^p \sum_{k=1}^d \hat{\beta}_{k,m,ij} m_{k,m,t}^Q
+
\sum_{m=2}^p \hat{\beta}_{M,m,ij}m_{M,m,t}^Q.
\end{align}
Equation \eqref{eq:cov_hat} delivers the $L^2(\omega)$-best linear combination of $\cF_p$ basis prices that targets $\Expect_t^Q[x_i x_j]$. Two sources of approximation error remain: a basis-misspecification error from projecting $x_i x_j$ onto the additive space $\cF_p$, which vanishes whenever $x_i x_j$ itself lies in $\cF_p$ (for instance, when $d=2$, by the variance addition identity in the proof of Proposition~\ref{prop:cov_addition}); and a finite-strike error in the univariate pricing of each $m_{k,m,t}^Q$ and $m_{M,m,t}^Q$, which inherits the finite-sample bound of the univariate projection estimator. Under the no-dividend assumption, $m_{k,1,t}^Q=0$, so the sector-level linear terms drop from \eqref{eq:cov_hat}. They are nevertheless retained in the projection \eqref{eq:poly_projection}, because they improve the weighted replication over the state space whenever $\omega$ is not centered exactly at the forward. For $p=4$, the inputs to the covariance estimator are the second, third, and fourth risk-neutral central moments of every sector and of the market portfolio.

Because the risk-neutral variance of the market is identified from index options, it is desirable that the covariance estimates satisfy the variance addition formula
\begin{equation}\label{eq:cov_sum}
\Var_t^Q R_{\ttt} = \sum_{i=1}^d w_i^2 \Var_t^Q R_{i,\ttt} + 2 \sum_{1 \le i < j \le d} w_i w_j \widehat{\Cov}_{ij,t}^Q.
\end{equation}
The next proposition shows that this accounting identity holds for any weighted projection, provided the projection space contains all univariate quadratic terms.

\begin{prop}[Variance addition under weighted projection]\label{prop:cov_addition}
Let $\cF$ be a linear function space such that $\set{x_1^2,\dots,x_d^2,x_M^2} \subset \cF$, and let $\widehat{\Pi}_{\cF}$ denote the $L^2(\omega)$-projection onto $\cF$. Define the covariance estimator for every pair using this same projection operator by
\begin{equation*}
\widehat{\Cov}_{ij,t}^Q = \Expect_t^Q\lr{\widehat{\Pi}_{\cF}[x_i x_j]}.
\end{equation*}
Then \eqref{eq:cov_sum} holds. In particular, the condition is satisfied by $\cF_p$ whenever $p\ge 2$.
\end{prop}

\begin{rem}\label{rem:x_M}
This result extends to the case with multiple index portfolios. Suppose there are two index excess returns $x_{M,1}=w_1\cdot x$ and $x_{M,2}=w_2\cdot x$ with corresponding options. Proposition~\ref{prop:cov_addition} holds simultaneously provided the projection space contains all univariate quadratic terms, including $x_{M,1}^2$ and $x_{M,2}^2$. Under this condition, \eqref{eq:cov_sum} holds for each weight vector $w_\ell$ ($\ell=1,2$), with the variance on the left-hand side taken for the corresponding portfolio.
\end{rem}

\subsection{Choice of weighting density and shrinkage interpretation}\label{sec:shrinkage}
The choice of weighting density determines where the projection is asked to fit the target payoff most accurately. In the multivariate setting, it also determines the dependence structure toward which the estimator is regularized. This point is closely related to Proposition~\ref{prop:additive_weights}. If the basis contains only individual-asset payoffs and the weighting density prices the additive basis correctly, then the projected price of a joint payoff equals its expectation under the weighting density. This condition holds, for example, in the dense-strike limit when the marginals of the weighting density coincide with the option-implied marginal distributions. Thus, for $g(x)=x_i x_j$,
\begin{equation*}
    \Expect_t^Q\lr{\widehat{\Pi}_{\cA}[x_i x_j]}
    =
    \Expect^{\omega}\lr{X_iX_j},
\end{equation*}
where $\cA$ denotes the additive span generated by payoffs that depend on one asset at a time. With product weights, $\omega(x)=\prod_{k=1}^d \omega_k(x_k)$, and centered excess returns, this expectation is zero. Hence, using only individual options and product weights amounts to imposing independence.

The covariance estimator in \eqref{eq:cov_hat} differs because the projection space also contains market-index payoffs, $x_M^m=(w'x)^m$. These payoffs depend jointly on all assets and are priced from index options. They therefore provide the source of joint information that is absent from individual options alone. To make the connection explicit, write the projection space as
\begin{equation*}
    \cF_p=\cA_p+\cZ_p,
\end{equation*}
where $\cA_p$ contains the sector-level polynomial payoffs and $\cZ_p$ contains the market-index polynomial payoffs. Let $\tilde z_r$ be an orthonormal basis for the residualized market-payoff space, then $\Expect^{\omega}[\tilde z_r]=0$ by construction, and under product weighting the projection price can be written as
\begin{equation*}
    \Expect_t^Q\lr{\widehat{\Pi}_{\cF_p}[x_i x_j]}
    =
    \Expect^{\omega}\lr{X_iX_j}
    +
    \sum_r \eta_{r,ij}
    \lr{
        \Expect_t^Q\lr{\tilde z_r(X)}
        -
        \Expect^{\omega}\lr{\tilde z_r(X)}
    },
\end{equation*}
where $\eta_{r,ij}=\seq{x_i x_j,\tilde z_r}$. The subtraction $\Expect^{\omega}[\tilde z_r]=0$ is kept to make the $Q$-vs-$\omega$ moment-gap interpretation transparent. The first term is the dependence implied by the weighting density. The second term is the correction implied by index option prices. Under product weights, the first term is zero, so the estimator is shrunk toward independence and all nonzero dependence comes from the residualized index payoffs. This interpretation is consistent with Proposition~\ref{prop:additive_weights}: individual options alone do not identify dependence, while index options add jointly informative prices.

A non-product weighting density changes the shrinkage target. For example, one could choose a multivariate Gaussian weighting density with marginal variances matched to the sector weighting densities and with common correlation $\rho_{\omega}$,
\begin{equation*}
    \omega_{\rho}(x)
    =
    \varphi_d\lr{
        x;0,\Sigma_{\rho}
    },
    \qquad
    (\Sigma_{\rho})_{ii}
    =
    \sigma_i^2,
    \qquad
    (\Sigma_{\rho})_{ij}
    =
    \rho_{\omega}\sigma_i\sigma_j
    \quad (i\ne j).
\end{equation*}
If $\rho_{\omega}$ is set equal to the equicorrelation estimate in \eqref{eq:rho_equi}, then the weighting density embeds the same common-correlation structure as the equicorrelation benchmark. In this case,
\begin{equation*}
    \Expect^{\omega_{\rho}}\lr{X_iX_j}
    =
    \rho_{\omega}\sigma_i\sigma_j,
\end{equation*}
so the projection estimator is tilted toward the equicorrelation-implied covariance rather than toward zero. 

In fact, this equivalence is exact in the quadratic Gaussian case. Let
\begin{equation*}
    \cF_2
    =
    \operatorname{span}
    \set{1,x_1,\dots,x_d,x_1^2,\dots,x_d^2,x_M^2}.
\end{equation*}
Suppose $\omega_{\rho}$ is a centered Gaussian density whose marginal variances equal the option-implied variances, $\sigma_i^2=\Var_t^Q(R_{i,\ttt})$, and whose common correlation $\rho_{\omega}$ is chosen so that the market variance under $\omega_{\rho}$ equals the option-implied market variance:
\begin{equation*}
    \Var_t^Q(R_{\ttt})
    =
    \sum_{i=1}^d w_i^2\sigma_i^2
    +
    2\rho_{\omega}
    \sum_{1\le i<j\le d}
    w_iw_j\sigma_i\sigma_j.
\end{equation*}
Solving this equation gives exactly the equicorrelation estimator in \eqref{eq:rho_equi}. Hence $\omega_{\rho}$ prices every payoff in $\cF_2$ correctly: constants, sector means, sector variances, and the market variance. Since the projection residual is orthogonal to constants under $L^2(\omega_{\rho})$, the same argument as in Proposition~\ref{prop:additive_weights} gives
\begin{equation*}
    \Expect_t^Q\lr{\widehat{\Pi}_{\cF_2}[x_i x_j]}
    =
    \Expect^{\omega_{\rho}}\lr{X_iX_j}
    =
    \rho_{\omega}\sigma_i\sigma_j.
\end{equation*}
Dividing by the option-implied marginal standard deviations yields
\begin{equation*}
    \widehat{\Corr}_{ij,t}^Q
    =
    \rho_{\omega}
    \quad
    \text{for all } i\ne j.
\end{equation*}
Thus, when the Gaussian weighting density is calibrated to the equicorrelation-implied covariance matrix, the projection estimator collapses exactly to the equicorrelation estimator. Intuitively, the index-option information has already been embedded in the weighting density through the calibration of $\rho_{\omega}$, leaving no residual correction for the projection to extract.

The same conclusion holds at any polynomial degree $p\ge 2$: even when $\cF_p$ contains the higher-degree basis elements $x_k^m$ and $x_M^m$ with $m\ge 3$, those elements enter the projection with zero coefficient under Gaussian $\omega_{\rho}$ and the quadratic argument above applies verbatim.

\begin{prop}[Gaussian equicorrelation collapse]\label{prop:gauss_equicor_collapse}
Under the assumptions of the quadratic case above with at least two entries of $w$ non-zero, and for every $p\ge 2$ and every pair $(i,j)$ with $i\ne j$,
\begin{equation*}
    \Expect_t^Q\lr{\widehat{\Pi}_{\cF_p}[x_i x_j]} = \rho_{\omega}\sigma_i\sigma_j.
\end{equation*}
\end{prop}

This result provides a useful way to view the weighting density in incomplete markets. Product weights give a conservative benchmark in which dependence is introduced only through traded index options. The Gaussian QMLE of \citet{engle2012dynamic} instead shrinks towards equicorrelation unless index-option prices provide information that pushes it away. To improve on the equicorrelation benchmark, the weighting density must therefore allow richer pair-specific dependence, or the projection basis and pricing inputs must include jointly informative moments not already matched by the equicorrelated Gaussian prior. A natural choice is the VG product weighting density, under which the cross-sectional information enters through the multinomial expansion of $\Expect^{\omega}[x_M^m]$ for $m\ge 3$ as well as $m=2$ (see Section \ref{sec:prod_weight_impl}). Hence,   the projection coefficients on $x_k^m$ for $m\ge 3$ are no longer forced to zero and the option-implied higher moments $m_{k,m,t}^Q$ enter the covariance estimate non-trivially.  Simulation evidence in Section~\ref{sec:equicor_simulation} illustrates a case in which product weighting beats the equicorrelation estimator.

\subsection{Simulation evidence for sector ETFs}\label{sec:equicor_simulation}
We evaluate the covariance estimator in \eqref{eq:cov_hat} on a Monte Carlo design: option prices are generated from a true joint risk-neutral distribution, marginal weighting densities are calibrated to these prices, and every input to both the projection and the equicorrelation benchmark is obtained from options only. Section \ref{sec:prod_weight_impl} discusses how to efficiently estimate the projection coefficients. 

\paragraph{Risk-neutral data-generating process.}
For $d=11$ sectors, let $Z\in\R^{11}$ denote log-returns and $R=\exp(Z)$ the corresponding gross returns. We simulate
\begin{equation*}
Z = Bf + \varepsilon,\qquad
f \sim \mathsf{N}\!\bigl(0,\diag(\sigma_{1,f}^2,\sigma_{2,f}^2)\bigr),\quad
\varepsilon \sim \mathsf{N}\!\bigl(0,\diag(\sigma_1^2,\ldots,\sigma_{11}^2)\bigr),\quad
f \perp \varepsilon,
\end{equation*}
with $B\in\R^{11\times 2}$, factor volatilities $\sigma_{i,f}\sim\mathsf{Unif}[0.15,0.30]$, idiosyncratic volatilities $\sigma_k\sim\mathsf{Unif}[0.03,0.20]$, and entries of $B$ drawn \iid\ from $\mathsf{Unif}[-0.4,1]$. The two-factor structure delivers cross-sectional heterogeneity in pairwise correlations. Gross returns $R$ are winsorized component-wise to $[0.4,1.5]$, and then rescaled by their component-wise sample mean so that $\Expect_t^Q R_{k,\ttt}=1$ holds exactly in each Monte Carlo replication; the corresponding sample risk-free rate is $R_{f,\ttt}=1$. Excess returns are $x_k=R_{k,\ttt}-1$ and the market excess return is $x_M=w'x$, with sector weights $w$ taken from a representative S\&P 500 sector market-capitalization snapshot. We draw $N_{\text{sim}}=2\times 10^5$ samples per replication.

\paragraph{Option pricing.}
For each sector $k=1,\dots,d$ and for the market portfolio, we choose $n_K=20$ strikes equally spaced between the 5th and 95th percentile of the simulated return distribution, and price each strike as its OTM option by sample averaging:
\begin{equation*}
C_{k,t}(K)=\Expect_t^Q\lr{\pmax{R_{k,\ttt}-K}} \approx \frac{1}{N_{\text{sim}}}\sum_{n=1}^{N_{\text{sim}}}\pmax{R_{k,\ttt}^{(n)}-K},\quad K>1,
\end{equation*}
and analogously for puts when $K\le 1$. No discounting is applied. The Monte Carlo sample size is large enough that finite-sample noise in these prices is negligible relative to the discretization induced by using only $n_K=20$ strikes.

\paragraph{Marginal calibration.}
For each underlying we fit the VG density to the simulated option prices. The fitted sector densities $\omega_1,\dots,\omega_d$ form the product weighting density $\omega(x)=\prod_{k=1}^d \omega_k(x_k)$ used in the projection. The fitted market density $\omega_M$ is used solely as the univariate $L^2$-weight for pricing the market polynomial payoffs $x_M^m$ from index options, not as a marginal of $\omega$.

\paragraph{Results.}
Table~\ref{tab:corr-summary} reports the distribution across 1{,}000 Monte Carlo replications of the MSE between the true vector of pairwise correlations (computed from the simulated joint sample) and each estimator. The projection-based estimator delivers a lower MSE across the bulk of the distribution: a 6.7\% reduction in median MSE, an 8.1\% reduction in mean MSE, and a 21.7\% reduction in maximum MSE. The within-replication correlation between the true and estimated correlation vectors is positive on average, indicating that the projection captures meaningful cross-sectional heterogeneity rather than tracking a common-correlation aggregate.

\begin{table}[htb!]
  \centering
  \begin{tabular}{lccccc}
    \toprule
    \midrule
    & Min & Median & Max & Mean & Std. dev. \\
    \midrule
    Equicorrelation        & 0.0232 & 0.1392 & 0.3868 & 0.1459 & 0.0483 \\
    Projection correlation & 0.0341 & 0.1299 & 0.3030 & 0.1341 & 0.0393 \\
    \bottomrule
  \end{tabular}
  \caption{\textbf{Summary statistics of MSE}. The equicorrelation row corresponds to the estimator \eqref{eq:rho_equi}; the projection row corresponds to the projection-based covariance estimator \eqref{eq:cov_hat}, converted to correlations by dividing by the option-implied marginal standard deviations.}
  \label{tab:corr-summary}
\end{table}

\subsection{Practical implementation}\label{sec:prod_weight_impl}
The coefficients in \eqref{eq:poly_projection} are determined by the weighted normal equations. Let $\phi_1,\dots,\phi_q$ denote the basis functions that span $\cF_p$, and write
\begin{equation*}
\widehat{\Pi}_{\cF_p}[x_i x_j](x)=\sum_{\ell=1}^q \hat\theta_{\ell,ij}\phi_\ell(x).
\end{equation*}
Then $\hat\theta_{ij}$ solves
\begin{equation}\label{eq:normal_product}
G_\omega \hat\theta_{ij}=b_{ij},
\qquad
(G_\omega)_{\ell r}=\Expect^{\omega}\lr{\phi_\ell(X)\phi_r(X)},
\qquad
(b_{ij})_\ell=\Expect^{\omega}\lr{X_iX_j\phi_\ell(X)}.
\end{equation}
The computational problem thus reduces to evaluating moments under the weighting density. For a general joint density $\omega$, these are $d$-dimensional integrals. The computation becomes low-dimensional once $\omega$ is chosen to be a product density,
\begin{equation*}
\omega(x)=\prod_{k=1}^d \omega_k(x_k).
\end{equation*}
Let
\begin{equation*}
\mu_{k,r}:=\int_{\R} x_k^r\omega_k(x_k)\diff x_k,
\qquad r=0,1,\dots,2p,
\end{equation*}
with $\mu_{k,0}=1$. These univariate moments are available in closed form for common choices of $\omega_k$,  and can otherwise be obtained by one-dimensional numerical integration. Product weights imply that every monomial moment factors. For any multi-index $\nu=(\nu_1,\dots,\nu_d)$,
\begin{equation}\label{eq:monomial_factorization}
\Expect^{\omega}\lr{X^\nu}
=
\prod_{k=1}^d \mu_{k,\nu_k},
\qquad
X^\nu:=\prod_{k=1}^d X_k^{\nu_k}.
\end{equation}
Terms involving the market excess return are handled by expanding $x_M=w'X$. For any integer $r\ge 0$,
\begin{equation}\label{eq:market_moment_product}
\Expect^{\omega}\lr{X_M^r}
=
\sum_{\alpha_1+\cdots+\alpha_d=r}
\binom{r}{\alpha_1,\dots,\alpha_d}
\prod_{k=1}^d w_k^{\alpha_k}\mu_{k,\alpha_k}.
\end{equation}
More generally, for any multi-index $\nu$ and any integer $r\ge 0$,
\begin{equation}\label{eq:mixed_market_moment_product}
\Expect^{\omega}\lr{X^\nu X_M^r}
=
\sum_{\alpha_1+\cdots+\alpha_d=r}
\binom{r}{\alpha_1,\dots,\alpha_d}
\prod_{k=1}^d w_k^{\alpha_k}\mu_{k,\nu_k+\alpha_k}.
\end{equation}
Equations \eqref{eq:monomial_factorization}--\eqref{eq:mixed_market_moment_product} give all entries of $G_\omega$ and $b_{ij}$. For example, if $\phi_\ell(x)=x_a^m$ and $\phi_r(x)=x_M^n$, then
\begin{equation*}
(G_\omega)_{\ell r}
=
\Expect^{\omega}\lr{X_a^m X_M^n}
=
\sum_{\alpha_1+\cdots+\alpha_d=n}
\binom{n}{\alpha_1,\dots,\alpha_d}
\prod_{k=1}^d w_k^{\alpha_k}\mu_{k,m\ind{k=a}+\alpha_k}.
\end{equation*}
Similarly, the right-hand side entry for a market basis function $\phi_\ell(x)=x_M^m$ is
\begin{equation*}
(b_{ij})_\ell
=
\Expect^{\omega}\lr{X_iX_jX_M^m}
=
\sum_{\alpha_1+\cdots+\alpha_d=m}
\binom{m}{\alpha_1,\dots,\alpha_d}
\prod_{k=1}^d w_k^{\alpha_k}\mu_{k,\ind{k=i}+\ind{k=j}+\alpha_k}.
\end{equation*}
For $p=4$, the highest required moment order is $2p=8$: the Gram matrix involves products of basis functions up to degree eight, while the right-hand side involves $x_i x_j$ times basis functions of degree at most four. With $d=11$, the largest multinomial sum has $\binom{d+2p-1}{2p}=\binom{18}{8}$ terms, which is small enough to evaluate directly. After forming $G_\omega$ and $b_{ij}$, the projection coefficients are obtained from the linear system in \eqref{eq:normal_product}. This step is repeated for each pair $i,j$, while the Gram matrix $G_\omega$ is common across pairs and can be factorized once.

\section{Empirical estimates of SVIX and VIX}\label{app:svix_vix}
According to the simulation results, the projection approach compares favorably to the CM formula especially when the number of observed option prices is small. When the number of observed options is large it is a priori not so clear whether a more refined approximation yields economically different results.  To investigate the benefits of the projection approach in the latter case, we estimate the SVIX and VIX from Examples \ref{exmp:var}--\ref{exmp:vix} using both methods. The calculation of both indexes requires options on the S\&P500, which is one of the most liquid option markets worldwide. The SVIX and VIX are therefore a natural test case.

The options data on the SP500 come from OptionMetrics and span the period January 4, 1996 until July 20, 2023. Several data cleaning procedures are applied before each volatility index is calculated. The procedure is almost identical to CBOE's method when it calculates the VIX. A detailed description of our procedure is included in Appendix \ref{app:data_clean}. To implement the projection estimator, we use the VG process as the weighting distribution.

First, consider the SVIX defined by 
\begin{equation}\label{eq:svix}
\mathrm{SVIX}_{\ttt}^2 = \frac{1}{T-t} \Var_t^Q\lro{\frac{R_{\ttt}}{R_{f,\ttt}}}.
\end{equation}
\citet{martin2017expected} derives conditions under which the conditional equity premium satisfies
\begin{equation*}
\frac{1}{T-t} \lro{\Expect_t R_{\ttt} - R_{f,\ttt}} \ge R_{f,\ttt} \mathrm{SVIX}_{\ttt}^2.
\end{equation*}
In fact, when running the regression
\begin{equation}\label{eq:svix_reg}
\frac{1}{T-t} \lro{R_{\ttt} - R_{f,\ttt}} = \beta_0 + \beta_1 R_{f,\ttt} \mathrm{SVIX}_{\ttt}^2 + \varepsilon_T,
\end{equation}
\citet{martin2017expected,martin2025information} cannot reject the null hypothesis that $\beta_0 = 0$ and $\beta_1 = 1$, thus suggesting that the lower bound is tight. This conclusion is particularly interesting as it gives a model-free way to measure the equity premium in real time. Given its importance, we reassess this claim by using our projection method to measure $\mathrm{SVIX}_{\ttt}^2$. Table \ref{tab:svix} shows the results. For the 30-day horizon the difference between the CM and projection method is small, consistent with Proposition \ref{prop:cm_proj_weights} that the choice of method is immaterial when many options are traded. The gap widens at longer horizons, where the projection delivers a smaller $\beta_1$, close enough to the null $\beta_1 = 1$ that a tight lower bound cannot be rejected.

\begin{table}[htbp]
\centering
\resizebox{\textwidth}{!}{%
\begin{tabular}{lcccccccc}
\toprule
\midrule
& \multicolumn{2}{c}{30 days} & \multicolumn{2}{c}{60 days} & \multicolumn{2}{c}{90 days} & \multicolumn{2}{c}{180 days} \\
\cmidrule(lr){2-3} \cmidrule(lr){4-5} \cmidrule(lr){6-7} \cmidrule(lr){8-9}
& Proj. & CM & Proj. & CM & Proj. & CM & Proj. & CM \\
\midrule
$\beta_0$
& $\underset{(0.0407)}{0.002}$ & $\underset{(0.0400)}{0.005}$
& $\underset{(0.0437)}{-0.005}$ & $\underset{(0.0427)}{-0.010}$
& $\underset{(0.0510)}{-0.002}$ & $\underset{(0.0504)}{-0.005}$
& $\underset{(0.0355)}{-0.038}$ & $\underset{(0.0365)}{-0.052}$ \\
$\beta_1$
& $\underset{(1.0133)}{1.432}$ & $\underset{(1.0816)}{1.493}$
& $\underset{(1.0682)}{1.487}$ & $\underset{(1.1424)}{1.750}$
& $\underset{(1.2528)}{1.377}$ & $\underset{(1.3602)}{1.589}$
& $\underset{(0.8086)}{2.235}$ & $\underset{(0.8371)}{2.865}$ \\
$R^2$ (\%)
& 1.13 & 1.08 & 1.86 & 2.19 & 2.08 & 2.35 & 5.98 & 7.94 \\
\# obs
& 6932 & 6932 & 6895 & 6895 & 6865 & 6865 & 6745 & 6745 \\
\bottomrule
\end{tabular}
}
\caption{\textbf{Equity premium regression.} This table reports estimates from regression~\eqref{eq:svix_reg} for return horizons of 30, 60, 90, and 180 days. Newey--West standard errors, using a bandwidth equal to the number of trading days in the horizon, are reported in parentheses below the coefficients.}
\label{tab:svix}
\end{table}

\begin{figure}[htb!]
\centering
\includegraphics[width=0.65\linewidth]{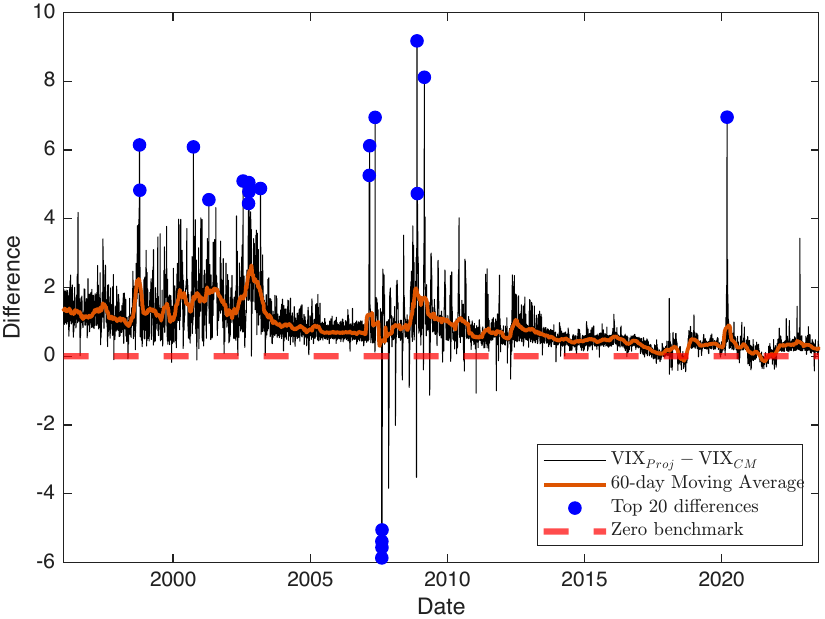}
\caption{\textbf{VIX estimate}. This figure shows the projection VIX estimate minus the VIX estimate obtained by CM. The solid orange line denotes the 60-day moving average of this difference. The blue dots indicate the 20 largest observed differences.}\label{fig:vix}
\end{figure}

In addition to SVIX, we also estimate the VIX. Figure~\ref{fig:vix} plots the time series of the difference between the two VIX estimates. The solid orange line is its 60-day moving average, which remains positive throughout the sample, consistent with the simulation. We mark the 20 largest differences in absolute value with blue dots. They concentrate in episodes of market stress: the Asian and dot-com crises (1998--2003), the run-up to and outbreak of the global financial crisis (2007--2009), and the onset of the COVID-19 pandemic (March 2020). The largest gap in absolute value reaches close to 9 percentage points, which is economically significant: portfolios with hundreds of VIX futures contracts can experience multi-million-dollar P\&L swings. This peak occurs on November 24, 2008, near the height of the global financial crisis, when the projection-implied VIX is 64.3\% while the CM approximation yields 55.1\%. During periods of heightened uncertainty, risk-neutral mass shifts to the left tail, which amplifies entropy because $\log(x)$ decays steeply near zero (see~\eqref{eq:rn_ent}). In such episodes the CM approximation, linearized around the risk-free rate, can understate the contribution of the deep left tail, whereas the projection method remains reliable because it approximates $\log(x)$ well over the entire domain. The exception is a cluster of large negative differences in mid-August 2007. These are an interpolation artifact rather than an economic feature: on the relevant days the 30-day VIX is extrapolated from two longer-dated maturities, one of which has an unusually thin option chain (only twelve OTM strikes covering a narrow range). The CM estimate at this thin maturity is biased downward by tail truncation, and linear backward extrapolation propagates the bias into the 30-day estimate. The projection method is unaffected and gives a stable estimate, consistent with the sparse-strike advantage documented in the simulations of Section~\ref{sec:simulate}.

\subsection{Option data preprocessing}\label{app:data_clean}
We use S\&P500 option data from OptionMetrics, covering the period January 4, 1996 to July 20, 2023. Following the CBOE procedure, we discard all in-the-money put and call options, as well as any option with a bid price of zero. When there are two consecutive strikes with a bid price equal to zero, all options with higher strikes (for calls) or lower strikes (for puts) are discarded. For each remaining option, the price is defined as the average of the bid and ask prices. In total, this filtering yields 11.738 million option prices. The risk-free rate for each return horizon is obtained from the zero-coupon yield curve dataset provided by OptionMetrics.

\section{Details on simulation}\label{app:simulation}
In the Monte-Carlo simulation, we use two different models to generate option prices. In both cases the time to maturity is 1 year. The first model is the
standard \citet{BlackScholes1973} model with a risk-free rate of $5\%$ and
volatility of $20\%$. The second model is a two-factor stochastic-volatility
model with double-exponential price jumps, calibrated to the risk-neutral
estimates of \citet{andersen2015risk} (henceforth AFT).\footnote{We hold the
jump intensities constant; AFT allow them to vary with the state, but a
constant-intensity specification is sufficient to generate a realistic
non-Gaussian risk-neutral density at a fixed horizon.} Under the risk-neutral measure, the log asset price
$\log S_t$ and its two variance factors $V_{1,t}, V_{2,t}$ follow
\begin{align*}
\diff \log S_t &= \lro{ r - q - \tfrac{1}{2}\lro{V_{1,t-} + V_{2,t-}}
      - \lambda^- \Expect^Q[\exp(\xi^-)-1]
      - \lambda^+ \Expect^Q[\exp(\xi^+)-1] } \diff t \\
   &\quad + \sqrt{V_{1,t-}}\, \diff W_{1,t} + \sqrt{V_{2,t-}}\, \diff W_{2,t}
      + \xi^- \diff N^-_t + \xi^+ \diff N^+_t, \\[4pt]
\diff V_{i,t} &= \kappa_i \lro{\theta_i - V_{i,t-}} \diff t
      + \sigma_i \sqrt{V_{i,t-}}\, \diff B_{i,t}, \qquad i = 1,2,
\end{align*}
where $V_{i,t-} = \lim_{s \uparrow t} V_{i,s}$ denotes the left limit. The
Brownian motions satisfy $\mathrm{corr}(\diff W_{i,t}, \diff B_{i,t}) = \rho_i$,
while the two factor pairs $(W_1, B_1)$ and $(W_2, B_2)$ are mutually
independent. Price jumps arrive as two independent Poisson processes $N^-_t$ and
$N^+_t$ with constant intensities $\lambda^-$ and $\lambda^+$. The associated
log-jump sizes are exponentially distributed, with $\xi^- < 0$ having density
$\eta^- e^{\eta^- x}$ on $(-\infty, 0)$ and $\xi^+ > 0$ having density
$\eta^+ e^{-\eta^+ x}$ on $(0, \infty)$, so that
$\Expect^Q[\exp(\xi^-)] = \eta^-/(\eta^- + 1)$ and
$\Expect^Q[\exp(\xi^+)] = \eta^+/(\eta^+ - 1)$. The drift compensation ensures
that the discounted price is a martingale, i.e. $\Expect^Q[S_T] = S_0 e^{(r-q)T}$.

For simulation, we only need to calibrate the model under the risk-neutral
measure. The diffusive parameters are the risk-neutral estimates of AFT, the
initial variance states $V_{1,0}$ and $V_{2,0}$ are set to their model-implied
means, and the jump intensities $\lambda^-, \lambda^+$ are held constant at
values matching the average jump frequencies in AFT (roughly $5.6$ downward and
$1.8$ upward jumps per year). The dividend yield is set to $q = 0$. All
parameters are summarized in Table \ref{tab:aft}.

\begin{table}[h!]
\centering
\begin{tabular}{l d{4}}
\toprule
\midrule
Parameter & \multicolumn{1}{c}{Value} \\
\midrule
$\kappa_1$    & 10.9890  \\
$\theta_1$    &  0.0030  \\
$\sigma_1$    &  0.2490  \\
$\rho_1$      & -0.9590  \\
$V_{1,0}$     &  0.0170  \\
\midrule
$\kappa_2$    &  1.8640  \\
$\theta_2$    &  0.0100  \\
$\sigma_2$    &  0.1700  \\
$\rho_2$      & -0.9790  \\
$V_{2,0}$     &  0.0120  \\
\midrule
$\eta^-$      & 25.9440  \\
$\eta^+$      & 36.6200  \\
$\lambda^-$   &  5.5980  \\
$\lambda^+$   &  1.7923  \\
\midrule
$r$           &  0.0500  \\
\bottomrule
\end{tabular}
\caption{\textbf{AFT model calibration}.}
\label{tab:aft}
\end{table}

\subsection{Robustness}
\begin{figure}[htb!]
\centering
\begin{subfigure}{0.48\linewidth}
\includegraphics[width=\linewidth]{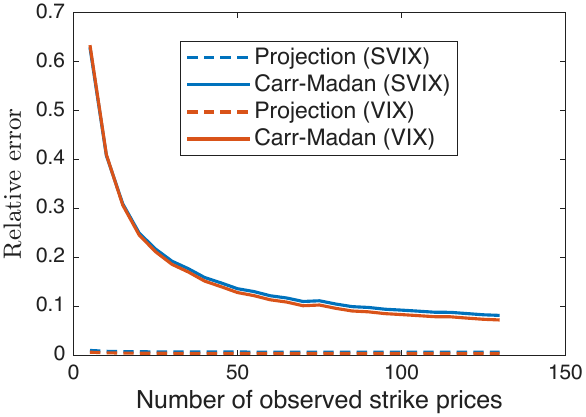}
\caption{Black-Scholes model}
\label{fig:bs_fixedRange}
\end{subfigure}
\begin{subfigure}{0.48\linewidth}
\includegraphics[width=\linewidth]{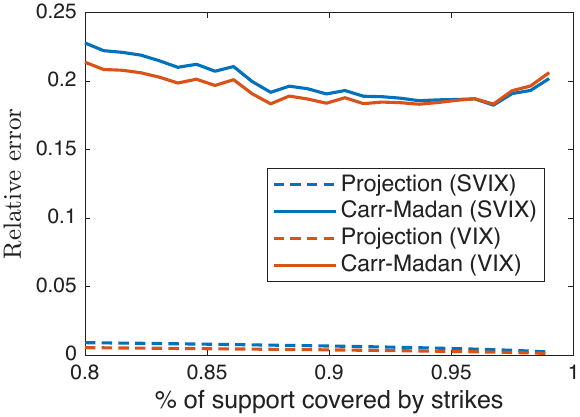}
\caption{Black-Scholes model}
\label{fig:bs_varRange}
\end{subfigure}
\begin{subfigure}{0.48\linewidth}
\includegraphics[width=\linewidth]{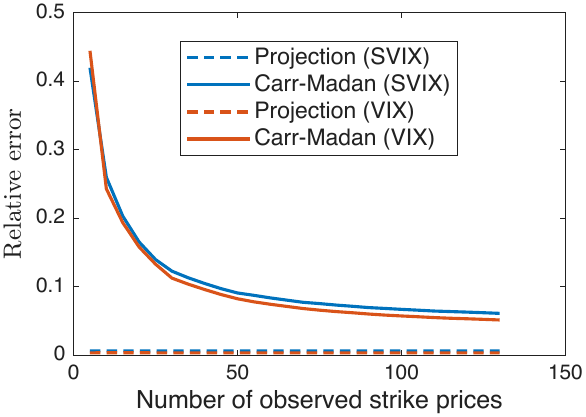}
\caption{Black-Scholes model (equal)}
\label{fig:bs_fixedRange_equal}
\end{subfigure}
\begin{subfigure}{0.48\linewidth}
\includegraphics[width=\linewidth]{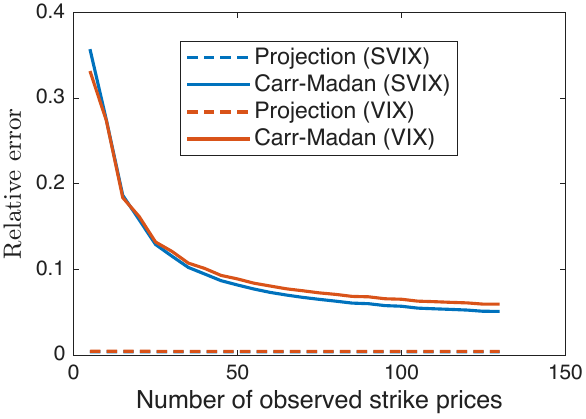}
\caption{AFT model (equal)}
\label{fig:aft_fixedRange_equal}
\end{subfigure}
\caption{\textbf{Relative approximation error}. The figure shows the convergence rate as a function of the number of strikes and as a function of the strike range in the Black-Scholes model (upper panels). The bottom panels show the convergence rate as a function of the number of strikes when they are equally spaced, both in the Black-Scholes model and the AFT model.}\label{fig:mse_robust}
\end{figure}

In this section we consider several changes to the simulation design to assess the robustness of the findings in Section~\ref{sec:uni_proj}. First, instead of the AFT model, we generate option prices from the \citet{BlackScholes1973} model. Because the VG family nests the Black-Scholes density as a limiting case, this design isolates the estimation error when the weighting density is correctly specified. Panels~\ref{fig:bs_fixedRange} and~\ref{fig:bs_varRange} show that projection remains very accurate\footnote{This is not surprising as the bound in Proposition \ref{prop:proj_errorBound} shows that the estimation error should be zero. In simulation, it is not exactly zero because the VG parameters are estimated from options, and estimation noise prevents the VG density to collapse to lognormal.}, and the results are quantitatively similar to those in Section~\ref{sec:uni_proj}. This similarity indicates that the projection's advantage over CM stems from the efficiency of the finite-strike replication rather than from the particular choice of the weighting density. Second, Panels~\ref{fig:bs_fixedRange_equal} and~\ref{fig:aft_fixedRange_equal} report the relative error when the strikes are equally spaced rather than uniformly distributed. Equally spaced strikes improve the accuracy of both estimators, but projection continues to dominate CM by roughly an order of magnitude.

\begin{figure}[htb!]
    \centering
    \begin{subfigure}{0.48\linewidth}
    \includegraphics[width=\linewidth]{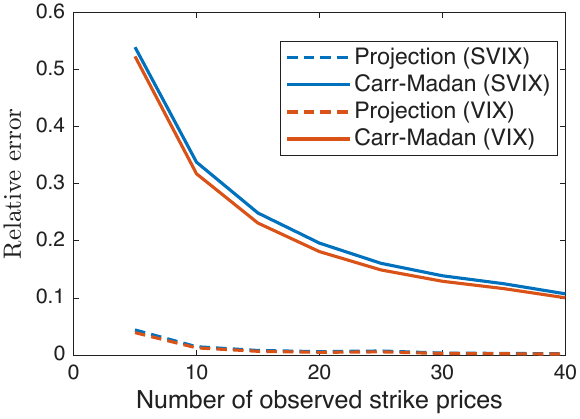}
    \caption{Target zone, number of strikes}
    \label{fig:tzone}
    \end{subfigure}
    \begin{subfigure}{0.48\linewidth}
    \includegraphics[width=\linewidth]{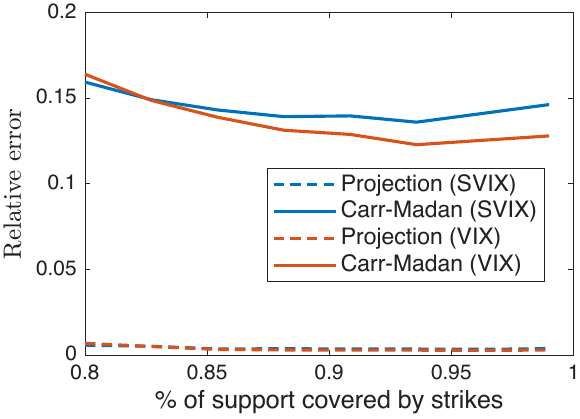}
    \caption{Target zone, strike range}
    \label{fig:tzone_range}
    \end{subfigure}
    \caption{\textbf{Relative approximation error in the target-zone model}. The figure plots the relative error of the projection estimator and the CM approximation for the SVIX and VIX under the bimodal risk-neutral density, as a function of the number of observed strikes (left) and the strike range (right).}
    \label{fig:mse_tzone}
\end{figure}

In addition to the equity-index models above, we consider a bimodal risk-neutral density that captures the target-zone regime of Section~\ref{sec:target_zone}. With small probability investors attach mass to the central bank abandoning the currency floor, which produces a lower crash mode, while the main mode corresponds to the floor being maintained. We model this as $S_T$ being a mixture of two lognormals: with probability $1-\pi$ the floor is abandoned and $S_T$ has forward $F_c = 0.80\,S_0$ (a 20\% devaluation) and volatility $\sigma_c$; with probability $\pi$ the floor holds and $S_T$ has forward $F_h$ and a tighter volatility $\sigma_h < \sigma_c$. The forward $F_h$ is pinned down by the martingale condition $\pi F_h + (1-\pi)F_c = S_0 e^{rT}$. We set $\pi = 0.85$, $\sigma_h = 0.03$, $\sigma_c = 0.07$, $r = 5\%$, and $T = 1$ year. Option prices are the probability-weighted average of the two Black-Scholes prices, $C(K) = \pi\,C^{\mathrm{BS}}(F_h,K,\sigma_h) + (1-\pi)\,C^{\mathrm{BS}}(F_c,K,\sigma_c)$. This bimodal shape makes the VG weighting density far more misspecified than in the AFT model. Panels~\ref{fig:tzone} and~\ref{fig:tzone_range} show that, despite this heavy misspecification, the projection estimator is substantially more accurate than the CM approximation.

\subsubsection{Stochastic Volatility Inspired CM}

An alternative to the widely used CM approach is to estimate the entire risk-neutral distribution from option prices, which avoids the tail-truncation problem inherent in the standard CM approximation. For example, \citet{beason2022} use the stochastic-volatility-inspired (SVI) parametrization of \citet{gatheral2004} to estimate the risk-neutral density, from which moments can be extracted directly. This reduces both the discretization and tail-truncation error, but introduces a parametric fitting bias, since the five-parameter SVI curve need not coincide with the true smile. We implement the SVI method in the simulation by fitting the parametrization to the observed implied volatilities, evaluating the fitted smile on a dense strike grid, converting it back to option prices, and applying CM to the resulting surface. Because this surface spans the full support, it removes the tail truncation of standard CM.

\begin{figure}[htb!]
    \centering
    \begin{subfigure}{0.48\linewidth}
    \includegraphics[width=\linewidth]{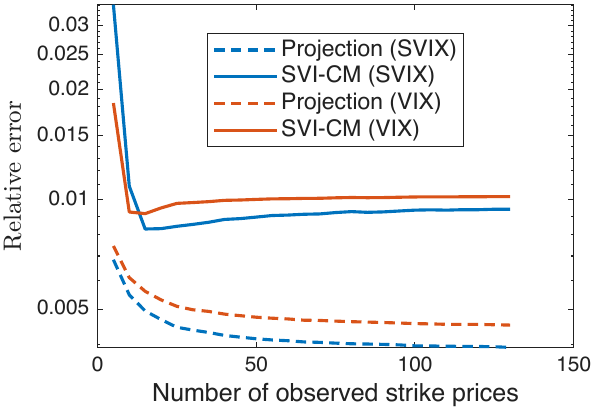}
    \caption{AFT, number of strikes}
    \end{subfigure}
    \begin{subfigure}{0.48\linewidth}
    \includegraphics[width=\linewidth]{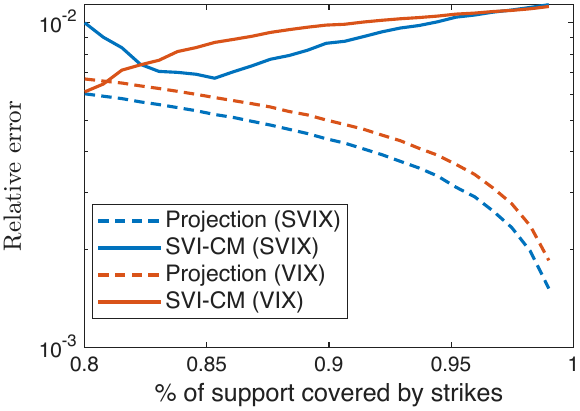}
    \caption{AFT, strike range}
    \end{subfigure}
    \begin{subfigure}{0.48\linewidth}
    \includegraphics[width=\linewidth]{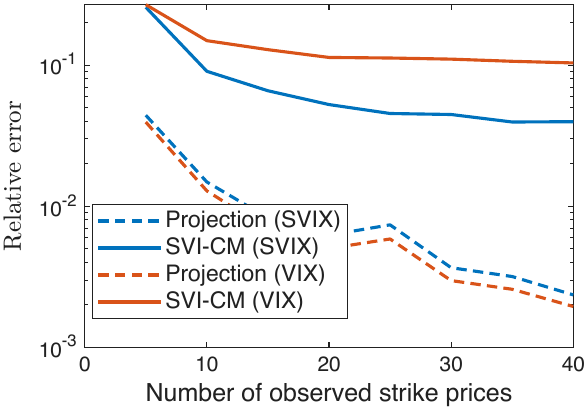}
    \caption{Target zone, number of strikes}
    \end{subfigure}
    \begin{subfigure}{0.48\linewidth}
    \includegraphics[width=\linewidth]{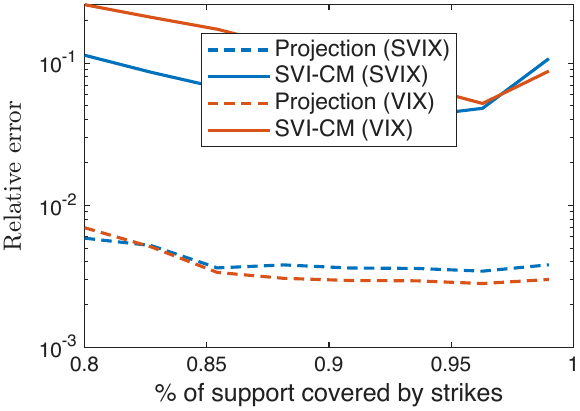}
    \caption{Target zone, strike range}
    \end{subfigure}
    \caption{\textbf{Projection versus SVI-CM}. Relative error (log scale) of the projection estimator and the SVI-CM benchmark for the SVIX and VIX, as a function of the number of observed strikes (left) and the strike range (right), for the AFT model (top) and the target-zone model (bottom).}
    \label{fig:mse_svicm}
\end{figure}

Figure~\ref{fig:mse_svicm} compares the projection estimator with SVI-CM. On the AFT model (top panels) both methods are accurate, but the projection error keeps declining with the number of strikes and the strike range, whereas SVI-CM plateaus at roughly $1\%$: its error is limited by the SVI fitting bias rather than by the data, so additional strikes no longer help. The projection is therefore several times more accurate. On the target-zone model (bottom panels) the fitting bias becomes severe, because a unimodal five-parameter smile cannot represent the bimodal density; SVI-CM plateaus at several percent (and even deteriorates as strikes are added), while the projection continues to converge and dominates by an order of magnitude. In both cases the projection avoids the parametric bias of a fixed smile, and the gap is widest precisely on the bimodal density where that bias is structural.

\FloatBarrier

\section{Additional information on SNB announcement}

\begin{figure}[!htb]
\centering
\includegraphics[width=0.6\linewidth]{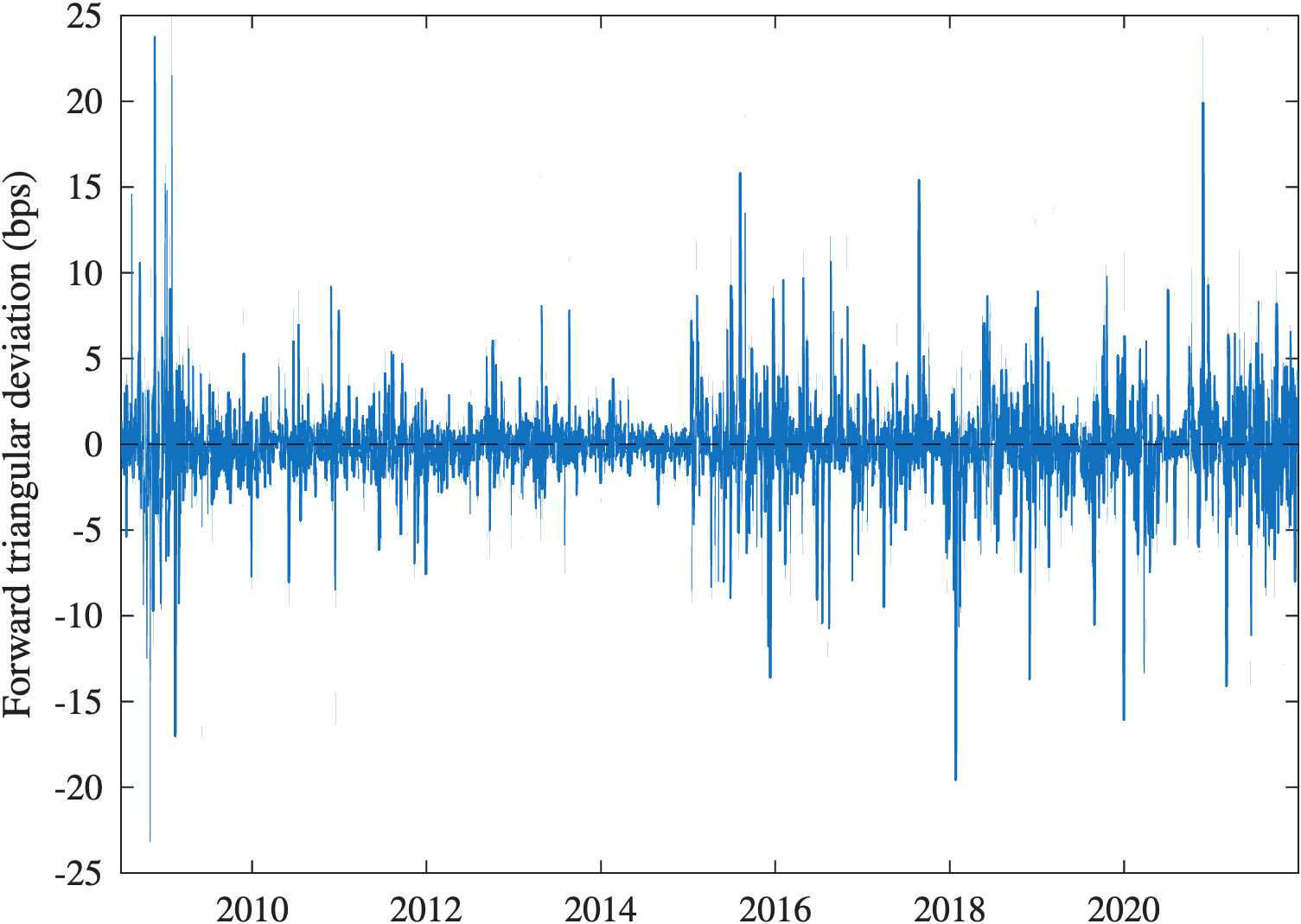}
\caption{\textbf{Triangular no-arbitrage diagnostic}. The figure plots the daily deviation between the observed EUR/USD forward rate and its implied
value from the EUR/CHF and USD/CHF forwards, i.e., $(F_{3,\ttt} - F_{1,\ttt} / F_{2,\ttt}) / F_{3,\ttt}$, expressed in basis points.}                 
\label{fig:triangle_deviation}
\end{figure} 

\begin{figure}[!htb]
\centering
\includegraphics[width=0.5\linewidth]{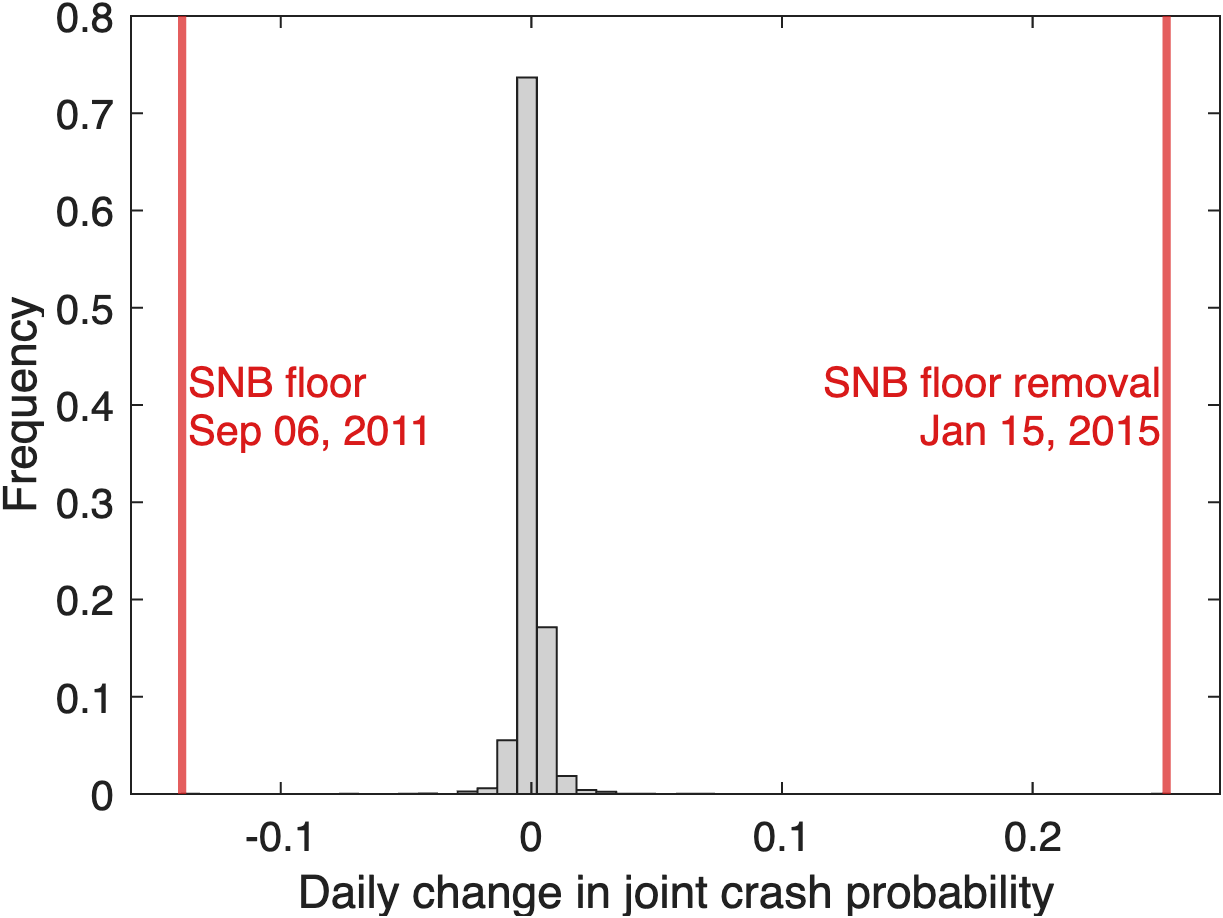}
\caption{\textbf{Empirical distribution of SNB event-day changes.} The histogram plots daily changes in the joint risk-neutral crash probability of EUR/CHF and USD/CHF over the sample (July 2008--December 2022). Red vertical lines mark the event-day changes. Both events lie at the extremes of the empirical distribution.} 
\label{fig:placebo}
\end{figure}

\end{document}